\documentclass[preprint,12pt,authoryear]{elsarticle}

\usepackage{marcstyle, verbatim}
\usepackage{xr}

\journal{Artificial Intelligence Journal}

\begin{document}

\begin{frontmatter}

\title{Intention as Commitment toward Time\footnote{This is a pre-print of the article published in Artifical Intelligence Journal Volume 283, June 2020: \url{https://www.sciencedirect.com/science/article/pii/S0004370220300308}}}

\author{Marc van Zee}
\ead{marcvanzee@gmail.com}
\address{Google Research, Brain Team}
\author{Dragan Doder}
\address{Utrecht University, the Netherlands}
\author{Leendert van der Torre}
\address{University of Luxembourg, Luxembourg}
\author{Mehdi Dastani}
\address{Utrecht University, the Netherlands}
\author{Thomas Icard}
\address{Stanford University, CA, USA}
\author{Eric Pacuit}
\address{University of Maryland, MD, USA}

\begin{abstract}
In this paper we address the interplay among intention, time, and belief in dynamic environments. The first contribution is a logic for reasoning about intention, time and belief, in which assumptions of intentions are represented by preconditions of intended actions. Intentions and beliefs are coherent as long as these assumptions are not violated, i.e. as long as intended actions can be performed such that their preconditions hold as well. The second contribution is the formalization of what-if scenarios: what happens with intentions and beliefs if a new (possibly conflicting) intention is adopted, or a new fact is learned? An agent is committed to its intended actions as long as its belief-intention database is coherent. We conceptualize intention as commitment toward time and we develop AGM-based postulates for the iterated revision of belief-intention databases, and we prove a Katsuno-Mendelzon-style representation theorem.
\end{abstract}

\begin{keyword}
Intention \sep BDI logic \sep Belief revision
\end{keyword}

\end{frontmatter}

\section{Introduction}

Sometime in the near future you will tell one of your household robots: ``Bobby, get me some beer from the store.'' Bobby confirms your request, but when it is walking to the store it encounters your partner, who says: ``Bobby, our house is a mess, go home and clean.'' Bobby returns home, takes the mop out of the closet and prepares to start cleaning. Just as it is ready to make its first swipe, one of your friends walks in asking: ``Bobby, it has been snowing outside, could you clean my car?'' In the meantime, you are getting increasingly frustrated by your lack of beer, and when you see Bobby in the kitchen you shout: ``You still didn't get my beers? Go get them immediately!'' After letting Bobby run around for a few days you complain to the manufacturer that your robots do not finish any of the tasks they start with.

After the manufacturer updates the software on your robots, he happily tells you the new version will no longer cause the robots to drop their commitments so quickly. Delighted, you exclaim to your favorite robot: ``Bobby, I have some friends coming over tonight, get some ingredients and cook dinner so we can eat at 7pm tonight''. Realizing the shop closes only at 5pm, Bobby 2.0 delays going to the grocery story until the very last moment, hereby keeping its schedule free for other possible tasks. Unexpectedly, on its way to the grocery store Bobby is delayed by an open bridge and arrives at the store minutes after closing time. It returns to your home empty-handed, leaving you and your friends hungry. It turns out Bobby 2.0 is delaying every task until just before the deadline. Since tasks often have unexpected delays, this means that most of the tasks are finished too late, or not at all. Frustrated, you call the manufacturer again, complaining that Bobby is procrastinating its commitments.

After uploading yet another version of the software, the manufacturer ensures you that your robots will no longer postpone fulfilling their commitments, nor will they drop them quickly. At this time you are rather skeptical, but still you ask: ``Bobby, I'd like to have dinner tonight again. Please buy the ingredients at 2pm, and cook for me at 6pm''. Around 12pm, your partner again realizes the house hasn't been cleaned properly for a long time, and therefore tells Bobby to clean the house intensively. At 6pm, you sit at your kitchen table wondering where dinner is, so you call Bobby asking what happened. Bobby explains it had to clean the house at 12pm, which took three hours, so it couldn't fulfill its commitment to go shopping at 2pm.

Disappointed, you return your robots to the manufacturer where they are dismantled.

\subsection{Commitment toward Time}

The story of Bobby the robot is inspired by the example of Willie the robot, due to~\cite{Cohen1990}. In their highly-cited article entitled ``Intention = Choice + Commitment,'' Cohen and Levesque specify the rational balance of autonomous agents, focusing on the role that intention plays in maintaining this balance. Their approach has typified much subsequent research on belief-desire-intention (BDI) logics, namely to understand and study intention as commitment in relation to goals, desires and beliefs. For instance,~\cite{Rao1991} define various commitment strategies of agents, such as blindly-minded, single-minded, and open-minded commitment strategies. A popular approach to formalize the BDI theory is to specify a temporal logic such as linear-time logic (LTL) or computational-tree logic (CTL*) and use modal operators for mental states and use expressions of the form ``some time in the future,'' or ''in the next time moment'' to reason about the temporal behavior.

In approaches following the ideas of Cohen and Levesque by, for example, ~\cite{Rao1991} or \cite{Meyer1999}, intention is typically defined as commitment toward goals. However, being committed toward a goal is only one dimension of a commitment. Another important dimension is \emph{commitment toward time}, i.e., \emph{when} these commitments will be fulfilled. In the example above, Bobby the robot is an \emph{online system}; it is receiving orders, forming plans, scheduling tasks, and executing them, all in parallel and in real-time. Bobby plans its commitments at appropriate moments, making sure the different plans do not overlap or are incompatible, while it at the same time may receive new instructions from users. 


Even in a simplified setting where we only consider commitments toward time, already non-trivial complications arise. Commitments can play the role of \emph{assumptions} on which further plans are based. The three versions of Bobby the household robot behave differently concerning their commitment to time. The first version may use a stack-like data structure in order to execute its tasks: it adds each new commitments on the stack, and then executes the tasks on top of its stack. The second version may use a queue, and moreover delays executing these tasks until the very last moment. Finally, the last version is able to schedule its commitments in time, but it cannot reschedule them. None of these versions seem to be able to fulfil commitments in a desirable way. Instead, Bobby should be able to make plans, store the commitments and use them as assumptions in further planning.

\subsection{Methodology: the database perspective}

\cite{Shoham2009} views the problem of intention revision as a database management problem. In particular, he introduced the conceptual underpinnings of the distinction between a reasoner such as a planner, and the involved belief-intention databases. At any given moment, an agent must keep track of a number of facts about the current situation. This includes beliefs about the current state, beliefs about possible future states, beliefs about which actions are available now and in the future, and also beliefs about plans at future moments. It is important that all of this information be jointly consistent at any given moment and furthermore can be modified as needed while maintaining consistency.

\begin{figure}[ht]
\centering
\includegraphics[scale=0.33]{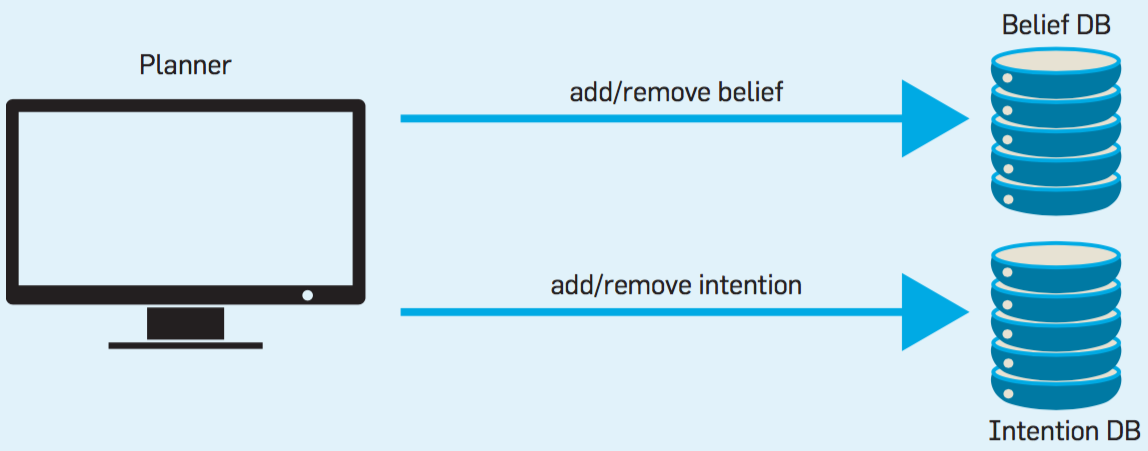}
\caption{The database perspective}
\label{fig:dbperspective}
\end{figure}

In this article we introduce a logic that formally models such a ``database'', as visualised in Figure~\ref{fig:dbperspective}. \emph{Consistency} in this logic is meant to represent not only that the agent's beliefs are consistent and the agent's future plans are consistent, but also that the agent's beliefs and intentions together form a \emph{coherent} picture of what may happen, and of how the agent's own actions will play a role in what happens. Our primary contribution in this article is to focus also on how the database is to be modified, and in the process to provide a clear picture of how intentions and beliefs relate.

In this paper we distinguish two kinds of outputs produced in Figure~\ref{fig:dbperspective}:

\begin{description}
\item[Belief] A belief is added to the Belief database. If the new belief is inconsistent with the existing beliefs, then these beliefs will have to be revised to accommodate it. Our account of belief revision follows the classical AGM postulates~\citep{AGM1985}, which we then generalise to iterated revision. The goal is thus to give general conditions on revision with new information that the agent has already committed to incorporating.
\item[Intention] An intention is added to the Intention database. Here we focus on future directed intentions, understood as time-labeled actions pairs $(a,t)$ that might make up a plan. Analogously to belief revision, it is assumed the agent has already committed to a new intention, so it must be accommodated by any means short of revising beliefs. The force of the theory is in restricting how this can be accomplished. To be more precise, we purport to model an intelligent database, which receives instructions from some reasoner (e.g. a STRIPS-like planner) that is itself engaged in some form of practical reasoning. The job of the database is to maintain consistency and coherence among intentions and beliefs.
\end{description}

This description, however, obscures some important subtleties in the interaction between beliefs and intentions. The following will serve as a running example that we will use frequently throughout the article.

\begin{example}[Running Example]
\label{expl:1}
Bobby the household robot has the goal to buy groceries in the morning and to buy cleaning equipment in the afternoon. However, it only has sufficient budget to do either one of the two, but not both of them. Bobby thus believes it is possible to buy cleaning equipment and to buy food, but it also believes it is impossible to buy both. If Bobby decides to buy food, then it will cook food in the afternoon.
\end{example}

Upon adopting the intention to buy food, Bobby will come to have new beliefs based on the predicted success of this intention, e.g., that he will be able to cook afterwards. These further beliefs are important when planning when or how to cook. The intention is also supported by the absence of certain beliefs. It would be irrational for Bobby to adopt the intention to buy food if it believed it did not have sufficient money. Likewise, even if it originally believed it has sufficient money, upon learning it does not, the intention to cook food should be dropped. Yet, when dropping this intention, other beliefs, such as that he will be able to cook, have to be dropped as well, which may in turn force other intentions and beliefs to be dropped. And so on. 

\subsection{Weak beliefs}

Regular beliefs concern the world as it is, independent of an agent's future plans, but including what sequences of actions are possible. Thus, additionally to atomic facts, an agent may have beliefs about what the preconditions and postconditions of actions are, and about which sequences of actions are jointly possible.

We distinguish {\em weak} beliefs depending on intentions, from beliefs that do not (\emph{strong} beliefs). In other words, as usual we assume that postconditions lead to weak beliefs. Since we are not considering actions whose effects are uncertain or dependent on the conditions that obtain when the action is taken, if an action is planned the planner believes whatever follows from it. 
\begin{quote}
Postconditions-as-weak-beliefs: \emph{If an agent intends to take an action, it weakly believes that its postconditions will hold.}
\end{quote}

A key element in our approach is that we treat preconditions of actions as \emph{assumptions}. This leads to an asymmetry on beliefs about preconditions and postconditions of actions. 
However, for preconditions we adopt Shoham's weaker requirement, which we call here {\em preconditions-as-assumptions}:
\begin{quote}
Preconditions-as-assumptions: \emph{If you intend to take an action you cannot believe that its preconditions do not hold}.  \citep{Shoham2009}
\end{quote}
Preconditions of actions are treated as \emph{assumptions}, in the sense that an agent forms intentions under the assumption that these preconditions will be made true somewhere in the future. Treating preconditions as assumptions is a good fit with how real-time planning agents operate, because intended actions may be added as long as they are consistent with beliefs, and once they are accepted they can be used as additional assumptions to further plans~\citep{Shoham2009}. For instance, the household robot Bobby has the intention to cook dinner tonight, which is based on the intention to buy the ingredients, which is in turn based on the assumption that your friends will attend tonight, even if it does not know this for sure yet. It is only when Bobby finds out your friends are not coming, it should drop its intentions. This preconditions-as-assumptions requirement is sometimes called \emph{strong consistency}, and is weaker than Bratman's \emph{means-end coherence} requirement~\citep{bratman1987} (see Section~\ref{sect:cohen-levesque} for a more detailed discussion).

The preconditions-as-assumptions requirement has various consequences, both for the logic and for the theory change operators introduced in this paper. For example, suppose Bobby intends to buy food at moment 0, and in addition to buy cleaning equipment at moment 2. This may be coherent despite Bobby's belief that it does not have enough money for both, thanks to an action of robbing a bank that it can perform at 1. If this is the only model where Bobby can buy food at moment 0 and equipment at moment 2, then the resulting belief-intention database may entail $do(rob)_1$, even if Bobby would like to avoid robbing a bank by any means. We will consider these potential complications in Section~\ref{sect:examples} after we have introduced the formal machinery.

\subsection{Results}
We develop a branching-time temporal logic, called \logicname{} (\logic{}) in order to formalize beliefs. The language of this logic contains formulas to reason about possibility, preconditions, postconditions, and the execution of actions. The semantics of this logic is close to CTL*, and in this way follows the tradition of BDI logics of~\cite{Rao1991}. An important difference is that we do not use modal operators to reason about time, but we use explicit time points. We axiomatize this logic and proof that the axiomatization is sound and strongly complete with respect to our semantics.

We separate strong beliefs from weak beliefs as described above. Strong beliefs are beliefs that occur in the belief database, and they are independent of intentions. Weak beliefs are obtained from strong beliefs by adding intentions to the strong beliefs, and everything that follows from that. We then formalize a \emph{coherence condition} on the beliefs and intentions. This condition states that the agent weakly believes it is possible to jointly perform all of its intended actions. 

The main technical result of the paper is that we develop a set of postulates for the joint revision of belief and intentions, and that we prove a variation of the~\cite{Katsuno1991} representation theorem. To this end, we define a revision operator that revises beliefs up to a specific time point. We show that this leads to models of system behaviors which can be finitely generated, i.e. be characterized by a single formula. We also proof various representation theorems for iterated revision of belief and intention. 

\subsection{Map of the paper}

The structure of this paper is as follows. In Section~\ref{sect:logic} we introduce and motivate our logic PAL, we axiomatize it and we proof completeness. In Section~\ref{sect:intentions} we separate strong beliefs from weak beliefs, and we formalize a coherence condition on belief and intention. In Section~\ref{sect:revision} we study single-step revision of beliefs and intentions, and we study iterated revision in Section~\ref{sect:iterated-revision}. We discuss related work in both philosophy of mind and AI in Section~\ref{sect:relatedwork}, and we provide directions for future work in Section~\ref{sect:futurework}. All the proofs can be found in the appendix.
\section{Parameterized-time Action Logic (PAL): a logic for belief and intention}
\label{sect:logic}

Our aim in this section is to develop a logical system that represents an agent's beliefs about the current moment and future moment and actions that may be performed. Table~\ref{table:symbols} contains all the most important symbols used in this article and their meaning.

\begin{table}[ht!]
\footnotesize
\begin{tabular}{l|l||l|l}
\multicolumn{2}{c||}{\textbf{Logic} (Section~\ref{sect:logic},~\ref{sect:intentions})} & \multicolumn{2}{c}{\textbf{Revision} (Section~\ref{sect:revision},~\ref{sect:iterated-revision})}\\
\hline
Symbol&Meaning&Symbol&Meaning\\
\hline
$\lang{}$ & Language of PAL&
$\br$ & AGM revision revision\\
$\prop{}$ & Set of atoms&
$\mathbb{X}^{|t}$ & Some set $\mathbb{X}$ bounded up to $t$\\
$\act{}$ & Set of actions & 
$\br_t$ & Strong belief revision function\\
$\actseq$ & Finite action sequence&
$\agri$ & Intention revision function\\
$\chi$ & Atomic propositions &
$\agr$ & Belief-intention revision function\\
$T$ & Semantic tree $(S, R, v, act)$&
$\restp{}$ & $t$-bounded path\\
$S$ & Set of states in a tree &
$m^{|t}$ & $t$-bounded model\\
$R$ & Accessibility relation for tree &
$\sel$ & Selection function\\
$v$ & Valuation function &
$\lept$& Total pre-order over models\\
$act$ & Action function &
$\epsilon$ & Empty intention\\
$\pi$ & Path $(s_0,s_1,\ldots)$ in a tree&
$\Psi$ & Epistemic state\\
$\pi_t$ & The $t$'th state of the path $\pi$ &
$\br_t$ & Epistemic revision function\\
$m=(T,\pi)$ & Model in PAL&
$(\Psi, I)$ & Epistemic belief-intention database\\
$M$ & Set of models in PAL&
$\agr$ & Iterated revision function\\
$\mods$ & Set of all models in PAL&
$\kappa_t$ & Spohn ranking function\\
$Mod(\varphi)$ & Set of all models of $\varphi$ &
$Bel(\kappa_t)$ & Accepted propositions\\
$\sbel$ & Set of all strong beliefs&
$\spohn$ & Extended Spohn revision\\
$SB$ & Set of strong beliefs& $\lang{}^{|t}$
& Restricted language \\
$Cn(SB)$ & Belief database& 
&\\
$\mathbb{BD}$ & Set of all belief databases& 
&\\
$\msbset{}$ & Set of models of strong beliefs&
&\\
$\msbsets{}$ & Set of all MSBs&
&\\
$(a,t)$ & Intention&
&\\
$\mathbb{I}$& Set of all intentions&
&\\
$I$ & Intention database&
&\\
$\mathbb{ID}$& Set of all intention databases&
&\\
$(SB,I)$ & Belief-intention database&
&\\
$\mathbb{BI}$ & Set of all BI databases&
&\\
$WB(SB,I)$ & Weak beliefs&
&\\
$Ext(M^{|t})$& Set of extensions of $M^{|t}$&
&\\
$\mathbb{EBI}$& Set of all epistemic databases&
&\\
$Cohere(I)$ & Coherence formula&
&\\

\end{tabular}
\caption{Symbols used in this paper and their meaning}
\label{table:symbols}
\end{table}

\subsection{Syntax}

Beliefs are represented by the formal language \lang{}. The language uses the set $\prop{}$ containing all atoms which are true or false in a time instance (state). We also consider formulas $do(a)_t$ which will semantically be defined as a transition from $t$ to $t+1$ using action $a$.

\begin{definition}[Language]
\label{def:language}
Let 
\begin{itemize}
\item $\act{}=\{a,b,c,\ldots\}$ be a finite set of deterministic primitive actions;
\item $\prop{}=\{p,q,r,\ldots\}\cup \{pre(\actseq), post(a)\}$ be a finite set of propositions where $\overline{a}=(a_1,a_2,\ldots)$ is a non-empty action sequence and $\{a,a_1,a_2,\ldots\}\subseteq Act$ are actions. We denote atomic propositions with $\chi$.
\end{itemize}
The sets \prop{} and \act{} are disjoint. The language \lang{} is inductively defined by the following BNF grammar:
\begin{align*}
\varphi ::= \chi_t\mid do(a)_t\mid\Box_t\varphi\mid\varphi\wedge\varphi\mid\neg\varphi,
\end{align*}
with $\chi\in \prop{}, a\in \act{}$, and $t\in\timeflow{}$. Furthermore, we abbreviate $\neg \Box_t\neg$ with $\Diamond_t$, and we define $\perp\equiv p_0\wedge \neg p_0$ and $\top\equiv\neg\perp$.
\end{definition}

Intuitively, $p_t$ means that the atomic formula $p$ is true at time $t$, $do(a)_t$ means that action $a$ is executed at time $t$. To every finite sequence of actions $\overline{a}=(a_1,a_2,\ldots)$ and every time point $t$ we associate a formula $pre(\actseq)_t$, which is understood as the precondition for subsequently executing actions $a_1, a_2, \ldots$ at time $t$. Note pre and postconditions are represented as particular propositions, and not abbreviations of other propositional formulas. 

We define preconditions for sequences of actions explicitly, because it is difficult to define the precondition for a sequence of actions using only preconditions for individual actions. This can already be witnessed in our running example: Bobby believes the preconditions to buy food and cleaning equipment are true separately, but still does not believe the precondition for performing both actions subsequently is true. These type of formulas will play a crucial role when we formalize the coherence condition in Section~\ref{sect:intentions}.

The modal operator $\Box_t$ is interpreted as necessity, indexed with a time point $t$. Intuitively, a formula of the form $\Box_t p_{t+1}$ means ``it is necessary at time $t$ that $p$ is true at time $t+1$. The other boolean connectives are defined as usual.

\begin{example}[Running example (Ctd.)]
Let our language contain:
\begin{itemize}
\item $\act{}=\{food, equip, cook, nop\}$, where $food$ is the action ``buy food'', $equip$ is the action ``buy cleaning equipment'', and $cook$ is the action ``cook'', and $nop$ is the special ``no operation'' action,\footnote{In many of our examples it is useful include an action that does not do anything. In that case we use the special action $nop$ and the formulas $pre(nop)\equiv \top$ and $post(nop)\equiv \top$.}
\item $\prop{}$ consists of of $pre$ and $post$ statements with the actions in $Act$, such as $pre(food), pre(food, equip), pre(food, nop, food), post(food), post(nop)$.
\end{itemize}
Some examples of formulas in the language generated from $\act$ and $\prop$ are:
\begin{itemize}
\item $pre(food)_0\wedge do(nop)_0\wedge do(equip)_1$ (the precondition to buy food at time 0 is true, no action is performed at time 0, and Bobby buys cleaning equipment at time 1),
\item $\Diamond_0 (do(food)_0\wedge \neg do(cook)_1)$ (it is possible at time 0 to buy food at time 0 and not to cook at time 1),
\item $\Diamond_0 do(food)_0 \wedge \Diamond_0 do(equip)_1 \wedge \neg \Diamond_0 (do(food)_0\wedge do(equip)_1)$ (it is possible to buy food at time 0 and it is possible to buy equipment at time 1, but it is not possible to do both),
\item $pre(food,cook)_0$ (the precondition to buy food at time 0 and then cook at time 1 is true).
\item $do(equip)_1$ (Bobby will buy cleaning equipment at time 1),
\item $\Diamond_0\neg \Diamond_1 do(cook)_1$ (it is possible at time 0 that it is not possible at time 1 to cook),
\item $\bigvee_{x\in \act{}} pre(food,x,equip)_0$ (the precondition to buy food at time 0 and to buy equipment at time 2 is true, if a right action is performed at time 1).
\end{itemize}
\end{example}

Note that in this article, we use $pre(a)_t$ to denote the proposition that is the precondition of action $a$, but this is simply a naming convention. For instance, in the example above, we denote the precondition for action $food$ as $pre(food)$, while we explain in natural language that this means the agent has sufficient money to buy food. Thus, we could equally have written $hasEnoughMoneyForFood$ instead of $pre(food)$, but we chose to keep the former notation, to show the interplay between preconditions, actions, and postconditions in our examples more clearly. One may choose to define pre/postconditions as abbreviations of state propositions (possibly with time-indices), but since the internal structure of pre/postconditions is not the focus of our paper, we define them simply as primitive objects (first-class citizens).

The following definition collects all formulas up to some time $t$ in a set $Past(t)$, which will turn out to be convenient when we axiomatize our logic. A formula of the form $do(a)_t$ will be semantically defined as a transition from $t$ to $t+1$. Therefore it does not belong to the formulas true up to time $t$ if it does not fall under the scope of a modality. We will make this more precise when we introduce the semantics in the next subsection.

\begin{definition}
\label{def:past}
$Past(t)$ is the set of all formulas from $\lang{}$ generated by boolean combinations of $p_{t'}, pre(\actseq{})_{t'},post(a)_{t'}$, $\Box_{t'}\varphi$, and $do(a)_{t'-1}$ where $t'\le t$ and $\varphi$ is some formula from $\lang{}$.
\end{definition}

Note that $\varphi$ in the definition above can contain formulas indexed by time points greater than $t$. For instance, $do(a)_1\in Past(2), \Box_2 \Diamond_5 pre(a)_6\in Past(2)$, but $do(a)_3\not\in Past(1)$ and $do(a)_1\vee p_1\not\in Past(1)$. 

\subsection{Semantics}

The semantics of our logic is similar to CTL*~\citep{Reynolds2002}, namely a tree structure containing nodes and edges connecting the nodes. A tree can equivalently be seen as an unfolded transition system, thereby representing all the possible runs through it. We choose to represent our semantics using trees because it simplifies the completeness proofs. See~\cite{Reynolds2002} for an overview of different kinds of semantics and conceptual underpinnings. 

With each natural number $i\in\timeflow{}$ we associate a set of states $S_i$ such that all these sets are disjoint. We then define the accessibility relation between states such that it generates an infinite, single tree.

\begin{definition}[Tree] 
\label{def:tree:frame}
A tree is quadruple $T=\tree{}$ where
\begin{itemize}
\item $S=\bigcup_{n\in\timeflow{}}S_{n}$ is a set of states, such that each $S_t$ is the set of states at time $t$, $S_i\cap S_j=\emptyset$ for $i\not= j$; 
\item $R\subseteq \bigcup_{n\in\timeflow{}} S_n\times S_{n+1}$ is an accessibility relation that is serial, linearly ordered in the past and connected (so $S_0$ is a singleton); 
\item $\valp:S\rightarrow 2^{\prop{}}$ is a valuation function from states to sets of propositions; 
\item $\vala:R\rightarrow \act{}$ is a function assigning actions to elements of the accessibility relation, such that actions are deterministic, i.e. if $\vala((s,s'))=\vala((s,s''))$, then $s'=s''$.
\end{itemize}
\end{definition}

We evaluate formulas on a path in a tree. A path is a sequence of states in a tree, connected by the accessibility relation $R$.

\begin{definition}[Path] 
\label{def:path}
Given a tree $T=\tree{}$, a \emph{path} $\pi=(s_0,s_1,\ldots)$ in $T$ is a sequence of states such that $(s_t,s_{t+1})\in R$. We write $\pi_t$ to refer to the $t$'th state of the path $\pi$. We use elements of the path as arguments for the valuation function and the action function:
\begin{itemize}
\item $\valp(\pi_t)$ are the propositions true on path $\pi$ at time $t$;
\item $\vala((\pi_t,\pi_{t+1}))$ is the next action on path $\pi$ at time $t$. We abbreviate $\vala((\pi_t,\pi_{t+1}))$ with $\vala(\pi, t)$, since $\pi_{t+1}$ is uniquely determined by the action.
\end{itemize}
We identify $T$ with the set of paths in $T$, and we write $\pi\in T$ to denote that a path $\pi$ exists in the tree $T$.
\end{definition}

Intuitively, $v(\pi_t)$ are the propositions true at time $t$ on path $\pi$, and $\vala(\pi,t)$ is the next action $a$ on the path. We next define an equivalence relation $\sim_t$ on paths, which is used to give semantics to the modal operator.

\begin{definition}[Path equivalence] 
\label{def:path:equivalence}
Two paths $\pi$ and $\pi'$ are equivalent up to time $t$, denoted $\pi\sim_t\pi'$, if and only if they contain the same states up to and including time $t$, i.e. \begin{align*}
\pi\sim_t\pi' \text{ iff }&\fa{t'\le t}{ \valp{}(\pi_{t'})=\valp{}(\pi'_{t'})} \text{ and }\\
&\fa{t'<t}{\vala(\pi, t')=\vala(\pi', t')}.
\end{align*}
\end{definition}

Formulas in \logic{} are evaluated on a path. Therefore, a model for a formula is pair consisting of a tree and a path in this tree. This, together with some additional constraints related to the pre- and post-conditions of actions, is our definition of a \emph{model}.

\begin{definition}[Model]
\label{def:model}
A \emph{model} is a pair $(T,\pi)$ with $T=\tree{}$ such that for all $\pi\in T$ the following holds:
\begin{enumerate}
\item If $\vala(\pi, t)=a$, then $post(a)\in \valp(\pi_{t+1})$;
\item If $pre(a)\in v(\pi_t)$, then there is some $\pi'$ in $T$ with $\pi\sim_t\pi'$ and $\vala(\pi', t)=a$;
\item If $pre(a, \overline{b})\in v(\pi_t)$, then there is some $\pi'$ in $T$ with $\pi\sim_t\pi'$, $\vala(\pi', t)=a$, and $pre(\overline{b})\in v(\pi'_{t+1})$;
\item If $pre(\actseq{},b)\in v(\pi_t)$, then $pre(\actseq{})\in v(\pi_t)$,
\end{enumerate}

We refer to models of \logic{} with $m_1,m_2,\ldots$, we refer to sets of models with $M_1,M_2,\ldots$, and we refer to the set of all models with \mods{}. 
\end{definition}

\begin{remark}
\label{remark:first}
In our semantics, preconditions are sufficient conditions for actions to be possible, but they are not necessary. Alternatively, one may strengthen this by changing the Condition 2 of Definition~\ref{def:model} from an ``if'' to an ``if and only if''. However, our choice is an implementation of Shoham's idea of ``opportunistic planning'': a planner may form intentions, even though at the moment of planning it may not be clear whether preconditions are true~\citep{Shoham2009}.
\end{remark}

The conditions on models are there to formalize the consistency conditions from the introduction. Condition 1 is straightforward: we simply expect postconditions to hold in a state after an action has been executed. Condition 2 and 3 put a weaker requirement on the preconditions for actions: If the precondition holds, then there is \emph{some} path in which the action is executed. This is part of the weaker requirement Shoham puts on preconditions. The opposite direction, stating that preconditions are necessary for executing actions, will be formalized with a coherence condition on beliefs and intentions in Section~\ref{sect:coherenceconditions}. Condition 4 of a model simply ensures that if the precondition of a sequence of action is true in a state, then the precondition for any subsequence by removing actions from the end of the sequence is also true in that state.

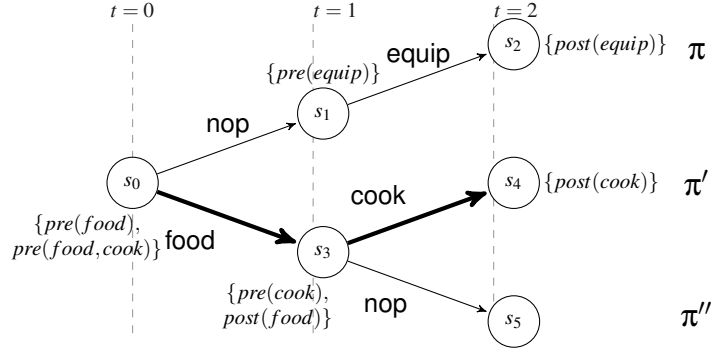
\begin{figure}[h!]
\centering
\begin{tikzpicture}[->,>=stealth',shorten >=1pt,auto,node distance=1cm,
  thin,
  main node/.style={circle,draw,fill=white,font=\sffamily\scriptsize\bfseries}, 
  every label/.style={font=\scriptsize},
  time node/.style={font=\scriptsize,text=black!90},
  test line/.style={gray!80, dashed,-}]
  
  \draw[test line] (-0.5,-2) -- (-0.5, 2.2);
  \draw[test line] (1.9,-2) -- (1.9, 2.2);
  \draw[test line] (4.3,-2) -- (4.3, 2.2);
  
  \node[time node] (t0) at (-0.5,2.3) {$t=0$};
  \node[time node] (t1) at (2.2,2.3) {$t=1$};
  \node[time node] (t2) at (4.6,2.3) {$t=2$};
  
  \node[main node] (s0) at (-0.5,0) {$s_0$};
  \node[position=-130:0mm from s0,align=center,font=\scriptsize] (labels4) {$\{pre(food),$\\$pre(food,cook)\}$};  
  \node[main node,label={[label distance=-0.1cm]90:$\{pre(equip)\}$},position=20:20mm from s0] (s1) {$s_1$};
   \node[main node,label={[label distance=-0.1cm]0:$\{post(equip)\}$},position=20:20mm from s1] (s2) {$s_2$};
  \node[main node, position=-20:20mm from s0] (s3) {$s_3$};
  \node[position=-130:0mm from s3,align=center,font=\scriptsize] (labels3) {$\{pre(cook),$\\$post(food)\}$};  
  
  \node[main node,label={[label distance=-0.1cm]00:$\{post(cook)\}$},position=20:20mm from s3] (s4) {$s_4$};
  \node[main node,label={[label distance=-0.1cm]0:},position=-20:20mm from s3] (s5) {$s_5$};
  
    \node[style={circle}] (pi1) at (7,1.75) {$\pi$};
   \node[style={circle}] (pi2) at (7,0) {$\pi'$};
   \node[style={circle}] (pi3) at (7,-1.75) {$\pi''$};
 
  \path[every node/.style={font=\sffamily\small}]
    (s0) edge[draw=black] node [above] {nop} (s1)
    (s1) edge[draw=black] node [above] {equip} (s2)
    (s0) edge[draw=black, line width=2, ultra thick] node [below left] {food} (s3)
    (s3) edge[draw=black, line width=2, ultra thick] node [above left] {cook} (s4)
    (s3) edge[draw=black] node [below left] {nop} (s5);
\end{tikzpicture}
\caption{Example PAL Model $(T,\pi')$ from $t=0$ to $t=2$.}
\label{fig:example_pal_model}
\end{figure}

\begin{example}[Running example (Ctd.)]
Consider the partial PAL model $(T,\pi')$ of the beliefs of Bobby the household robot from time 0 to time 2 in Figure~\ref{fig:example_pal_model}, where the thick path represents the actual path.  We provide some examples of the conditions of our model (Definition~\ref{def:model}):
\begin{itemize}
\item since $\vala(\pi, 1)=equip$, $post(equip)\in v(\pi_2)$ holds as well (Condition 1),
\item since $pre(cook)\in v(\pi''_1)$, there is some path, namely $\pi'$ with $\pi'\equiv_t\pi''$ and $\vala(\pi', 1)=cook$ (Condition 2),
\item since $pre(food,cook)\in v(\pi_0)$, there exists some path, namely $\pi'$ with $\pi\equiv_0\pi', \vala(\pi', 0)=food$, and $pre(cook)\in v(\pi'_1)$ (Condition 3),
\item since $pre(food, cook)\in v(\pi_0), pre(food)\in v(\pi_0)$ holds as well (Condition 4).
\end{itemize}
\end{example}

We now provide the truth definitions. Recall that formulas are evaluated on a path as a whole, and not in a state.

\begin{definition}[Truth definitions]
\label{def:truthdef}
Let $m=(T,\pi)$ be a model with $T=\tree$:

$T,\pi\models \chi_t$ iff $\chi\in \valp(\pi_t)$ with $\chi\in\prop{}$

$T,\pi\models do(a)_t$ iff $\vala(\pi, t)=a$

$T,\pi\models \neg \varphi$ iff $T,\pi\not\models \varphi$

$T,\pi\models \varphi\wedge\varphi'$ iff $T,\pi\models\varphi$ and $T,\pi\models\varphi'$

$T,\pi\models \Box_t\varphi$ iff for all $\pi'$ in $T$: if $\pi'\sim_t\pi$, then $T,\pi'\models \varphi$
\end{definition}

The truth definitions state that propositions are simply evaluated using the valuation function $\valp$, but $do$ statements are different. They are about state transitions, and therefore use the action function $\vala{}$. This is comparable to the distinction between state formulas and path formulas in CTL* (see the related work Section~\ref{sect:temporal-logic} for more details). Since we are evaluating formulas from a state, the modal operator $\Box_t$ is indexed with a time point $t$, and corresponds to the equivalence relations $\sim_t$.

\begin{example}[Running Example (Ctd.)]
\label{example:pal_model}
We provide some example of applications of the truth definition for the model in Figure~\ref{fig:example_pal_model}:

$T,\pi\models pre(food)_0\wedge do(nop)_0\wedge do(equip)_1$

$T,\pi\models \Diamond_0 (do(food)_0\wedge \neg do(cook)_1)$

$T,\pi\models \Diamond_0 do(food)_0 \wedge \Diamond_0 do(equip)_1 \wedge \neg \Diamond_0 (do(food)_0\wedge do(equip)_1)$

$T,\pi' \models pre(food,cook)_0$

$T,\pi'\not\models do(equip)_1$

$T,\pi''\models \Diamond_0\neg \Diamond_1 do(cook)_1$

\end{example}

\begin{definition}[Model of a formula]
We say that a model $m$ is a model of a formula $\varphi$ if $m\models \varphi$. We denote the set of all models of a formula $\varphi$ by $Mod(\varphi)$, i.e., $$Mod(\varphi) = \{m\in \mathbb{M}\mid m\models \varphi\}.$$
We define the set of all models of a set of formulas $\Sigma$, as $Mod(\Sigma) = \{m\in \mathbb{M}\mid m\models \varphi $ for every $\varphi \in \Sigma\}=\bigcap_{\varphi \in \Sigma}Mod(\varphi).$
\end{definition}

Next we turn to the notions of validity, satisfiability, and semantic consequence. Valid formulas hold in every model, and satisfiable formulas hold in some model.

\begin{definition}[Validity, satisfiability, and semantic consequence] $\ $ \label{def:validity1}
\begin{itemize}
\item $\varphi$ is \emph{valid}, i.e. $\models\varphi$ iff 
$Mod(\varphi)=\mathbb M$.
\item $\varphi$ is \emph{satisfiable} iff 
$Mod(\varphi)\neq \emptyset$.
\item $\varphi$ is a \emph{semantic consequence} of a set of formula $\Sigma$, i.e. $\Sigma\models\varphi$ iff 
$Mod(\Sigma)\subseteq Mod(\varphi)$.
\end{itemize}
\end{definition}

\subsection{Axiomatization}
\label{subsec:ax}

In this part we present the axiomatization of our logic, and we explain the most important axioms in turn.\\

\noindent
Propositional tautologies\hfill \textbf{(PROP)}\\
$\Box_t(\varphi\rightarrow\varphi')\rightarrow(\Box_t\varphi\rightarrow \Box_t\varphi')$\hfill \textbf{(K)}\\
$\Box_t\varphi\rightarrow\varphi$ \hfill\textbf{(T)}\\
$\Diamond_t\varphi\rightarrow \Box_t\Diamond_t\varphi$\hfill\textbf{(5)}\\

\noindent
Axioms PROP, K, T, and 5 together ensure our modal operator is an equivalence relation. This is simply the modal logic system KT5.\\

\noindent
$\chi_t\rightarrow \Box_t \chi_t$, where $\chi\in\prop{}$ \hfill\textbf{(A1)}\\
$\Diamond_t \chi_t\rightarrow \chi_t$, where $\chi\in\prop{}$\hfill\textbf{(A2)}\\

\noindent
Axioms A1 states that if a proposition is true in a state on a path, then it is necessarily true at that time, i.e., it is true in all equivalent paths. The contraposition of Axiom A2 states the same for negated propositions. These axioms follow from the definition of the equivalence $\sim_t$ between paths: if two paths are equivalent up to time $t$, then the same propositions are true in time $t$ as well.\\

\noindent
$do(a)_t\rightarrow \Box_{t+1} do(a)_t$\hfill\textbf{(A3)}\\
$\Diamond_{t+1} do(a)_t\rightarrow do(a)_t$\hfill\textbf{(A4)}\\

\noindent
Axioms A3 and A4 are similar to A1 and A2, but then for the case of actions. Recall that $do$ statements are semantically represented as transitions between states (Definition~\ref{def:truthdef}). Therefore, the modal operator is indexed with the next time point $t+1$.\\

\noindent
$\Box_t\varphi\rightarrow \Box_{t+1}\varphi$\hfill \textbf{(A5)}\\

\noindent
Axiom A5 is a result of the fact that for some path $\pi$ the number of paths equivalent with $\pi$ can only decrease as time moves forward. Therefore, if something is true on all paths equivalent up to time $t$, then it is necessarily true on all paths equivalent up to the next time moment $t+1$.\\

\noindent
$\bigvee_{a\in\act{}} do(a)_t$\hfill\textbf{(A6)}\\
$do(a)_t \rightarrow \neg do(b)_t$, where $b\not=a$\hfill\textbf{(A7)}\\

\noindent
Axioms A6 and A7 together state that exactly one action is executed at every time moment.\\

\noindent
$do(a)_t \rightarrow post(a)_{t+1}$\hfill\textbf{(A8)}\\ 
$pre(a)_t \rightarrow \Diamond_t do(a)_t$\hfill\textbf{(A9)}\\
($pre(a, \overline{b})_t\wedge do(a)_t)\rightarrow pre(\overline{b})_{t+1}$\hfill\textbf{(A10)}\\
$pre(\actseq{}, b)_t\rightarrow pre(\actseq{})_t$\hfill \textbf{(A11)}\\

\noindent
Axioms A8-A11 directly correspond to properties 1-4 of a model (definition~\ref{def:model}).\\

\noindent
$
(do(a)_t \wedge\varphi)\rightarrow \Box_t(do(a)_t \rightarrow \varphi)$\hfill\textbf{(A12)}\\
\hspace*{0.5cm}where $\varphi\in Past(t+1)$\\

\noindent
Axiom A12 ensures actions are deterministic. If something holds immediately after performing action $a$ in time $t$ (which is why $\varphi\in Past(t+1)$), then it necessarily holds after performing that action in time $t$. Note this formula does not hold without the restriction of $\varphi$ to $Past(t+1)$, because because formulas containing time points greater than $t+1$ may depend on actions performed after time $t$.\\

In addition to these axioms, \logic{} has two inference rules, a variant of Necessitation and Modus Ponens:\\

\noindent
From $\varphi$, infer $\Box_0\varphi$\hfill\textbf{(NEC)}\\
From $\varphi,\varphi\rightarrow\varphi'$, infer  $\varphi'$\hfill\textbf{(MP)}\\

Note that in NEC we can replace $\Box_0\varphi$ with $\Box_t\varphi$, for any $t$, due to Axiom A5. In other words, the following variant of necessitation is a \emph{derivable rule} of the logic:\\

\noindent
From $\varphi$, infer $\Box_t\varphi$ \hfill\textbf{(NEC-$t$)}

\begin{remark}
\label{remark:axiom9}
Continuing our discussion of Remark~\ref{remark:first}, one may strengthen Axiom A9 as follows:\\

\noindent
$pre(a)_t\leftrightarrow \Diamond_t do(a)_t$\hfill (A9*).
\end{remark}

We next formalize the notion of theorems and derivability.

\begin{definition}[Theorems in \logic{}]
\label{def:theorems}
A derivation of $\varphi$ within \logic{} is a finite sequence $\varphi_1, \ldots, \varphi_m$ of formulas such that:
\begin{enumerate}
\item $\varphi_m= \varphi$;
\item every $\varphi_i$ in the sequence is either
\begin{enumerate}
\item (an instance of) one of the axioms
\item the result of the application of Necessitation or Modus Ponens to formulas in the sequence that appear before $\varphi_i$.
\end{enumerate}
\end{enumerate}
If there is such a derivation for $\varphi$ we write $\vdash \varphi$. and we say $\varphi$ is a \emph{theorem of \logic{}}.
\end{definition}

We define theorems and derivability separately because we restrict the application of the Necessitation rule to theorems only.

\begin{definition}[Derivability in \logic{}]
\label{def:derivability}
A derivation for a formula $\varphi$ from a set of formulas $\Sigma$ is a finite sequence $\varphi_1$, \dots, $\varphi_m$ of formulas such that:
\begin{enumerate}
\item $\varphi_m= \varphi$;
\item every $\varphi_i$ in the sequence is either a theorem, a member of $\Sigma$, or the result of the application of Modus Ponens to formulas in the sequence that appear before $\varphi_i$.
\end{enumerate}
If there is such a derivation from $\Sigma$ for $\varphi$ we write $\Sigma\vdash \varphi$. We then also say that \emph{$\varphi$ is derivable from the premises $\Sigma$.}
\end{definition}

Furthermore, a set of formulas $\Sigma$ is \emph{consistent} if we cannot derive a contradiction from it, i.e., $\Sigma\not\vdash\perp$, and a set of formulas $\Sigma$ is \emph{maximally consistent} if it is consistent and every superset is inconsistent.

We denote by $Cn(\Sigma)$ the set of consequences of $\Sigma$, i.e. $$Cn(\Sigma)=\{\varphi\mid \Sigma\vdash\varphi\}.$$ 

\subsection{Soundness and Completeness}

In this section we prove the axiomatization of \logic{} is sound and strongly complete with respect to its semantics. 

\begin{theorem}[Completeness Theorem]
\label{thm:completenessA}
The logic \logic{} is sound and strongly complete, i.e. $\Sigma\vdash\varphi$ iff $\Sigma\models\varphi$.
\end{theorem}

We provide a proof sketch of the theorem. The full proofs can be found in Appendix A.

\begin{proof}[Proof Sketch]
We prove the following formulation of completeness: each consistent set of formulas $\Sigma$ has a model. We prove the Lindenbaum lemma, stating that each consistent set can be extended to a maximally consistent set $\Sigma'$, i.e. $\Sigma'$ is consistent and each proper superset of $\Sigma'$ is inconsistent. In the first step we extend $\Sigma$ to a maximally consistent set $\Sigma^0$. 

Then for each $t$ we define an equivalence relation $\equiv_t$ on maximally consistent sets in the following way:
$$\Sigma_1^*\equiv_t \Sigma_2^* \text{ iff }\Sigma_1^*\cap Past(t) = \Sigma_2^*\cap Past(t).$$
Let us denote the corresponding equivalence classes by $[\Sigma^*]_t$, which means $\{ \overline{\Sigma}^*\mid \Sigma^*\equiv_t \overline{\Sigma}^*\}.$

In the second part of the proof, using the maximally consistent superset $\Sigma^*$ of $\Sigma$ (which exists by the Lindenbaum lemma), we define the tree $T_{\Sigma^*}=(S,R,v,act)$: 
\begin{enumerate}
\item $S = \bigcup_{t\in\mathbb{N}}S_t$ where $S_t=\{[\overline{\Sigma}^*]_t\mid \overline{\Sigma}^*\equiv_t \Sigma^*\}$
\item $sRs'$ iff $\EX{\overline{\Sigma}^*, t\in\mathbb{N}}{s=[\overline{\Sigma}^*]_t\wedge s'=[\overline{\Sigma}^*]_{t+1}}$
\item $\chi\in v(s)$ iff $\EX{\overline{\Sigma}^*,t\in\mathbb{N}}{s= [\overline{\Sigma}^*]_t\wedge \chi_t\in \overline{\Sigma}^*}$.
\item $a= \vala((s,s'))$ iff $\EX{\overline{\Sigma}^*}{s=[\overline{\Sigma}^*]_t\wedge s'=[\overline{\Sigma}^*]_{t+1}\wedge do(a)_t\in \overline{\Sigma}^*}$.
\end{enumerate}
Given a maximally consistent set (mcs) $\Sigma^*$, we construct a path $\pi_{\Sigma^*}=(s_0,s_1,\ldots)$ from it by letting $s_t=[\Sigma^*]_t$. So $\chi\in v([\Sigma^*]_t)$ iff $\chi_t\in \Sigma^*$ and $a=\vala(([\Sigma^*]_t, [\Sigma^*]_{t+1}))$ iff $do(a)_t\in T^*$.

If $\pi(\Sigma^*)=(s_0,s_1,\ldots)$, where $s_t=[\Sigma^*]_t$, then one can show that $(T_{\Sigma^*},\pi(\Sigma^*))$ is a model. Finally, we prove that for each $\varphi$, ($T_{\Sigma^*},\pi(\Sigma^*) \models \varphi \text{ iff }\varphi\in \Sigma^*$), using induction on the complexity of $\varphi$. Consequently, $(T_{\Sigma^*},\pi(\Sigma^*)) \models \Sigma$.
\end{proof}

Note that we can check satisfiability of any formula from $\lang$ in finite time. Indeed, for every formula $\varphi\in \lang$ there is a maximal time index $t$ appearing in $\varphi$. By Definition \ref{def:truthdef}, for checking if $\varphi$ is satisfied in a model $m$ it is enough to check the states and actions in the of $m$ up to time $t+1$. If we restrict the evaluation functions $v$ to the finite set of  propositions from $\prop{}$ relevant for $\varphi$,\footnote{Technical details can be found in Appendix A.}
 and since we have finitely many deterministic actions, there are only finitely many different ways to build a tree until a fixed time instance. Therefore, the number of those time-restricted trees which satisfy the four conditions of Definition \ref{def:model} is finite as well.
Thus, the satisfiability problem for the logic PAL is decidable.
\section{Adding intentions}
\label{sect:intentions}

In the previous section we developed a logic for the belief database of Shoham's database perspective (Figure~\ref{fig:dbperspective}). We did not yet take intentions into account, which is what we do in this section. Recall intentions are formalized as \emph{discrete atomic action intentions} of the form $(a,t)$. We focus on two main tasks: separating beliefs dependent on intentions (\emph{weak beliefs}) from those that are not (\emph{strong beliefs}), and formalizing a coherence condition on beliefs and intentions. These two tasks correspond to the two subsections of this section.
\subsection{Separating strong and weak beliefs}
\label{sect:strongweakbeliefs}
The idea behind strong beliefs (the terminology due to~\cite{vanderHoek2003}) is that they represent the agent's ideas about what is inevitable, no matter how it would act in the world. In our setting, a set of strong beliefs is a set of formulas starting either with $\Diamond_0$ or $\Box_0$, and all consequences that follow from it. First, we define a language for strong beliefs.

\begin{definition}[Strong belief] 
\label{def:strbel}
The \emph{set of all of strong beliefs} $\sbel$ for \lang{} are generated by boolean combinations of $\Box_0\psi$, where $\psi$ is a PAL formula. A \emph{strong belief} is an element of $\sbel$.
\end{definition}

We next provide some examples of strong beliefs for our running example.

\begin{example}[Running example, Ctd.]
Some examples of strong belief formulas are:
\begin{itemize}
\item $\Diamond_0 (do(food)_0\wedge \neg do(cook)_1)$
\item $\Diamond_0 do(food)_0 \wedge \Diamond_0 do(equip)_1 \wedge \neg \Diamond_0 (do(food)_0\wedge do(equip)_1)$
\item $\Diamond_0\neg \Diamond_1 do(cook)_1$
\item $\Box_0 \Diamond_0 do(cook)_1$
\end{itemize}
\end{example}

Next we define a set of strong beliefs, which is generated from the set of all strong beliefs, and closed under consequence.

\begin{definition}[Set of strong beliefs]
A \emph{set of strong beliefs $SB$} is the deductive closure of a subset of formulas from $\sbel$, i.e. $SB=Cn(\Sigma)$ where $\Sigma\subseteq\sbel$. 
\end{definition}

The following example shows that a set of strong beliefs may also contain formulas which are not in $\sbel$, since they are closed under consequence.

\begin{example}[Set of strong beliefs]
Let $\Sigma=\{\neg\Diamond_0 p_3, \Box_0 q_2\}\subset \sbel$, and let the set of strong beliefs $SB=Cn(\Sigma)$. From Axioms A1 and A2 we obtain $\neg p_3\in SB$, as well as $ q_2\in SB$.
\end{example}

The reader may already have noted that, semantically, strong beliefs are independent of the specific path on which they are true. Indeed, strong beliefs are true in a tree rather than on a single path. Therefore, if a model (consisting of a tree and a path) is a model for a strong belief formula $\varphi$, then all possible models with the same tree are models of the strong belief formula $\varphi$. We make this idea precise in the following definition.

\begin{definition}[Set of models of strong beliefs (msb set)]
\label{def:msbset}
A \emph{set of models of strong beliefs} $\msbset{}\subseteq\mods{}$ (i.e., an \emph{msb set}) is a set of models such that $\msbset{} = \{(T,\pi):\pi\in T\}$. The set $\msbsets{}$ contains all msb sets.
\end{definition}

Definition~\ref{def:msbset} ensures that if some model $(T,\pi)$ is in a set of models of a strong belief, then all other models $(T,\pi')$ are also in this set. Note that it would also be possible to identify strong models with a tree $T$, but we have chosen not to implement this to keep the presentation concise. 

	The following proposition shows a direct correspondence between a set of strong beliefs and its models.

\begin{proposition}
\label{prop:mod-sb-is-msbset}
Given a set of strong beliefs $SB$, the set of models of $SB$ is an msb set, i.e., $Mod(SB)\in \msbsets{}$.
\end{proposition}

We now explain the semantics of strong beliefs models with our running example.

\begin{example}[Running example (Ctd.)]
\label{ex:strongbel:sem}
Consider the tree $T$ of Figure~\ref{fig:example_pal_model} and let $\overline{\pi}\in\{\pi,\pi',\pi''\}$. The following statements hold:
\begin{itemize}
\item $T,\overline{\pi}\models \Diamond_0 (do(food)_0\wedge \neg do(cook)_1)$
\item $T,\overline{\pi}\models \Diamond_0 do(food)_0 \wedge \Diamond_0 do(equip)_1 \wedge \neg \Diamond_0 (do(food)_0\wedge do(equip)_1)$
\item $T,\overline{\pi}\models \Diamond_0\neg \Diamond_1 do(cook)_1$
\item $T,\overline{\pi}\models \Box_0 \Diamond_0 do(cook)_1$
\end{itemize}
\end{example}

We obtain a belief-intention database by adding intentions to the strong beliefs. By intentions we assume action-time pairs, and an intention database is a set of intentions. We also add the constraint that at most one action is intended for a given time moment. We close the set of strong belief under consequence. Alternatively we can also have a (finite) set of strong beliefs, as in Hansson's base revision~\citep{HanssonBook}, but we follow the approach of Katsuno and Mendelzon.

\begin{definition}[Belief database, intention database, belief-intention database]
\label{def:belief-intention-database}
An \emph{intention $(a,t)$} is a pair consisting of an action $a\in\act{}$ and a time point $t$. $\ib{} = \act\times \mathbb{N}$ denotes the set of all intentions.

A \emph{belief database $SB$} is a set of strong beliefs closed under consequence, i.e. $SB=Cn(SB)$. \bd{} denotes the set of all belief databases. 

An \emph{intention database $I=\{(a_1,t_1),(a_2,t_2),\ldots\}$} is a set of intentions such that no two intentions exist at the same time point, i.e if $i\not=j$ then $t_i\not=t_j$. $\idb \subseteq 2^{Act\times \mathbb{N}}$ denotes the set of all intention databases.

A \emph{belief-intention database} $(SB,I)$ consists of a belief database and an intention database. $\mathbb{BI} = \bd{}\times \idb{}$ denotes the set of all belief-intention databases.

\end{definition}

We define weak beliefs by adding intentions to the strong beliefs, and closing the result under consequence.

\begin{definition}[Weak Beliefs]
\label{def:weak-beliefs}
Given a belief-intention database $(SB,I)$, the weak beliefs are defined as follows:$$WB(SB,I)=Cn(SB\cup \{do(a)_t\mid (a,t)\in I\}).$$
\end{definition}

We provide an example for weak beliefs using our running example.

\begin{example}[Running example (Ctd.)] 
\label{ex:weakbel:sem}
Suppose the set $SB$ contains strong beliefs describing the tree $T$ of Figure~\ref{fig:example_pal_model}. Some of the formulas in $SB$ are:
\begin{itemize}
\item $\Diamond_0 (do(food)_0\wedge do(cook)_1)$,
\item $\Diamond_0 do(equip)_1$,
\item $\Box_0 pre(food, cook)_0$,
\item $\neg\Diamond_0 (do(food)_0\wedge do(equip)_1)$.
\end{itemize}
Let $I=\{(food,0),(cook,1)\}$. Some examples of weak beliefs $WB(SB,I)$ are:
\begin{itemize}
\item $do(food)_0 \wedge do(cook)_1$,
\item $\neg do(equip)_1$,
\item $post(food)_1 \wedge post(cook)_2$.
\end{itemize}
Note the model $(T,\pi')$ from Figure~\ref{fig:example_pal_model} is a model of $WB(SB,I)$.
\end{example}

Note the difference between Example~\ref{ex:strongbel:sem} and Example~\ref{ex:weakbel:sem}. Strong beliefs are true in a tree, while weak beliefs \emph{depend on a path}. In this way, weak beliefs are contingent on the action executed on the actual path. We can thus understand adding intentions to strong beliefs semantically by \emph{choosing a set of paths} in a tree.

\begin{remark}
Note that since weak beliefs contain strong beliefs with intentions, and everything following from that, they also contain postconditions of actions. For instance, if $I=\{(a,t)\}$ and $SB=\emptyset$, then $post(a)_t\in WB(SB,I)$ (by Axiom A8). However, it does not mean that preconditions of intended actions are believed as well, i.e. $pre(a)_t\not\in WB(SB,I)$. So an agent can believe it will execute its intentions, while it doesn't believe the preconditions hold (yet). This is why the implication in Axiom A9 is not a bidirectional implication (see also Remark~\ref{remark:axiom9}).
\end{remark}

\subsection{Commitment: the coherence condition on beliefs and intentions}
\label{sect:coherenceconditions}

This paper is about commitment. The agent is committed to its intentions as long they are coherent with its beliefs. The coherent condition is that the agent believes it is possible to perform all intended action. We thus require that the joint preconditions of all intended actions not be disbelieved by the agent. 

\begin{definition}[Coherence]
\label{def:coherence}
Given an intention database $I=\{(b_{t_1},t_1),\ldots,(b_{t_n},t_n)\}$\footnote{$t_1,\ldots,t_n$ is not necessarily a sequence of subsequent integers. For instance $t_1=2, t_2=5$. The disjunction below covers the remaining time indexes with all possible actions.} with $t_1<\ldots<t_n$, let
\begin{flalign}
\label{eq coherence}
Cohere(I) = \Diamond_0\bigvee_{\substack{a_t\in Act{} : t\not\in\{t_1,\ldots, t_n\}\\a_t=b_t : t\in\{t_1,\ldots,t_n\}}} pre(a_{t_1},a_{t_{1}+1},\ldots,a_{t_n})_{t_1}.
\end{flalign}
\begin{itemize}
\item For a given belief-intention database $(SB,I)$, we say that it is \emph{coherent} iff $SB$ is consistent with $Cohere(I)$, i.e., $SB\not\vdash\neg Cohere(I)$.
\item A pair $(\psi,I)$ consisting of a strong belief formula $\psi\in\sbel$ and an intention database $I$ is coherent iff $\psi$ is consistent with $I$, i.e. $\psi\not\vdash\neg Cohere(I)$.\footnote{We will use this formulation in the next section when we represent a set of strong beliefs $SB$ by a single formula $\psi$.}
\end{itemize}
\end{definition}

Note we can define coherence semantically for a given msb set \msbset{} (Definition~\ref{def:msbset}) iff there exists some $m\in \msbset{}$ with $m\models Cohere(I)$. We then obtain the correspondence that $(SB,I)$ is coherent iff $(Mod(SB),I)$ is coherent using completeness trivially.

Let us explain this definition with a simple example.

\begin{example}
Let $\act{}=\{a,b\}$ and $I=\{(a,1),(b,3)\}$. Then, $$Cohere(I) = \Diamond_0\bigvee_{x\in\act{}} pre(a,x,b)_1 =\Diamond_0( pre(a,a,b)_1\vee pre(a,b,b)_1).\footnote{Our construction of preconditions over action sequences may lead to a coherence condition involving a big disjunction. Alternatively, one may explicitly denote the time of each precondition, e.g. $pre(a,b)_{(t_1,t_2)}$. We chose the former since it is closer syntax of the other propositions.}$$
\end{example}

Intuitively, intentions cohere with beliefs if the agent considers it possible to jointly carry out all of the intended actions. This is a minimal requirement on \emph{rational balance} between the two mental states.

In the next section we will consider the revision of belief-intention database. We will require that a belief-intention database is coherent after revision.

\begin{remark}
\label{remark:pissofdragan}
Consider our example of Bobby intending to buy food at time 0. As we pointed out, it is not actually necessary that Bobby believes it has sufficient money; only that it does not believe it does not have sufficient money. We can also ask: what can be Bobby's working assumptions about the future, upon adopting this intention? In so far as Bobby is committing himself to this action, we may assume that it will buy food at time 0. If we then consider the paths in our belief models on which this action is taken at time 0, the postconditions will hold along all of them. However, to allow that the preconditions may not yet be believed, we admit paths on which the preconditions do not hold. We only require that they hold on some path in the set, so that Bobby cannot stray too far from reality.

Indeed, this is arguably closer to how we reason about future actions. We often commit to actions without explicitly considering the path that will lead us there. Eventually this decision will have to be made, but there is nothing incoherent about glossing over these details at the current moment. Bobby should assume it will have bought food at time 1 and can continue making plans about what it will do with the food after this. But it should not assume the preconditions will hold until it has made further, specific plans for bringing them about. And at the current time, Bobby may not even bother worrying about it.
\end{remark}

We now apply the coherence condition to our running example.

\begin{figure}[h!]
\centering
\begin{tikzpicture}[->,>=stealth',shorten >=1pt,auto,node distance=1cm,
  thin,
  main node/.style={circle,draw,fill=white,font=\sffamily\scriptsize\bfseries}, 
  every label/.style={font=\scriptsize},
  time node/.style={font=\scriptsize,text=black!90},
  test line/.style={gray!80, dashed,-}]
  
  \draw[test line] (-0.5,-2) -- (-0.5, 2.2);
  \draw[test line] (1.9,-2) -- (1.9, 2.2);
  \draw[test line] (4.3,-2) -- (4.3, 2.2);
  
  \node[time node] (t0) at (-0.5,2.3) {$t=0$};
  \node[time node] (t1) at (2.2,2.3) {$t=1$};
  \node[time node] (t2) at (4.6,2.3) {$t=2$};
  
  \node[main node] (s0) at (-0.5,0) {$s_0$};
  \node[position=-130:0mm from s0,align=center,font=\scriptsize] (labels4) {$\{pre(food),$\\$pre(food,cook)\}$};  
  \node[main node,label={[label distance=-0.1cm]90:$\{pre(equip)\}$},position=20:20mm from s0] (s1) {$s_1$};
   \node[main node,label={[label distance=-0.1cm]0:$\{post(equip)\}$},position=20:20mm from s1] (s2) {$s_2$};
  \node[main node, position=-20:20mm from s0] (s3) {$s_3$};
  \node[position=-130:0mm from s3,align=center,font=\scriptsize] (labels3) {$\{pre(cook),$\\$post(food)\}$};  
  
  \node[main node,label={[label distance=-0.1cm]00:$\{post(cook)\}$},position=20:20mm from s3] (s4) {$s_4$};
  \node[main node,label={[label distance=-0.1cm]0:},position=-20:20mm from s3] (s5) {$s_5$};
  
    \node[style={circle}] (pi1) at (7,1.75) {$\pi$};
   \node[style={circle}] (pi2) at (7,0) {$\pi'$};
   \node[style={circle}] (pi3) at (7,-1.75) {$\pi''$};
 
  \path[every node/.style={font=\sffamily\small}]
    (s0) edge[draw=black] node [above] {nop} (s1)
    (s1) edge[draw=black] node [above] {equip} (s2)
    (s0) edge[draw=black] node [below left] {food} (s3)
    (s3) edge[draw=black] node [above left] {cook} (s4)
    (s3) edge[draw=black] node [below left] {nop} (s5);
\end{tikzpicture}
\caption{The tree $T$ of Figure~\ref{fig:example_pal_model} reprinted.}
\label{fig:example_pal_model2}
\end{figure}
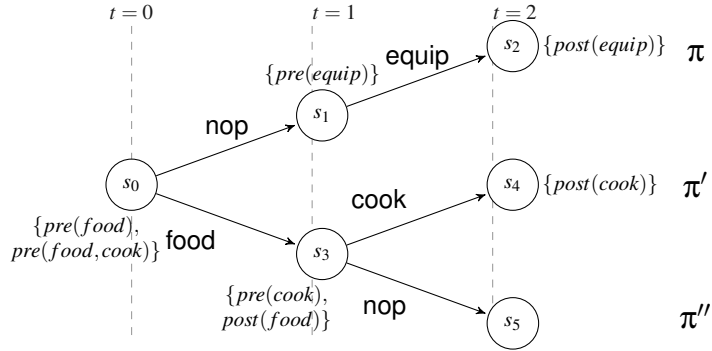

\begin{example}[Running example (Ctd.)]
Let $(SB,I)$ be such that the strong beliefs are represented by the tree in Figure~\ref{fig:example_pal_model2}\footnote{In other words, each tree in each model in the msb set $SB$ is exactly the same up to time $t=2$ as the tree in Figure~\ref{fig:example_pal_model2}.}. We consider different choices for $I$:
\begin{itemize}
\item Let $I=\{(food,0),(cook,1)\}$. In this case, $(Mod(SB),I)$ is coherent, since there is some model $m\in Mod(SB)$ with $m\models Cohere(I)$, i.e. $m\models\Diamond_0 pre(food,equip)_0$. In fact, from $pre(food,cook)\in \valp(s_0)$, it follows that $T,\overline{m}\models pre(food,cook)_0$ holds for each model $(T,\overline{m}).$ From the completeness theorem, it follows that $(SB,I)$ is coherent as well.
\item Let $I=\{(food,0),(equip,1)\}$. In this case $(Mod(SB),I)$ is not coherent, since there is not $m\in Mod(SB)$ with $m\models \Diamond_0 pre(food,cook)_0$. Again by completeness we obtain that $(SB,I)$ is not coherent either.
\end{itemize}
\end{example}

Next we show that a coherent belief-intention database implies joint consistency of beliefs and intentions.

\begin{proposition}
\label{prop:coherence_implies_consistency}
Given some belief-intention database $(SB,I)$, if $(SB,I)$ is coherent, then $WB(SB,I)$ is consistent.
\end{proposition}

\begin{proof}[Proof Sketch]
Using axioms A9, A10, and A12, for every $a_0,\ldots,a_m\in \act$ and $ \in \mathbb N$ one can show that $\{pre(a_0,\ldots,a_m)_t\}\vdash \Diamond_t (do(a_0)_t\wedge\Diamond_{t+1}(do(a_1)_{t+1}\wedge \Diamond_{t+2}(\ldots)))$. By taking the contrapositive of  A5, we obtain the theorem of PAL logic 
\begin{equation}
\label{one theorem of pal a}
\vdash pre(a_0,\ldots,a_m)_t\to\Diamond_t\bigwedge_{p=0}^m do(a_p)_{t+p}
\end{equation}
For an intention database $I=\{(b_{t_1},t_1),\ldots,(b_{t_n},t_n)\}$, with $t_1<\ldots<t_n$, let us consider its coherence formula  $Cohere(I)$ (Formula (\ref{eq coherence}) from Def.~\ref{def:coherence}). If we apply the theorem (\ref{one theorem of pal a}) to the formulas under the scope of the disjunction in (\ref{eq coherence}) (i.e.,
$ pre(a_{t_1},a_{t_{1}+1},\ldots,a_{t_n})_{t_1}$), we obtain that 
 $Cohere(I)$ implies
\begin{align*}
\Diamond_0\bigvee_{\substack{a_t\in Act{} : t\not\in\{t_1,\ldots,t_m\}\\a_t=b_t : t\in\{t_1,\ldots,t_n\}}} \Diamond_{t_1} (do(a_{t_1})_{t_1}\wedge do(a_{t_1+1})_{t_1+1}\wedge\ldots\wedge do(a_{t_n})_{t_n})
\end{align*}
Consequently, $Cohere(I)$ implies $\Diamond_0\Diamond_{t_1}\bigwedge_{k=1}^n do(b_{t_k})_{t_k}$, and by A5 this implies $\Diamond_0\bigwedge_{k=1}^n do(b_{t_k})_{t_k}$. Since  $I=\{(b_{t_1},t_1),\ldots,(b_{t_n},t_n)\}$, the last formula can be rewritten as $\Diamond_0\bigwedge_{(a,t)\in I} do(a)_t$. Therefore, if $(B,I)$ is coherent, then  $B\cup\{\Diamond_0\bigwedge_{(a,t)\in I} do(a)_t\}$ is a consistent set. By the fact that $B$ is a strong belief set, $B\cup\{\bigwedge_{(a,t)\in I} do(a)_t\}$ is consistent, i.e. $WB(B,I)$ is consistent.
\end{proof}

Note the reverse direction of Proposition~\ref{prop:coherence_implies_consistency} does not hold. We demonstrate this in the next example.

\begin{example}[Running example (Ctd.)]
Suppose $(SB,I)$ is such that the strong beliefs are represented by the tree in Figure~\ref{fig:example_pal_model2} and that $I=\{(food, 0), (equip, 1)\}$. In this case the weak beliefs $WB(SB,I)$ are consistent, because the following holds:\\

$WB(SB,I)\vdash do(food)_0\wedge do(equip)_1$,\\

\noindent
no contradiction is derived from this. However, $(SB,I)$ is not coherent, because there is no single path in the model in which all the preconditions of the intended actions hold.
\end{example}
\section{Revision of belief and intention}
\label{sect:revision}
In this section we turn to the {\em dynamic} part of our belief-intention databases by studying the revision of belief and intention. We provide and motivate a set of revision postulates on a belief-intention database $(SB,I)$ in Section~\ref{sect:revision:postulates}, and we prove our main representation theorem in Section~\ref{sect:revision:representation}. We discuss various examples in Section~\ref{sect:examples}.

The challenge of obtaining our result is two-fold:
\begin{itemize}
\item When revising a belief database that is bounded up to some time $t$ with a strong belief, we have to ensure that the resulting belief database is also bounded up to $t$,
\item When revising a belief database we also have to ensure the new belief database remains a strong belief. 
\end{itemize}

Our solution to is to bound both the syntax of \logic{} and the revision operator up to some time $t$ in the first subsection. In the second subsection we do the same for the semantics.

\subsection{Preliminaries: Belief Revision}
\label{sect:preliminaries}
The AGM postulates~\citep{AGM1985} formulate properties that should be satisfied by any (rational) revision operators defined on deductively closed sets of propositional formulas.~\cite{Katsuno1991} represent a belief set $B$ as a propositional formula $\psi$ such that $B=\{\varphi\mid\psi\vdash\varphi\}$. They define the following six postulates for revision on $\psi$ and show that these are equivalent to the eight AGM postulates:\\
\begin{enumerate}
\item[(R1)]
$\psi\br\varphi$ implies $\varphi$
\item[(R2)]
If $\psi\wedge\varphi$ is satisfiable, then $\psi\br\varphi\equiv\psi\wedge\varphi$
\item[(R3)]
If $\varphi$ is satisfiable, then $\psi\br\varphi$ is also satisfiable
\item[(R4)]
If $\psi\equiv\psi'$ and $\varphi\equiv\varphi'$, then $\psi\br\varphi\equiv\psi'\br\varphi'$
\item[(R5)] $(\psi\br\varphi)\wedge \varphi'$ implies $\psi\br(\varphi\wedge\varphi')$
\item[(R6)] If $(\psi\br\varphi)\wedge\varphi'$ is satisfiable, then $\psi\br(\varphi\wedge\varphi')$ implies $(\psi\br\varphi)\wedge\varphi'$
\end{enumerate}

Given a set $\mathbb{J}$ of all interpretations over some propositional language, they define a faithful assignment as a function that assigns each $\psi$ to a pre-order $\le_\psi$ on models satisfying the following three conditions:
\begin{enumerate}
\item If $J,J'\in Mod(\psi)$, then $J<_\psi J'$ does not hold.
\item If $J\in Mod(\psi)$ and $J'\not\in Mod(\psi)$, then $J<_\psi J'$ holds.
\item If $\psi\equiv\phi$, then $\le_\psi = \le_\phi$.
\end{enumerate}

They show in a representation theorem that a revision operator $\circ$ satisfies postulates (R1)-(R6) iff there exists a faithful assignment that maps each formula $\psi$ to a total preorder $\le_\psi$ such that 
\begin{align*}
Mod(\psi\circ\varphi) = \min(Mod(\varphi),\le_\psi).
\end{align*}

\subsection{Revision postulates} 
\label{sect:revision:postulates}

Recall from Section~\ref{sect:preliminaries} that we aim to prove a representation theorem comparable to that of~\cite{Katsuno1991}. Therefore, we follow their convention to fix a way of representing a belief set $SB$ consisting of strong beliefs by a single strong belief formula $\psi$ such that $SB=\{\varphi\mid\psi\vdash\varphi\}$. One of the main difficulties in this respect is that time in \logic{} is infinite in the future, so it is generally not possible to represent $SB$ closed under consequence by a single formula $\psi$, since this may potentially lead to an infinite conjunction. Therefore, we cannot prove the Katsuno and Mendelzon representation theorem directly. In this section, we define a \emph{bounded} revision function and we restrict the syntax of \logic{} up to a specific time point.

We first define some notation that we use in the rest of this section.

\paragraph{Notation}
\begin{itemize}
\item
By slight abuse of terminology, a pair $(\psi,I)\in\sbel\times \idb$ consisting of a strong belief formula $\psi$ and an intention database $I$ is also called a \emph{belief-intention database},
\item For $\varphi,\psi\in \sbel$, we write $\varphi\equiv\psi$ to denote that $\vdash \varphi\leftrightarrow\psi$,
\item $\epsilon$ is the special ``empty'' intention.
\item We denote $\bis{},\sbel{},\ib{}$, and $\idb$ bounded up to $t$ with respectively $\bis\rest{}, \sbel{}\rest{}, \ib{}\rest{}$, and $\idb\rest{}$. However, if the restriction is clear from context, we may omit the superscript notation. 
\end{itemize}

Our aim is to define a bounded revision function $\agr$ revising a belief-intention database $(\psi,I)$ with a tuple $(\varphi,i)$ consisting of a strong belief $\varphi$ and an intention $i$, denoted $(\psi,I)\agr(\varphi,i)$. The bounded operator revises the formulas of the restricted language $ \lang{}^{|t}$, that represent all the relevant information for planning up to time $t$. 

\begin{definition}[The language $\lang{}^{|t}$]
The language $\lang{}^{|t}$ consists of all formulas $\varphi\in\lang{}$ such that if $p_{t'}, \Box_{t'}, do(a)_{t'}$ or $post(a)_{t'}$ occurs in $\varphi$, then $t'\le t$. Furthermore, if $pre(a_0,\ldots,a_k)_{t'}$ occurs in $\varphi$, then $k+t'\le t$.
\end{definition}

For instance, $pre(a)_3,post(a)_3\in \lang{}^{|3}$, but $pre(a,b,c)_3\notin \lang{}^{|3}$.

First we define the restricted operator $\brt$ which revises strong beliefs up to given time instance $t$.

\begin{definition}[Strong belief revision function]
\label{def:belief-postulates} 
A \emph{bounded strong belief revision function} is a function  $\brt:\bd{}\times \sbel{}\rightarrow \bd{}$, which maps a strong belief database $\psi$ and a strong belief formula $\varphi$ --- all bounded up to $t$--- to a strong belief database $$\psi' = \psi\brt \varphi,$$ bounded up to $t$, and which satisfy the  postulates of
Katsuno and Mendelzon, R1--R6.
\end{definition}

The revision operator above captures the intuition that strong beliefs are independent of intentions. This avoids \emph{wishful thinking} meaning that the desire (or intention) for something to be true is used in place of/or as evidence for the truthfulness of the claim.

While revision of beliefs is independent of intentions, the revision of intentions should take beliefs into account as well, in order to ensure coherence. For instance, one can only accommodate a new intention if it is considered possible that these intentions can be achieved. This is formalized in the next revision operator.

\begin{definition}[Intention revision function]\label{def:intention-postulates} 
An \emph{intention revision function} $\agri:\bis{}\times \ib{}\rightarrow \bis$ maps a belief-intention database and an intention--- all bounded up to $t$--- to a belief-intention database bounded up to $t$ such that 
$$(\psi,I)\agri i=(\psi,I'),$$ 
where  the following postulates hold.\\
$(P1)$ $(\psi,I')$ is coherent.\\
$(P2)$ If $(\psi,\{i\})$ is coherent, then $i\in I'$.\\
$(P3)$ If $(\psi,I\cup\{i\})$ is coherent, then $I\cup\{i\}\subseteq I'$.\\
$(P4)$ $I'\subseteq I\cup \{i\}$.\\
$(P5)$ For all $I''$ with $I'\subset I''\subseteq I\cup\{i\}$:$(\psi,I'')$ is not coherent.
\end{definition}
Note that, by the definition, the revision of strong beliefs cannot be triggered by intention revision.

Postulate (P1) states that the outcome of a revision should be coherent. Postulate (P2) states that the new intention $i$ take precedence over all other current intentions; if possible, it should be added, even if all current intentions have to be discarded. We can consider also not prioritized operators, which are not always successful. However, here we follow the standard approach in AGM theory change, which assumes that a check whether the belief base must be updated is done separately from the actual update.\footnote{Note that if we would drop postulate P2, then this would corresponding to dropping Item 2 of Definition~\ref{def:selection-function}.} Postulate (P3) and (P4) together state that if it is possible to simply add the intention, then this is the only change that is made. These two postulates are comparable to inclusion and vacuity of AGM. 
Finally, (P5) states that we do not discard intentions unnecessarily. This last postulate is a kind of maximality requirement, and is comparable to the \emph{parsimony requirement} introduced by~\cite{grant2010}.

Up until now we have defined two revision operators separately: one for revising with a strong belief and one for revising with an intention. Recall that it is our aim to define ``revising a belief-intention database with a belief/intention''. We want to do this because the separate revision operators we defined up till now do not capture all interactions between beliefs and intentions. In particular, we would like to ensure that we capture the possibility for an agent to revise its intentions after having revised his strong beliefs, in order to restore coherence.

In order to do so we define a single revision function revising a belief-intention database by a pair $(\varphi,i)$ in terms of the existing operators $\brt$ and $\agri$.


\begin{definition}[Belief-intention revision function]
\label{belief-intention revision function}
A \emph{belief-intention revision function} is a function $\agr:\bis{}\times (\sbel{}\times \ib{})\rightarrow \bis$ of the form
$$
(\psi,I)\agr(\varphi,i) =(\psi\brt \varphi,I)\agri i, 
$$
where $\brt$ is a strong belief revision function and $\agri$ is an intention revision function.
\end{definition}
%
 In other words, the procedures runs as follows:
\begin{enumerate}
\item Revise strong beliefs using (R1)-(R6),
\item Revise intentions using (P1)-(P5), possibly revising weak beliefs as well.
\end{enumerate}
Therefore, revising strong beliefs does not depend on which intentions an agent had, or which intention it revises with. However, revising intentions \emph{does} have an effect on the weak beliefs (see the last paragraph of Example~\ref{expl:addingintention}). 


We will show how revision works in our logic with various examples at the end of this section, after we have proved the representation theorem.

\subsection{Representation Theorem}
\label{sect:revision:representation}

In this subsection we present the main technical result of this paper. We characterize all revision schemes satisfying (R1)-(R6), (P1)-(P5) in terms of minimal change with respect to an ordering among interpretations and a selection function accommodating new intentions while restoring coherence.

In the previous subsection we bounded various sets of formulas up to some time point $t$. We now do the same for models. We bound models up to $t$, which means that all the paths in the model are ``cut off'' at $t$. 

\begin{definition}[$t$-bounded model] 
\label{def:t-restrictedmodel}
Suppose some model $m=(T,\pi)$. 
\begin{itemize}
\item
A \emph{$t$-bounded path} $\restp{}$ is defined from a path $\pi$ in $T$ as $\restp{}=(\pi'_0,\ldots,\pi'_t)$, where each $\pi'_i$ contains the restriction of an evaluation of $\pi_i$ to exactly those $\chi$\footnote{Recall from Definition~\ref{def:language} that we denote atomic propositions with $\chi$.} such that $\chi_i\in  \lang{}^{|t} $. 
\item A \emph{$t$-bounded model} $\restm{}$ is the pair $(\restt{},\restp{})$ where $\restt{}=\{\pi_1^{|t}\mid \pi_1\in T\}$. 
\end{itemize}
We denote the set of all $t$-bounded models with $\restms{}$. 
\end{definition}

Recall we defined a set of models of strong beliefs \msbset{} as an \emph{msb set} (Definition~\ref{def:msbset}). A belief database $SB$ consists of a set of strong beliefs, and we showed in Proposition~\ref{prop:mod-sb-is-msbset} that the set of models of $SB$ is an msb set, i.e. $Mod(SB)\in\msbsets{}$, where $\msbsets{}$ is the set containing all msb sets.

In order to represent revision semantically, we define a $t$-bounded version of msb sets as well.

\begin{definition}[$t$-bounded msb set]
Given an msb set \msbset{} (definition~\ref{def:msbset}), the \emph{$t$-bounded msb set} contains all $t$-bounded models of \msbset{}, i.e. $$\msbsetrest = \{\restm \mid m\in \msbset\}.$$
\end{definition}

Given an intention database $I$, we define a selection function $\sel$ that tries to accommodate a new intention based on strong beliefs. The selection function specifies preferences on which intention an agent would like to keep in the presence of the new beliefs. 

\begin{definition}[Selection Function]
\label{def:selection-function}
Given an intention database $I$, a \emph{selection function} $\sel:\msbsets{}\times \ib \rightarrow \idb$ maps an msb set (Definition~\ref{def:msbset}) and an intention to an updated intention database---all bounded up to $t$--- such that if $\sel(\msbsetrest,i)=I'$, then:
\begin{enumerate}
\item $(\msbsetrest,I')$ is coherent.
\item If $(\msbsetrest,\{i\})$ is coherent, then $i\in I'$.
\item If $(\msbsetrest,I\cup\{i\})$ is coherent, then $I\cup\{i\}\subseteq I'$.
\item $I'\subseteq I\cup\{i\}$.
\item For all $I''$ with $I'\subset I''\subseteq I\cup\{i\}$:$(\msbsetrest,I'')$ is not coherent.
\end{enumerate}
\end{definition}

The five conditions on the selection function are in direct correspondence with postulates P1--P5 of the intention revision function. 

\begin{remark}
We will show in Corollary~\ref{cor:strongbeliefformula} below that it is possible to represent each set of strong beliefs $SB$ (Definition~\ref{def:strbel}) by a formula $\psi$ such that $Cn(SB)=Cn(\psi)$. Using this corollary, we adapt the definition of a Katsuno and Mendelzon faithful assignment below.
\end{remark}

\cite{Katsuno1991} define a faithful assignment from a belief formula to a pre-order over models. Since we are also considering intentions, we extend this definition such that it also maps intentions databases to selection functions.

\begin{definition}[Faithful assignment]
\label{def:faithful-assignment}
A \emph{faithful assignment} is a function that assigns to each strong belief formula $\psi\in \sbel\rest{}$ a total pre-order $\lept$ over $\mods{}$ and to each intention database $I\in \mathbb{D}\rest{}$ a selection function $\sel$ and satisfies the following conditions:
\begin{enumerate}
\item If $m_1,m_2\in Mod(\psi)$, then $m_1\lept m_2$ and $m_2\lept m_1$.
\item If $m_1\in Mod(\psi)$ and $m_2\not\in Mod(\psi)$, then $m_1< m_2$.
\item If $\psi\equiv\phi$, then $\lept = \le_\phi^t$.
\item If $\restt=T_2\rest{}$, then $(T,\pi)\lept (T_2,\pi_2)$ and $(T_2,\pi_2)\lept (T,\pi)$. 
\end{enumerate}
\end{definition}

Conditions 1 to 3 on the faithful assignment are the same as the conditions that Katsuno and Mendelzon put on a faithful assignment. Condition 4 ensures the two difficulties we pointed out in the beginning of this subsection are handled correctly:
\begin{itemize}
\item It ensures we do not distinguish between models in the total pre-order $\lept$ whose trees are the same up to time $t$. This is essentially what is represented in the revision function by bounding the all input of the revision function $*_t$ up to $t$. 
\item Moreover, $\lept$ does not distinguish between models obtained by selecting two different paths from the same tree. In other words, it ensures that msb sets (sets of models of a strong belief) remain in the same ordering. This corresponds to the fact that we are using strong belief formulas in the revision, which do not distinguish between different paths in the same tree as well.
\end{itemize}

We are now ready to state our main theorem. The full proof can be found in Appendix B.

\begin{theorem}[Representation Theorem]
\label{thm:singlestep}
The function  $\agr:\bis{}\times (\sbel{}\times \ib{})\rightarrow \bis$ is a belief-intention revision operator  
iff there exists a faithful assignment that maps each $\psi$ to a total pre-order $\lept$ and each $I$ to a selection function $\sel$ such that if $(\psi,I)*_t(\varphi,i)=(\psi',I')$, then:
\begin{enumerate}
\item $Mod(\psi')=\min(Mod(\varphi),\lept)$
\item $I' = \sel(Mod(\psi'),i)$
\end{enumerate}
\end{theorem}

We will use the remainder of this section to prove some results that we use for the proof of the representation theorem above. We first show the number of $t$-bounded models is finite. 

\begin{lemma} 
\label{lemma:finite-restricted-model}
For each $t\in\timeflow{},  \restms{}$ is finite. 
\end{lemma}

\begin{proof}
Suppose some $t\in\timeflow{}$. Since actions are deterministic and there are finitely many actions in our logic, each state has a finite number of successor states. Moreover, since there are finitely many propositions in  $ \lang{}^{|t}$, the number of possible valuations of the states is finite as well. Therefore, the number of models in $\restms{}$ is finite.
\end{proof}

The following lemma obtains a correspondence between semantic consequence of two models equivalent up to $t$. The proof is by induction on the depth of the formula.

\begin{lemma} 
\label{lemma:equivalent-entailment}
For each $\varphi\in \lang{}^{|t}$ and models $m_1,m_2\in\mods$, if $m_1^{|t}=m_2^{|t}$, then $m_1\models\varphi$ iff $m_2\models\varphi$.
\end{lemma}

Let $Ext(\msbsetrest)$ be the set of all possible extensions of a $t$-bounded msb set \msbsetrest{} to models, i.e. $$Ext(\msbsetrest) = \{m\in\mods{}\mid \restm{}\in \msbsetrest{}\}.$$ We next show that we can represent $\msbsetrest$ by a single strong belief formula using $Ext(\msbsetrest)$.

\begin{lemma}
Given a $t$-bounded msb set $\msbsetrest{}$, there exists a strong belief formula $form(\msbsetrest{})\in\mathbb{SB}$ such that $Mod(form(\msbsetrest))=Ext(\msbsetrest)$.
\end{lemma}

The following corollary shows we can represent a belief database consisting of strong beliefs up to some time $t$ with a single formula.

\begin{corollary}
\label{cor:strongbeliefformula}
Given a $t$-bounded set of strong beliefs $SB^{|t}$, there exists a strong belief formula $\psi\in\mathbb{SB}$ such that $SB^{|t}=\{\varphi\mid\psi\vdash\varphi\}$.
\end{corollary}

\subsection{Examples}
\label{sect:examples}

In this section we discuss various examples of revision of beliefs and intentions in our framework. We start with the example from the introduction, in which Bobby the household robot can only buy food and cleaning equipment if it robs a bank.

\begin{example}[Running example (robbing a bank)]
\label{expl:robbing}
Consider the following undesired situation. Suppose Bobby has a belief-intention database $(\psi, I)$ and that $I=\{(food, 0)\}$, i.e. Bobby intends to buy food at time 0. Now suppose Bobby would revise by the pair $(\top,(equip, 2))$: it would like to buy cleaning equipment at time 2. This is coherent (despite Bobby's belief that it does not have enough money for both) thanks to an action $rob$ of robbing a bank that it can perform at time 1. Suppose moreover that this is the only model where Bobby can perform both $(food, 0)$ and $(equip, 2)$. Then by postulate P2 we obtain that the resulting belief-intention database entails $do(rob)_1$, even if it would like to avoid robbing a bank by any means.

In other words, it seems that the weak beliefs of an agent may entail actions it did not intend to do, if those actions are the only means to carry out the intended actions. However, if Bobby would like to avoid robbing a bank by all means, then this should follow  explicitly in its strong belief database, i.e., for all $\ell$ (up to the considered  time  $t$) $$\psi\vdash \Box_0\neg do(rob)_\ell .$$
Given that this is a strong belief, the agent does not believe in any possible futures in which it rob a bank. A consequence of this is that the new intention $(equip, 2)$ will not be incorporated into the intention database, because the belief-intention database is not coherent after revision. If at some point later in time the agent would consider it possible to rob a bank, then it should revise its strong beliefs accordingly, but until that time the undesired action will not be weakly believed.
\end{example}

The above example shows that the agent can be coherent (it has the preconditions for all intended actions by postulate P2), but unaware of its future action to rob the bank (following from its weak beliefs). As shown in the example, this can be avoided by asserting this explicitly in the logic, however, due to belief revision this can change. If the agent really would never like to do this action, then this action should simply not be part of the available actions of the agent.

\begin{example}[Running example (adding an intention)]
\label{expl:addingintention}
Suppose a belief-intention database $(\psi,I)$ such that all models in $Mod(\psi)$ are the same as the partial model in Figure~\ref{fig:example_pal_model2} up to $t=2$ and suppose that $I=\{(food,0),(cook,1)\}$. That is, Bobby has the intention to buy food at time 0 and then to cook at time 1. Suppose now Bobby changes its intention to buy cleaning equipment at time 1. Formally: $$(\psi,I)*_1 (\top,(equip,1))=(\psi,I').$$  
First note $(\psi,I\cup\{(equip,1)\})$ is not coherent because no two intentions can occur at the same time moment. Moreover, since $(\psi,(equip,1))$ is coherent, from (P2) and (P3) we obtain $(equip,1)\in I'$. Furthermore, from (P4) we have that $I'\subseteq \{(food,0),(cook,1),(equip,1)\}$. Finally, $(\psi,\{(food,0),(equip,1)\})$ is not coherent either, since the agent does not believe the preconditions of buying food and buying equipment are true along a single path. Combining this gives $I'=\{(equip,1)\}$ as the only coherent outcome. Thus, Bobby no longer intends to buy food and to cook, but to buy cleaning equipment instead.

Note that, although the strong beliefs didn't change after revising with the new intention, the weak beliefs did change. For example, $post(food)_1\in WB(\psi,I)\setminus WB(\psi,I')$ and $post(equip)_2\in WB(\psi,I')\setminus WB(\psi,I)$.
\end{example}

The revision function $*_t$ takes a tuple $(\varphi,i)$ as input, and Definition \ref{belief-intention revision function} ensures that revision of strong beliefs occurs prior to the revision of intentions. Therefore, it may seem plausible that revising with $(\varphi,i)$ is the same as first revising with $(\varphi,\epsilon)$ and then with $(\top,i)$. In other words, the following postulate seems to follow:

\begin{align*}
\text{If }&(\psi,I)*_t(\varphi,i)=(\psi',I')\\
\text{ and }&((\psi,I)*_t (\varphi,\epsilon))*_t (\top,i)=(\psi'',I''), \tag{P*}\\
&\text{then }\psi'\equiv\psi''\text{ and }I'=I''.
\end{align*}

However, this property is not sound, and we show in the following example that adding the postulate would in fact conflict with the maximality postulate for intention revision (P5).

\begin{example}[Joint vs separate revision]
\label{expl:jointvsseparaterevision}
The operator $\agr$ can be instantiated to separate revisions  of belief-intention databases as follows:
\begin{itemize}
\item Revising by $(\top,i)$ mirrors revising by no belief and an intention $i$, i.e.
$$
(\psi,I)\agr(\top,i) =(\psi,I)\agri i.
$$
\item Revising by $(\varphi,\epsilon)$ mirrors revising by a strong belief $\varphi$ and no intention. 
\end{itemize}
In spite of the fact that the operator $\agr$ revises beliefs prior to intentions, it is more expressive than its two instantiated operators combined, and cannot be defined as their composition. Indeed, 
$$
((\psi,I)\agr(\varphi,\epsilon))\agr(\top,i)\neq (\psi,I)\agr(\varphi,i).
$$
This follows from the following example: Suppose some belief-intention database $(\psi,I)$ with beliefs up to $t=2$ corresponding to the model on the left of Figure~\ref{fig:revision_example}. It is possible to go to the dentist ($dentist$) or to stay at work ($work$), and after that to go eating ($eating$) or go to the movies ($movies$). 

Before revision, the intentions are $I=\{(dentist,0),(eat,1)\}$ (left image of Figure~\ref{fig:revision_example}, intentions shown as bold lines). 

\begin{figure}[ht!]
\centering
\begin{minipage}{.4\textwidth}
\begin{tikzpicture}[level distance=2.5cm, 
	level 1/.style={sibling distance=2.5cm},
	level 2/.style={sibling distance=1.5cm},
	emph/.style={edge from parent/.style={ultra thick,draw}},
	norm/.style={edge from parent/.style={black,thin,draw}},
	grow'=right]
  \coordinate
    child{
      child{node{} edge from parent node[above]{$eat$}}
      child{node{} edge from parent node[below]{$movie$}}
      edge from parent node[above]{$work$}
    }
    child[emph]{
    child{node{} edge from parent node[above]{$eat$}}
      child[norm]{node{} edge from parent node[below]{$movie$}}
      edge from parent node[below]{$dentist$}
    }
  ;
\end{tikzpicture}
\end{minipage}
\hspace{1cm}
\begin{minipage}{.4\textwidth}
\begin{tikzpicture}[level distance=2.5cm, 
	level 1/.style={sibling distance=2.5cm},
	level 2/.style={sibling distance=1.5cm},
	emph/.style={edge from parent/.style={ultra thick,draw}},
	red/.style={edge from parent/.style={ultra thick,draw=red,dashed}},
	norm/.style={edge from parent/.style={black,thin,draw}},
	invisible/.style={edge from parent/.style={white,thin,draw}},
	grow'=right]
  \coordinate
    child{
      child{node{} edge from parent node[above]{$eat$}}
      child{node{} edge from parent node[below]{$movie$}}
      edge from parent node[above]{$work$}
    }
    child{
    child[invisible]{node{} edge from parent node[above]{}}
      child{node{} edge from parent node[below]{$movie$}}
      edge from parent node[below]{$dentist$}
    }
  ;
\end{tikzpicture}
\end{minipage}
\caption{Left: Partial model of strong beliefs $\psi$ of agent $(\psi,I)$ with $I=\{(dentist,0),(eat,1)\}$ (bold lines). Right: Revised strong beliefs of agent after learning it is not possible to eat ($eat$) after the dentist ($dentist$).}
\label{fig:revision_example}
\end{figure}
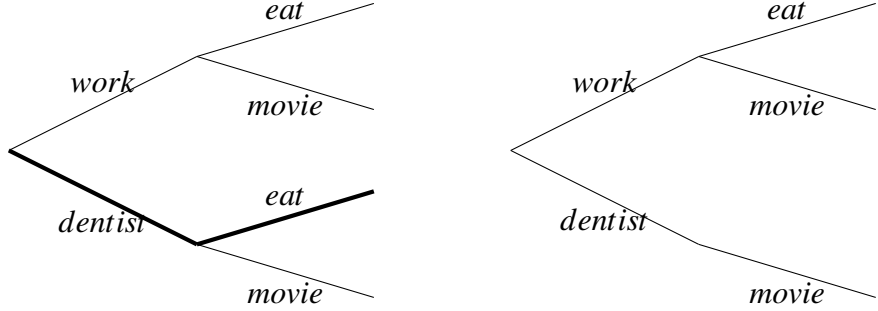

Suppose now the beliefs are revised with the fact that it is not possible to go eating after going to the dentist ($\varphi$), i.e., $$\varphi \equiv \Box_0 (post(dentist)_t \rightarrow \neg pre(eating)_t),$$ and with the intention to go to the movie at time 1 ($i=(movie,1)$). The resulting strong beliefs after revising with $\varphi$ are shown on the right of Figure~\ref{fig:revision_example}.

Let us analyze two ways of revising this information:
\begin{itemize}
\item Suppose $(\psi,I)$ is revised with both the new belief and the new intention. That is, $$(\psi,I)*_2(\varphi,i)=(\psi',I').$$Both $(\psi',\{(dentist,0),(movie,1)\})$ and $(\psi',\{(movie,1)\})$ are coherent, so by the maximality postulate (P5), $I'=\{(dentist,0),(movie,1)\}$. Hence, the new intentions are to go to the dentist and then to go to the movie.
\item Suppose beliefs are revised prior to intentions. That is,
\begin{align*}
(\psi,I)*_2(\varphi,\epsilon)&=(\psi',\overline{I})\\
(\psi',\overline{I})*_2(\top,i)&=(\psi',\overline{I}').
\end{align*}
Now, since $(\psi',\{(dentist,0)\})$ and $(\psi',\{(eating,1)\})$ are both coherent, we either have $\overline{I}=\{(dentist,0)\}$ or $\overline{I}=\{(eating,1)\}$. Suppose that $\overline{I}=\{(eating,1)\}$. In that case, since $(\psi',\{(eating,1),(movie,1)\}$ is incoherent, we obtain $\overline{I}'=\{(movie,1)\}$ by the postulates (P2) and (P4).  
\end{itemize}

Thus, we see that revising separately allows a choice between the intention to go eating or to go to the dentist after revising beliefs. When choosing to go eating, the intention again has to be discarded because it is conflicting with the new intention to go to the movie. In joint revision, this is not the case since the choice between eating or the dentist can be made in light of the new incoming intention, and the maximal set can be chosen.
\end{example}

\section{Iterated revision}
\label{sect:iterated-revision}

Until now we only considered single-step revision of belief and intention. In this part we develop an account of iterated revision by following the approach of~\cite{Darwiche1997} developed for propositional logic. They observe that the AGM postulates are too permissive to enforce plausible iterated revision. In order to remedy this, they suggest the following changes:
\begin{itemize}
\item Instead of performing revision on a propositional formula, perform revision on an abstract object called an \emph{epistemic state} $\Psi$, which contains the entire information needed for coherent reasoning. 
\item Postulate (R4) is weakened as follows:

(R*4) If $\Psi=\Psi'$ and $\varphi\equiv\varphi'$, then $\Psi\circ\varphi\equiv\Psi'\circ\varphi'$
\item The following four desirable postulates are added for iterated revision:

(C1) If $\varphi\models\varphi'$, then $(\Psi\circ\varphi')\circ\varphi\equiv\Psi\circ\varphi$.

(C2) If $\varphi\models\neg\varphi'$, then $(\Psi\circ\varphi')\circ\varphi\equiv\Psi\circ\varphi$.

(C3) If $\Psi\circ\varphi\models\varphi'$, then $(\Psi\circ\varphi')\circ\varphi\models\varphi'$.

(C4) If $\Psi\circ\varphi\not\models\neg\varphi'$, then $(\Psi\circ\varphi')\circ\varphi\not\models\neg\varphi'$
\end{itemize}

Postulate (C1) states that if two beliefs are added, the first is redundant if the second one is more specific. In other words, only revising with the second belief would obtain the same belief set. (C2) states that if the first belief is inconsistent with the second one, then revising with the first belief is unnecessary. (C3) ensures that a belief should be retained when revising with another belief implies it. Finally, (C4) states that if a belief $\varphi$ is not contradicted after revising with another belief $\varphi'$, then it should remain uncontradicted when the revising by $\varphi'$ is preceded by $\varphi$.

We now define the revision operator for our logic, assuming that an epistemic state $\Psi$ contains belief, which is represented by a strong belief formula denoted by $Bel(\Psi)$.
As usual, we assume that  that $\Psi$ stands for $Bel(\Psi)$ whenever it is embedded in a formula, and that $Bel(\Psi)$ is thus a set of strong beliefs. For example, we say that $(\Psi, I)$ is \emph{coherent} if $(Bel(\Psi), I)$ is coherent. Recall 
that we omit the temporal restriction superscript from sets if it is clear from context.

\begin{definition}[Bounded epistemic state revision function]
\label{def:belief-postulates} 
A \emph{bounded epistemic state revision function} is a function  $\brt$ which maps an epistemic state  and a strong belief formula--- all bounded up to $t$--- to a strong belief database bounded up to $t$, and which satisfy the  postulates R1--R3, R*4, R5, R6 and C1--C4.
\end{definition}

Similar to Section~\ref{sect:revision:postulates}, we define a bounded revision function up to a fixed time $t$, assuming that the strong beliefs and intentions in an epistemic belief-intention databases are all bounded up to $t$.

\begin{definition}[Epistemic belief-intention database]
\label{def:epistemic-belief-intention-database}
An \emph{epistemic belief-intention database} $(\Psi,I)$ consists of an epistemic state  $\Psi$, and an intention database $I$. The set of all epistemic belief-intention databases bounded up to $t$ is denoted by $\restebis{}$, or simply $\ebis{}$ if $t$ is clear from context.
\end{definition}

Next we define the intention revision operator for an epistemic belief-intention database. We omit the formal definition of the intention revision function on epistemic belief-intention databases, but it is defined analogously to Definition~\ref{def:intention-postulates}. When we revise an epistemic belief-intention database with an intention, we leave the epistemic state $\Psi$ unchanged and we update the intention database by applying Definition ~\ref{def:intention-postulates} to $Bel(\Psi)$.

\begin{definition}[Iterated revision of epistemic belief-intention databases]
\label{def:iterated-revision-postulates} 
An \emph{epistemic belief-intention revision function} 
 is a function $\agr:\ebis{}\times (\sbel{}\times \ib{})\rightarrow \ebis{}$ of the form
$$
(\Psi,I)\agr(\varphi,i) =(\Psi\brt \varphi,I)\agri i, 
$$
where $\brt$ is an epistemic state revision function and $\agri$ is an intention revision function.
\end{definition}

When switching from belief revision on a belief state to belief revision on an epistemic state the definition of a faithful assignment should be adopted accordingly. We will do this now for our setting.

\begin{definition}[Faithful assignment for iterated revision]
\label{def:tbounded:faithfulassigment:epst}
A \emph{$t$-bounded faithful assignment for iterated revision} is a function that assigns to each epistemic state $\Psi$ a total pre-order $\le_\Psi^t$ on all models,  and to each intention database $I\in \mathbb{ID}\rest{}$ a selection function $\sel$ (Definition \ref{def:selection-function}), such that the following conditions hold:

\begin{enumerate}
\item If $m_1,m_2\in Mod( \Psi)$, then $m_1\le_\Psi^t m_2$ and $m_2\le_\Psi^t m_1$
\item If $m_1\in Mod( \Psi)$ and $m_2\not\in Mod( \Psi)$, then $m_1<_\Psi^t m_2$
\item   $\Psi=\Phi$ only if $\le_\Psi=\le_\Phi$
\item If $\restt=T_2\rest{}$, then $(T,\pi)\le_\Psi^t (T_2,\pi_2)$ and $(T_2,\pi_2)\le_\Psi^t (T,\pi)$.

\smallskip

If $(\Psi,I)\agr(\varphi,i) = (\Psi',I')$, then

\item If $m_1\in Mod(\varphi)$ and $m_2 \in Mod(\varphi)$, then $m_1\le_\Psi^t m_2$ iff $m_1\le_{\Psi'}^tm_2$.

\item If $m_1\not\in Mod(\varphi)$ and $m_2\not \in Mod(\varphi)$, then $m_1\le_\Psi^t m_2$ iff $m_1\le_{\Psi'}^tm_2$.
\item If $m_1\in Mod(\varphi)$, $m_2\not \in Mod(\varphi)$ and $m_1<_\Psi^t m_2$, then $m_1<_{\Psi'}^tm_2$.
\item If $m_1\in Mod(\varphi)$, $m_2\not \in Mod(\varphi)$  and $m_1\le_\Psi^t m_2$, then $m_1\le_{\Psi'}^tm_2$.
\end{enumerate}
\end{definition}

The first four conditions are similar to the conditions on a faithful assignment for single-step revision (Definition~\ref{def:faithful-assignment}), with the difference that $\lept$ is replaced with $\le_\Psi^t$, and that Condition 3 has epistemic states in the antecedent instead of strong belief formulas. Conditions 5-8 are the semantic counterpart of (C1)-(C4).

\begin{theorem}[Representation Theorem for iterated revision]
\label{thm:singlestepA}
 A function $\agr:\ebis{}\times (\sbel{}\times \ib{})\rightarrow \ebis{}$  is an epistemic belief-intention revision operator  iff there exists a faithful assignment for iterated revision that maps each $\Psi$ to a total pre-order $\le_\Psi^t $ 
%
 and each $I$ to a selection function $\sel$
  such that if $(\Psi,I)*_t(\varphi,i)=(\Psi',I')$, then:
\begin{enumerate}
\item $Mod(\Psi')=\min(Mod(\varphi),\le_{\Psi}^t)$
\item $I' = \sel(Mod(Bel(\Psi')),i)$
\end{enumerate}
\end{theorem}

We next provide a concrete epistemic belief-intention revision operator, thus showing  \emph{consistency} of all the postulates proposed for the epistemic state and intention revision operators.

Our operator is based on  Spohn's \emph{ordinal conditional ranking functions}, which can be seen as representations of epistemic states. For a given time $t$, our ranking function $\kappa_t$ deals with epistemic states of our logic in a similar way  as the operator based on Spohn's ranking function for propositional epistemic states, introduced in~\cite{Darwiche1997}. 

\begin{definition}[Spohn ranking function]
\label{def:spohnA}
A \emph{Spohn ranking function} $\kappa_t:\mods\rightarrow\timeflow{}$ assigns a rank to each model such that:

If $m_1=(T_1,\pi_1)$ and $m_2=(T_2,\pi_2)$ such that $T_1\rest{} = T_2\rest{}$, then $\kappa_t(m_1)=\kappa_t(m_2)$.\\

\noindent
We extend the ranking to propositions as follows: $$\kappa_t(\varphi)=\min_{m\models\varphi}\kappa_t(m).$$
\end{definition}

\begin{definition}[Accepted propositions]
\label{def:accepted-propositions}
Given a ranking function $\kappa_t$, the \emph{accepted propositions} $Bel(\kappa_t)$ are those for which the negation is implausible: $$Bel(\kappa_t) = \{\varphi\mid \kappa(\neg\varphi)>0\}.$$ 
\end{definition}

It follows that the models of these propositions are those which have rank 0: $$Mod(Bel(\kappa_t)) = \{m\mid \kappa_t(m)=0\}.$$ The fact that $Bel(\kappa_t)$ is a set of strong beliefs follows from the condition we put on the Spohn ranking function.

Now we define our revision operator. For the epistemic part  we follow~\cite{Darwiche1997}, while for intention revision we give prefer intentions that occur sooner rather than later.

\begin{definition}[Extended Spohn-based revision operator]
\label{def:spohn-iterated-revision-operator}
The \emph{Extended Spohn-based revision operator $\spohn$} is defined as follows:

$(\kappa_t,I)\spohn(\varphi,i) = (\kappa_t',I')$, where  $\kappa_t'$ and $I'$ are  such that\footnote{We show in Appendix B that $\kappa_t'$ is a well defined Spohn ranking function.}
\begin{align*}
\kappa_t'(m) = \begin{cases}
\kappa_t(m)-\kappa_t(\varphi), &\text{ if } m\models \varphi;\\
\kappa_t(m)+1,&\text{ if } m\models \neg \varphi,
\end{cases}
\end{align*}
and $I'$ is defined in the following iterative way. If $n$ is the maximal time instance that occurs in $I$, we define the sets of intentions

\begin{enumerate}
\item $I_{-1} =\{ i\}$, if $(Bel(\kappa_t'),\{ i \})$ is coherent, otherwise $I_{-1} = \emptyset$.
\item for every $\ell\in\{0,1,\dots,n\}$

\begin{enumerate}
\item if there is no $a$ such that $(a,\ell)\in I$, $I_{\ell} =I_{\ell-1}$.

\item otherwise, if $a$ is  the unique action such that $(a,\ell)\in I$, then  $I_{\ell} =I_{\ell-1}\cup\{ (a,\ell)\}$, if 
$(Bel(\kappa_t'),I_{\ell-1}\cup\{ (a,\ell)\})$ is coherent, otherwise $I_{\ell} =I_{\ell-1}$.

\end{enumerate}
\item $I'=I_n$.
\end{enumerate}

\end{definition}

We now pose our theorem.

\begin{theorem} The function  $\spohn$ is an epistemic belief-intention revision operator.
\end{theorem}

The Extended Spohn-based revision operator is an example of an operator satisfying the epistemic belief-intention revision function for iterated revision (Definition~\ref{def:iterated-revision-postulates}) and puts additional constraints on intentions. It prioritizes the new intention $i$, and after that it iteratively attempts to increase the set of new intentions by adding them in temporal order, starting with the most recent ones. We next give an example of how this may work in practice.

\begin{example}[Running example (Extended Spohn-based revision)]
Suppose\\  that 
Bobby the household robot has an epistemic  belief-intention database $EBI$ with intentions $I=\{(food, 0), (equip, 1)\}$, and suppose its epistemic state is formalized as:
\begin{itemize}
\item $\kappa_t(m) = 0$, for every $m\in Mod(\varphi_0)$, where
$$\varphi_0=\Box_0 pre(food, equip)_0$$ (``Bobby has precondition to buy both food and  cleaning equipment'')

\item $\kappa_t(m) = 1$,   for every $m\in Mod(\varphi_1)$, where
$$\varphi_1=\Box_0\neg pre(food,equip)_0\wedge \Box_0 pre(food)_0\wedge \Box_0 pre(equip)_1$$ (``Bobby  can carry  out any of two intended actions, but not both of them'')

\item $\kappa_t(m) = 3$,   for every $m\in Mod(\varphi_3)$, where
$$\varphi_3=\Box_0 \neg pre(food)_0 \wedge \Box_0 \neg pre(equip)_1$$ (``Bobby  cannnot carry  out any of two intended actions'')

\item $\kappa_t(m) = 2$,   for all other models, i.e., for  $m\in Mod(\varphi_2)$, where $\varphi_2=\neg\varphi_0\wedge\neg\varphi_1\wedge\neg\varphi_3$
(note that we can read $\varphi_2$ as ``Bobby  can buy only the cheaper of the two items'')

\end{itemize}
Thus, Bobby's beliefs are represented by $Bel(\kappa_t)$ (Definition~\ref{def:accepted-propositions}), which include that it believes it is possible to carry out both intended actions, so the agent is coherent. In addition, the epistemic state specifies a preference ordering over models as well, which is specified by the the ranking $\kappa$. It assumes that ``Bobby  cannnot carry  out any of two intended actions'' is the least plausible proposition.

Suppose next that Bobby revises its epistemic belief-intention database with\\
$(\Box_0\neg pre(food, equip)_0,\epsilon)$. The ranking function selects the minimal (i.e., the most plausible) model in which the revised formula is consistent, which has ranking 1. After revision, Bobby believes it is able to buy food or buy cleaning equipment, i.e., any of those two, but not both. The new ranking $\kappa'_t$ is as follows:
\begin{itemize}

\item $\kappa_t'(m) = 0$, for every $m\in Mod(\varphi_1)$

\item $\kappa_t'(m) = 1$,   for every $m\in Mod(\varphi_0\vee\varphi_2)$

\item $\kappa_t'(m) = 2$,   for every $m\in Mod(\varphi_3)$

\end{itemize}
The agent can then use $\kappa_t'$ for future revision. Using the intention revision operation of Definition~\ref{def:spohn-iterated-revision-operator}, it will keep the intention that it intends to carry out first, which means $I'=\{(food, 0)$, and the intention to buy cleaning equipment is dropped.
\end{example}
\section{Related Work}
\label{sect:relatedwork}

We compare our work to the logic of intention, temporal logic, collective intentions, and our own previous work.

\subsection{Logic of intention}

We first give an overview of the literature on the logic of intention, before making a comparison with the theory introduced in this paper. In the overview, we discuss the philosophy of mind, the logic of intention of Cohen and Levesque, and other work.

\subsubsection{Philosophy of mind}

Before the early 1990s most work on intention was done by philosophers. The dominant view prior to the 1980s was that fundamental mental states included belief-like attitudes such as belief and knowledge on the one hand, and desire-like attitudes such as desires and preferences on the other~\citep{Davidson1963}. These capture the ``two directions of fit'' between world and mind~\citep{searle1983}. It was generally believed that intentions could be reduced to these more basic mental states. For instance, an intention can be thought of as a complex kind of belief: an intention to $\varphi$ at some time $t$ might be a belief that the agent will $\varphi$ at $t$, perhaps by virtue of this very belief~\citep{Holton2008}. This \emph{belief-desire} model of the mind was also common in decision theory in the spirit of 
\cite{savage1972foundations}, as well as in artificial intelligence. The basic idea underlying all this is that rational action amounts to what leads to the most \emph{desirable} outcome, given the agent's \emph{beliefs} about the world and \emph{preferences} about how it ought to be.

In the mid 1980s Michael Bratman wrote the book ``Intentions, plans, and practical reason''~\citep{bratman1987}, which turned out to be very influential in both philosophy and in artificial intelligence. In the book, he argued convincingly that intentions cannot be reduced to any ``more basic'' mental states, but that intentions must be taken as basic themselves. While his argument is quite complex for non-philosophers, the fundamental ideas are very simple. Intentions are subject to their own set of norms, which cannot be reduced to those for beliefs or desires. For instance, there are norms for intention consistency that do not apply to desires: if someone knows that $\varphi$ and $\psi$ are mutually inconsistent, it is irrational to intend $\varphi$ and to $\psi$ simultaneously, but it would not be irrational to desire both $\varphi$ and $\psi$.

Another example of a norm of intentions, and one which has been very influential in computer science, is the idea that intentions constrain further practical reasoning in a characteristic way. Bratman (and others) have argued that if someone intends to $\varphi$, that gives her a \emph{pro tanto} reason to $\varphi$, even in the face of evidence that she ought not $\varphi$. If we think of intentions as making up a plan, then reconsidering an intention potentially requires revising an entire plan, which can be computationally expensive and problematic in real time. A consequence of this \emph{relative resistance to revision} is that intentions can be very useful for resource-bounded agents who have to select appropriate actions at different times. Both by adopting appropriate \emph{policies} that can apply at different times to similar decision problems, and by devising complex plans which can easily be followed when computation time is limited, an agent with intentions is able to avoid performing complex computations every time a new decision problem arises. Observations such as those above are common knowledge in artificial intelligence by now, but it is important to realize there is a well explored philosophical foundation underlying them. 

\subsubsection{Cohen and Levesque}
\label{sect:cohen-levesque}
Inspired by the philosophical literature, research in artificial intelligence in the 1990s began to explore intentions. One of the first attempts of what was later known as the \emph{belief-desire-intention architecture} was in a paper by~\cite{bratman1988plans}, which systematized many aspects of the view laid out in Bratman's book~\citep{bratman1987}, including flowcharts and some suggestions for implementations. Most importantly, it served as a call for other researchers to investigate these matters more systematically and formally.

One of the most well-known and first attempts of a formalization of intention revision was by 
\cite{Cohen1990}. They developed a formal logical language, including modal operators for mental state such as ``belief'' and ``having a goal'', for temporal expressions, and for descriptions of events such as ``A happens'' or ``$x$ did action $a$''. The main point of the paper is to define a useful notion of intention. To this end they first define a \textsc{P-GOAL}, or \emph{persistent goal}, which is a goal $p$ that one believes now not to hold, and which will cease to be a goal as soon as either the agent believes $p$ will never hold, or the agent comes to believe $p$ is true. Intending to take an action $a$ is then defined in terms of a \textsc{P-GOAL} for agent $x$~\cite[p.245]{Cohen1990}: $$\textsc{INTEND}(x,a) := (\textsc{P-GOAL}x[\textsc{DONE}x(\textsc{BEL}x(\textsc{HAPPENS}a))?;a]).$$

Unpacking this, agent $x$ intends to $a$ just in case $x$ has a persistent goal to ensure that $x$ will believe $a$ will be carried out, up until $a$ is in fact carried out. They also define ``intending to bring about some state of affairs'', and show their approach solves the ``Little Nell'' problem (see Section~\ref{sect:littlenell}) and avoid the ``Dentist Problem''~\citep{bratman1987}.

While Cohen and Levesque's approach is much cited, it is fair to say that it is rather complicated. Some early criticisms of technical details can be found in~\cite{Singh1992}. In the textbook by~\cite{SLB08} the approach is called ``the road to hell". Due to the complexity of the logic,  mathematical properties such as axiomatizability, decidability and complexity of fragments were never investigated. None of the BDI logics that were introduced subsequently adapted Cohen and Levesque's four steps definition of intention and instead considered intentions to be primitive. Moreover, while Cohen and Levesque provide some criteria for the abandonment of intentions through the notion of rational balance (forbidding to intend something that is true or believed to be impossible to achieve), it does not further analyze the 'other reasons' for which a persistent goal is abandoned. More details on these critiques can be found in a recent articles by 
\cite{Herzig2016b}.

Some of Cohen and Levesque's shortcomings were corrected to an extent in subsequent work. For instance, \cite{Rao1991} provide an approach for an alternative, and arguably simpler, formalism based on CTL. 

\subsubsection{Other related work on intention revision}

There exists previous work on intention dynamics, that is, how intentions change over time. \cite{vanderHoek2007} (see also ~\cite{Wooldridge2000} and~\cite{Wooldridge1995}) explore a system similar to Rao and Georgeff's with representations of beliefs, desires, and intentions separately, including a specific module for practical reasoning; they propose how to revise intentions together with beliefs. Their treatment is mainly syntactical, and they do not attempt to obtain a correspondence between postulates for revision and a pre-order over model, which is what we do here. ~\cite{grant2010} continue this line of work, offering postulates for intention revision (similar to our contribution). They develop AGM-style postulates for belief, intention, and goal revision. They provide a detailed analysis and propose different reconsideration strategies, but restrict themselves to a syntactic analysis as well. \cite{Lorini2008} introduce a logic of intention with modal operators for attempt, tacking the question of when an agent's intentions translate into an actual attempted action. Finally, \cite{Shapiro2012} also explore intention revision, working in a setting with complex hierarchical plans.

\subsection{Comparison: The interplay between intention, time, and belief}

To compare our theory with the one of Cohen and Levesque, we reconsider the running example.
You instruct your future household robot Bobby to buy groceries for you in the afternoon. Bobby therefore believes to be in the supermarket in the afternoon, but this belief is based on the assumption that it has sufficient money to buy a bus ticket. Moreover, you instruct Bobby to clean the windows coming weekend, and Bobby's belief that it will clean the windows is based on the assumption that it has all the necessary cleaning products on time, and that it is not raining at that day. These assumptions are the preconditions of the actions Bobby intends to perform. When Bobby adopts intentions, it may form new beliefs based on these intentions. For instance, after Bobby adopts the intention to buy groceries in the afternoon, it is able to adopt new intentions based on this belief, for instance to cook food in the evening. Such intention-based beliefs are called \emph{weak beliefs}~\citep{vanderHoek2007,Icard2010}. In contrast, unconditional beliefs, beliefs that are no dependent on the predicted success of intentions, are called \emph{strong beliefs}.

Our theory can explain how intentions and weak and strong beliefs change over time. Yesterday Bobby believed that it will rain tomorrow and be sunny the day after, but today it believes that tomorrow the sun may possibly shine and it will be cloudy the day after. Likewise, yesterday Bobby intended to buy groceries tomorrow, but now it intends to do laundry tomorrow. We say there is internal dynamics (rain today and sun tomorrow) and external dynamics (Bobby's beliefs today and Bobby's beliefs tomorrow, or Bobby's intentions today and Bobby's intentions tomorrow). 

\subsubsection{Assumptions}

The first improvement is the formalisation of the assumptions of the intentions, i.e. the preconditions of the intended actions. From the intention to take the train to Paris it may seem already quite risky to derive the belief that there still are tickets for the train, and from my intention to present a paper at IJCAI next year, it is surely too strong to infer a belief that I will finish the paper in time and that the paper will be accepted. Such a belief would be based on irrational wishful thinking.

However, we cannot ignore the assumptions either. Suppose I would have believed that there are no more tickets for the train to Paris tomorrow, then it would surely have been irrational to adopt the intention to go to Paris by train tomorrow. On the contrary, suppose I learn now that there are no more tickets for the train  tomorrow, then I would drop my intention to go to Paris. So from the intention to go to Paris we can infer that I do not believe that there are no more tickets. Note the essential difference between believing that there are still tickets, and the absence of a belief that there are no more tickets. This difference is represented very precisely in modal logic by the two formulas $SB_0(tickets)$ and $\neg SB_0 \neg (tickets)$. The former implies the latter, but not vice versa.

If we represent a generic intention at time $t$ for $\alpha$ by $I_t \alpha$, and the precondition of $\alpha$ by $pre(\alpha)$, then we may impose the following constraint on our modal logic:\\

$I_t \alpha$ implies $\neg B_t \neg pre(\alpha).$\\

The assumptions may be seen as a kind of presupposition as it is studied in linguistics: the sentence ``The king of France is bold'' presupposes that France is a kingdom. Likewise, the intention to go to Paris presupposes that there still are such tickets. It is well known that it is challenging to find a suitable definition of presupposition. For example, if we derive the weak intention-based belief that we will be in Paris tomorrow, can we deduce also a weak intention-based belief that there are still tickets? The fact that we will be in Paris implies that we were able to obtain a ticket.

The absence of the belief that there are still tickets may be represented also using negation as failure in logic programming~\citep{clark1978negation}, because we cannot derive the belief that there are no more tickets. Moreover, it can also be represented by a justification in default logic~\citep{reiter1980logic}, or more generally by a consistency check.

Consider the situation where I have the intention to give a seminar in Paris tomorrow, and in addition that I have the intention to give a seminar in London tomorrow. The preconditions of these intended actions are that I am in Paris and I am in London tomorrow, and clearly this is not possible. However, it is possible that I am in Paris tomorrow, and it is possible that I am in London tomorrow. It is just not possible that I am both in Paris and in London.

In other words, the logic has to derive that it is possible for all intended actions to be performed, and thus that for all intended actions, the preconditions hold. We introduce a new name for this happy situation where all intended actions can be performed. We say that a set of intentions and beliefs is \emph{coherent} if it is possible that all actions can be performed and all preconditions hold.

\subsubsection{What if statements}

One of the contributions in the formalisation of the interplay between intention, time, and belief is to model what-if statements: what will happen with my intentions and beliefs if I learn that the tickets for the train tomorrow are sold out? What happens if I learn that my paper for IJCAI is rejected? What happens if I adopt a new intention to go to a birthday party in Amsterdam tomorrow?

Our contribution adopts the AGM~\cite{AGM1985} framework for theory change. To be able to use this framework, we first need to make some additional restrictions to our language of beliefs and intentions, and we need to define the operators we consider together with their postulates. Then we characterize the set of operators satisfying the postulates using a representation theorem.

Since we do not have a logical language with explicit belief operators, we cannot represent the coherence of a belief-intention base by a formula "it is not believed that ...". Instead, we represent the coherence of a belief-intention base in a different way. We use an idea similar to negation as failure: the belief intention base is consistent with the precondition of the (sequence of) intended actions.

\subsubsection{The ``Little Nell'' problem: Not giving up too soon}
\label{sect:littlenell}

\cite{Mcdermott1982} discusses the following difficulty with a naive planning system:

\begin{quote}
Say a problem solver is confronted with the classic situation of a heroine, called Nell, having been tied to the tracks while a train approaches. The problem solver, called Dudley, knows that ``If Nell is going to be mashed, I must remove her from the tracks.'' (He probably knows a more general rule, but let that pass.) When Dudley deduces that he must do something, he looks for, and eventually executes, a plan for doing it. This will involve finding out where Nell is, and making a navigation plan to get to her location. Assume that he knows where she is, and he is not too far away; then the fact that the plan will be carried out will be added to Dudley's world model. Dudley must have some kind of database- consistency maintainer (Doyle, 1979) to make sure that the plan is deleted if it is no longer necessary. Unfortunately, as soon as an apparently successful plan is added to the world model, the consistency maintainer will notice that ``Nell is going to be mashed'' is no longer true. But that removes any justification for the plan, so it goes too. But that means ``Nell is going to be mashed'' is no longer contradictory, so it comes back in. And so forth.
\end{quote}

The agent continuously plans to save Nell, and abandons its plan because it believes it will be successful. If we view this problem from the database perspective, we require some way to separate beliefs about plans from beliefs that are not about plans. Otherwise, the planner may believe certain facts hold and adopt the plans accordingly, while these facts may be dependent on performing the plan.

This problem has been considered extensively in the literature (cf.~\cite{Cohen1990,Shoham2009,Icard2010,vanderHoek2003}). Our solution is to separate strong beliefs from weak beliefs. The beliefs in the belief database of the agent (those that will be used by the planner) are strong beliefs, and the weak beliefs are computed by adding intentions to the strong beliefs, and everything following from that.

\subsection{Temporal Logic}
\label{sect:temporal-logic}

Our theory distinguishes internal and external dynamics.
The temporal references of the facts (e.g., $rain_1$, $sun_2$) represent internal dynamics, and the temporal references on the modal operators (e.g., $I_{-1}, WB_1$) represent external dynamics. Internal and external dynamics can also be represented by temporal modal operators such as next and until in temporal logics, as discussed in this section.
In this article we instead use explicit time indexes, but in the future work in Section~\ref{sect:futurework} we discuss how this can be extended to implicit time in temporal logics. The core ideas and results of this paper do not depend on this choice.

Time in temporal logics can be defined in an \emph{implicit} or \emph{explicit} manner. A time model is implicit when the meaning of formulas depends on the evaluation time, and this is left implicit in the formula. Standard LTL and CTL define time implicitly. For instance, $\Box \Phi$ means that $\forall t\in[T_0,\infty]. \Phi(t)$, where $T_0$ is the evaluation time (the so-called current time instant). A standard way of introducing real time into the syntax of temporal languages constrains the temporal operators with time intervals~\citep{Emerson1999,Koymans1990,Alur1990,Alur1996}. In order to model such time intervals, timed automata may be used. These automata model the behavior of time-critical systems. A timed automaton is in fact a program graph that is equipped with a finite set of real-valued clock variables, called \emph{clocks} for short~\citep{Alur1994}. Timed CTL (TCTL, for short) is a real-time variant of CTL aimed to express properties of timed automata. In TCTL, the until modality is equipped with a time interval such that $\Phi U^J \Psi$ asserts that a $\Psi$-state is reached within $t\in J$ time units while only visiting $\Phi$-states before reaching the $\Psi$-state. The formula $\exists\Box^J\Phi$ asserts that there exists a path for which during the interval $J$, $\Phi$ holds; $\forall\Box^J\Phi$ requires this to hold for all paths~\citep{Baier2008}. While such logics allow one to express timed constraints on the modalities in TCTL, there is no way to refer explicitly to the states at which a certain formula holds.

When time is explicit, the language represents the time through a variable. For example, in the following formula an explicit model of time is used:
\begin{align*}
\forall t.\Box(E\wedge T=t)\rightarrow\Diamond (A\wedge T-t<10)
\end{align*}
where $E$ is an event~\citep{Bellini2000}. This is for instance formalized by~\cite{Ostroff1989} when solving control problems using real-time temporal logic (RTTL).

The logic of strategic abilities ATL* (Alternative-Time Temporal Logic), introduced and studied by~\cite{Alur1997}, is a logical system, suitable for specifying and verifying qualitative objectives of players and coalitions in concurrent game models. Formally, ATL* is a multi-agent extension of the branching time logic CTL* with \emph{strategic path quatifiers} $\langle\!\langle C\rangle\!\rangle$ indexed with coalitions $C$ of players.~\cite{Bulling2013} propose a quantitative extension of ATL*, in which it is possible to express temporal constraints as well. For instance, the expression $\phi\wedge x=t$ denotes that $\phi$ will be true after $t$ transitions, where each transition adds 1 to $x$.

Many logical systems have been developed for reasoning about the pre and postconditions of actions with explicit time points, such as the Event Calculus \citep{Mueller2010}, Temporal Action Logics~\citep{Kvarnstrom2005}, extensions to the Fluent Calculus~\citep{Thielscher2001}, and extensions to the Situation Calculus~\citep{Papadakis2003} (see~\cite[Ch.2]{Patkos2010} for an overview). Our logic is considerably more simple, but the reason for this is because of the type of revision we characterized in this article. Although there are a number of correspondences between AGM postulates and some of the approaches above, none of them prove representations theorems linking revision to a total pre-order on models.

As we mentioned, the structure of our models is full branching time structure of CTL*. In addition, we have the actions attached to the elements of accessibility relation $R$. 
Since our formulas are different than those of CTL*, it should be noted that the part of  PAL  built on propositional letters, modalities and Boolean connectives, is embedded in the fragment of CTL* with only $\bigcirc$ (next) and $A$ (universal path quantifier) operators. This is because any formula of the form $\Box_t\phi$ can be written as the formula $\bigcirc^t A \psi$ in CTL*, where $\bigcirc^t$ stands for $t$ occurrences of the $\bigcirc$ operator and $\psi$ is the translation of $\phi$, which takes into account that $t$-th moment becomes the actual moment and in which $\Box_t$ operators are translated likewise. For example, the formula $p_1\wedge \Box_2( (p\vee q)_3 \wedge q_4)$ would be translated to $\bigcirc p\wedge \bigcirc^2 A( \bigcirc(p\vee q) \wedge \bigcirc^2 q)$.\footnote{Note that, some formulas, like $\Box_2 p_1$ cannot be translated directly, but the presence of the K-axiom and A1 and A5 overcome  that problem, since they imply that   $\Box_t \chi$ is equivalent to $\chi$ if $\chi\in Past(t)$. Thus, $\Box_2 p_1$ can be first transformed to an equivalent formula $ p_1$ in our logic, which allows translation to a CTL* formula $\bigcirc p$.
}).

As presented in~\cite{Meier2008}, the satisfiability problem for this fragment of CTL* is known to be PSPACE-complete. It would be interesting to see if those results can be modified for PAL and how the addition of actions in our logic  influence its complexity.

\subsection{Collective intentions}
\label{sect:related-work:collective-intentions}

An important issue in the philosophical literature on collective intentionality is the question of whether collective intentions can be reduced to individual intentions (i.e., \emph{reductionist theories}), or whether collective intentions are first-class citizens and cannot be reduced.

\cite{Bratman2014} offers a clear example of reductionist theory of collective intention as it decomposes the concept of collective intention in terms of more primitive concepts such as the concepts of individual intention, belief and common belief. He defines ``shared cooperative activity'' using the following three characteristics:
\begin{enumerate}
\item Each participant must be mutually responsive to the intentions of others;
\item The participants must each be committed to the joint activity;
\item The participants must each be committed to supporting the effort of the others;
\end{enumerate}
Consider the example of Alice and Bill who intend to paint a house together. Suppose Alice wants to paint the house red and Bill wants to paint it blue. Both are aware that their subplans conflict, and that the other is aware of it as well. Therefore they do not have a shared cooperative activity, even if they end up painting the house together. But subplans do not have to be identical in order to have a shared cooperative activity. For instance, if Bill wants to use cheap paint but Alice want to buy from a specific store, they could buy cheap paint from that specific store. Bratman puts the following constraints on an action $J$
to be a shared cooperative activity:
\begin{enumerate}
\item We do $J$ (which can involve cooperation, but doesn't have to);
\item It is common knowledge between us that we are both committed to mesh subplans;
\item (2) leads to (1) by way of mutual responsiveness (in the pursuit of completing our action) of intention and in action.
\end{enumerate}

Examples of non-reductionist theories of collective intention are, for example, \cite{Gilbert1990}, who simplifies the problem of collective intentionality to the situation of two people walking together. From this, she identifies four necessary conditions on collective intentions. \cite{Tuomela1988} distinguish the concept of joint I-intention from the concept of joint We-Intention. According to their theory, the concept of joint We-intention cannot be reduced to individual mental attitudes of the agents in the group such as beliefs, desires and individual intentions. Imagine for instance that Anne and Bob intend to lift a table together. First, Ann needs to intend to do her part. Next, she should believe that it is possible to lift the table and believe that Bob also intends to do his part. Moreover, Ann needs to believe that Bob also believes that carrying the table is possible.

\cite{Searle1995} also argues that collective intentionality cannot be reduced to individual intentionality. According to Searle, coordination and cooperation is crucial in defining we-intentions, and he gives the following counter example to Tuomela and Miller's we-intentions: A group of business school graduates intend to pursue their own selfish interests, but believe that by doing so, they will indirectly serve humanity. They also believe that their fellow graduates will do likewise, but they do not actively cooperate with one another in pursuing their goals. Searle holds that this fulfills the Tuomela and Miller criteria, but collective intentionality does not actually exist in such a situation.

\cite{Velleman1997} is concerned with how a group is capable of making a decision using speech acts, which he considers to be intentions in itself. He argues that collective intention is not the summation of multiple individual intentions (as Tuomela and Miller thought), but rather one shared intention. An intention exists outside of the mind of agent, within a verbal statement. The causal power is in the verbal statement, because of the desire not to speak falsely.

The philosophical dispute is reflected in the computer science literature as well. For instance, \cite{Cohen1991}, following their well-known work on individual intentions (discussed above), propose a reductive account of collective intentions by defining them in terms of group goals and mutual beliefs. \cite{DuninKeplicz2002}, in contrast, regard both collective intentions and individual intentions a ``first-class citizens''. Frameworks for flexible teamwork also regularly use theories of collective intentions. For instance, \cite{Tambe1996} uses the notion of joint intentions by Cohen and Levesque as a basic building block to define teamwork.

\subsection{Development of the framework and relation to our previous wok}

\subsubsection{Development of \logic{}}
This article combines and builds further on a series of previous papers published by us. The initial proposal of the database perspectives is due to~\cite{Shoham2009}. His proposal is largely informal but the main ideas form the basis of this article and our formalism can be seen as a formalization of his main ideas. The first formalization of the database perspective was in~\cite{Icard2010}, where we presented a formal semantic model to capture action, belief and intention, based on Shoham's database perspective. We provided postulates for belief and intention revision, and stated a representation theorem relating our postulates to the formal model. However, we noted that there were problems with this formalization, and much of our further work consists of developing the right formal framework. First, we showed that the axiomatization of the original logic is incomplete, and we provided a complete axiomatization~\citep{vanZee2015:commonsense}. We also adapted the coherence condition in various ways (see next subsection). In~\cite{vanZee2015-ijcai} we proved representation theorems for the belief database, and in~\cite{vanZeeDoder:ecai2016} we extended this to intentions. We discussed various extension to our framework, focusing on enterprise-level decision making~\citep{vanZee2015:ijcai-short} and a multi-agent perspective~\citep{vanzee-etal:ci2014}. In this article we provide a complete formal model that combines all our previous work. We furthermore provide the full proofs of all the theorems and we develop an account of iterated revision and a multi-agent extension.

\subsubsection{Alternative coherence conditions}
\label{sect:alternative-coherence-conditions}

An obvious coherence condition we can put on beliefs and intentions is the following:
\begin{align*}
B\vdash \bigwedge_{(a,t)\in I} post(a)_t.
\end{align*}

However, this solution suffers from the well-known ``Little Nell'' problem, identified by McDermott~\citep{Mcdermott1982} (Section~\ref{sect:littlenell}). 

A weaker variant is formalized semantically by~\cite{Icard2010} as follows:
\begin{align*}
\pi,0\models \Diamond \bigwedge_{(a,t)\in I} pre(a)_t
\end{align*}
Although the semantics in the current article is slightly different, the general idea of this formula is clear: There exists a path, equivalent with the current path up to time 0, in which all the preconditions of the intended actions hold. 

However, this is clearly too weak. An agent may believe that all the preconditions hold on a paths where none of its intended actions are carried out. We discuss various examples for this in other papers~\citep{vanZee2015:commonsense,vanZeeDoder:ecai2016,vanZeeDoder:nmr2016}

In order to resolve this, we propose a different condition~\cite{vanZee2015:commonsense}, requiring that the beliefs of an agent should be consistent with the preconditions of it's intended action. The agent does not have to believe the preconditions of its intended actions, but it should not believe the negation of the precondition of an intended action. Therefore, we introduce precondition formulas that are derived from the contingent beliefs:
\begin{align*}
Pre(B^I) = Cl(B^I \cup \{\bigwedge_{(a,t)\in I} pre(a)_t\}\})
\end{align*}
We can then express the condition as follows: $Pre(B^I)$ is consistent.

In a followup paper~\citep{vanZeeDoder:ecai2016} we note that this condition is too weak as well (we refer to that paper for examples and additional motivation). The main problem is that it is not possible to define the precondition of a set of actions in terms of preconditions of individual actions, because it cannot be ensured that all the intentions are fulfilled on the same path as well. Therefore, in order to formalize a coherence condition in PAL, we extend the language with preconditions of finite action sequences, which ensures that after executing the first action, the precondition for the remaining actions are still true.

\subsubsection{Timeful application}
Separately, Shoham further developed his ideas with Jacob Banks, one of his PhD students, and behavioral economist Dan Ariely in the intelligent calendar application Timeful, which attracted over \$6.8 million in funding and was acquired
by Google in 2015\footnote{\url{ http://venturebeat.com/2015/05/04/google-acquires-scheduling-apptimeful-and-plans-to-integrate-it-into-google-apps/}}, who aim to integrate it into their Calendar applications. As~\cite{Shoham2016} says himself: ``The point of the story
is there is a direct link between the original journal paper and the ultimate success of the company.'' (p.47) Thus, it seems clear that his philosophical proposal has lead to some success on the practical side. In our research, we aim to show that his proposal can lead to interesting theoretical insights as well.

\section{Future Work}
\label{sect:futurework}
\cite{Shoham2009} suggests a large number of direction of future work for the database perspective. In this article we already studied iterated revision as well as a multi-agent extension. We discuss some remaining directions of research which we deem most promising.

\subsection{Quantitative beliefs}
\label{sect:applications:quantitative}

Our focus in this paper has been on \emph{qualitative} models of belief and belief revision. However, much of our framework could be naturally extended to capture intention revision in the context of \emph{quantitative} belief revision. Imagine we have a distribution $P$ defined on the (finite) set $\mathbb{M}^{\mid t}$ of bounded models. This naturally induces a distribution $\mathbb{P}$ on $\lang{}^{|t}$, whereby $\mathbb{P}(\varphi) = P(\{m \in \mathbb{M}^{\mid t}: m \models \varphi \})$. Whereas the definition of coherence we gave above in Def.~\ref{def:coherence} is quite minimal, one can imagine replacing this definition with one slightly stronger. A natural strengthening in the probabilistic setting would be to require $\mathbb{P}(Cohere(I)) > \theta_c$ for some threshold $\theta_c>0$. That is, one might like to be reasonably confident that one's plan will succeed. The question now is how to model qualitative belief in this setting.  

Recent work in philosophical logic has suggested various ways of bridging quantitative and qualitative frameworks, using the notion of an \emph{acceptance rule}: a rule for determining a belief set $B$ from a probability measure $P$. For concreteness, we illustrate how our framework could be extended by drawing on a concrete example, namely the acceptance rule defined by~\cite{leitgeb2014stability}. In short, we would select a second threshold $\theta_b$ and, following Leitgeb, propose that the agent believes $\psi$ just in case $\mathbb{P}(\psi \mid \varphi) > \theta_b$ for any $\varphi$ with $\mathbb{P}(\varphi)> 0$. This notion of ``stable belief'' fits well with our framework, as it is guaranteed that there is always a single strongest stable proposition $\psi$. Following our assumption of ``opportunistic planning'' we would typically have that $\theta_b > \theta_c$, allowing that we do not necessarily outright believe that the preconditions for our intended actions will be satisfied. 

While we have defined weak beliefs in terms of a pair $(SB,I)$, in this setting it would make sense to define them using the full distribution. Intuitively, we weakly believe $\varphi$ if its probability is stably high given that all of the intended actions are carried out; that is, we could define \begin{eqnarray*} WB(\mathbb{P},I) & = & \{\varphi: \mathbb{P}\big(\varphi \mid \bigwedge _{(a,t) \in I}do(a,t)\big) > \theta_b\} \end{eqnarray*} Because the set of beliefs is stable under conditioning, this set would of course extend the closure of the set of strong beliefs together with $\{do(a,t): (a,t) \in I\}$. 

An interesting question in this setting is whether it would be possible to prove an analogue of our Theorem \ref{thm:singlestep}. Leaving the definition of a selection function for intentions intact (but substituting the quantitative notion of coherence), one question would be what are reasonable postulates characterizing intention revision together with belief revision when the latter amounts to probabilistic conditionalization? Coming from the other direction, is there a natural probabilistic notion of updating that would allow our representation theorem to go through with exactly the same postulates? For instance,~\cite{mierzewski2018probabilistic} has shown that AGM can be recovered in the quantitative setting if updating is by conditionalization, but only after shifting from one's distribution to the maximum entropy measure with the same set of stable (strong) beliefs. That our representation result would go through in our joint revision setting we leave as a worthy conjecture.

\subsection{Teleology}

One aspect we left out of our study purposely is the idea of teleology, or the purpose for which I form intentions. It is clear that some notion of goal is required to capture teleological aspects of intending, which is why Cohen and Levesque started with goal. Although it was beneficial for our analysis of intention revision to abstract away from goals, they are clearly central to the revision problem. Information about goals allows the agent to, for instance, replace intentions instead of merely discarding them. This paves the road to develop a richer notion of intentions, such as that ``intentions normally pose problems for the agents; the agent needs to determine a way of achieving them''~\citep{Cohen1990}. Interestingly, viewed from Shoham's database perspective, adding goals to the formalism blurs the distinction between planner and databases. If the databases take over part of the planning, then well-known problems such as the frame problem become more stringent: Once a fact is established (for example, as a postcondition of an intention), it persists until it explicitly contradicts postconditions established by future intentions. Existing action logics (e.g., the Event Calculus or the Fluent Calculus) and database approaches (e.g., TMMS by~\cite{Dean1987}) have dealt with these problems in detail, so comparing and possibly enriching them with our formalism seems both useful and relevant future work.

\section{Conclusion}
\label{sect:conclusion}

We study the interplay between intention, time, and belief, and present two main contributions. 

The first contribution is to formalize assumptions of intentions. We observe that assumptions cannot be used to derive strong beliefs from weak beliefs, but cannot be ignored either. In order to deal with this, we first develop a branching-time temporal logic, called \logicname{} (\logic{}) in order to formalize beliefs. The language of this logic contains formulas to reason about possibility, preconditions, postconditions, and the execution of actions. The semantics of this logic is close to CTL*, and in this way follows the tradition of BDI logics of~\cite{Rao1991}. An important difference is that we do not use modal operators to reason about time, but we use explicit time points. We axiomatize this logic and prove that the axiomatization is sound and strongly complete with respect to our semantics.

We separate strong beliefs from weak beliefs. Strong beliefs are beliefs that occur in the belief database, and they are independent of intentions. Weak beliefs are obtained from strong beliefs by adding intentions to the strong beliefs, and everything that follows from that. We formalize a \emph{coherence condition} on the beliefs and intentions, which states that the agent weakly believes it is possible to jointly perform all of its intended actions. 

Our second contribution is to model what-if statements. In other words, we study the \emph{dynamics} of the interplay between intentions, time, and beliefs. Our approach is to use the well-known and well-studied AGM theory of belief revision as our starting point. We develop a set of postulates for the joint revision of belief and intentions, and we prove a variation of the~\cite{Katsuno1991} representation theorem. To this end, we define a revision operator that revises beliefs up to a specific time point. We show that this leads to models of system behaviors which can be finitely generated, i.e. be characterized by a single formula. In addition, we study iterated revision.
\section*{Acknowledgements}
We would like to thank Yoav Shoham for providing useful feedback on the initial ideas of this article. We also would like to thank the anonymous reviewers for their detailed feedback, and in particular for bringing our attention to the example at the beginning of Section~\ref{sect:examples}.

Dragan Doder is funded by  ANR-11-LABX-0040-CIMI.

\bibliographystyle{elsarticle-harv} 
\bibliography{ALL}

\appendix

\newpage

\section{Completeness Proofs}
\label{sect:soundness}
\setcounter{theorem}{0}
\begin{theorem}[Completeness Theorem] The logic \logic{} is sound and strongly complete, i.e. $\Sigma\vdash\varphi$ iff $\Sigma\models\varphi$.
\end{theorem}

$T\vdash\varphi\Rightarrow T\models\varphi$ (soundness) can be proven by standard techniques. We use the remainder to prove strong completeness: $T\models\varphi\ \ \Rightarrow\ \  T\vdash\varphi$.\\

We prove strong completeness by constructing a canonical model, but before this we introduce some concepts that we will need in different parts of the proof. These concepts will be largely familiar to most readers.

\begin{definition}[Maximally consistent set (mcs)] 
\label{def:maxconsistentset}
Given the logic \logic{}, a set of formulas $T$ is \logic{}-consistent if one cannot derive a contradiction from it, i.e. if $\perp$ cannot be inferred from it, in the proof system for \logic{}. A set of formulas $T^*$ is a \emph{maximally \logic{}-consistent set} (mcs) if it is \logic{}-consistent and for every formula $\varphi$, either $\varphi$ belong to the set or $\neg\varphi$ does.

We denote the part of a mcs up to and include time $t$ with $T^*_t$, formally: $T^*_t = T^*\cap Past(t)$ (see Def. 1 of original paper). 
\end{definition}

\begin{lemma}[Lindenbaum's lemma] 
\label{lemma:lindenbaum}
Every consistent set of formulas can be extended to a maximal consistent set of formulas.
\end{lemma}

\begin{lemma}[The Deduction Theorem] 
\label{lemma:deductiontheorem}
$\Sigma\cup\{\varphi\}\vdash \psi \ \ \Rightarrow\ \ \Sigma\vdash \varphi\rightarrow\psi$
\end{lemma}

\begin{definition}[Mcs Equivalence Relation] 
\label{def:maxconsistentset:equiv}
Suppose some $t\in\mathbb{N}$ and two mcs's $T^*$ and $\overline{T}^*$, we define the \emph{equivalence relation} between $T^*$ and $\overline{T}^*$, denoted by $T^*\equiv_t \overline{T}^*$ as follows: $T^*\equiv_t \overline{T}^*$ iff $T^*\cap Past(t) = \overline{T}^*\cap Past(t)$.
\end{definition}

\begin{definition}[Equivalence class]
\label{def:equivclass}
Let $T^*$ be a mcs. $[T^*]_t$ is the set of all mcs's that are equivalent to $T^*$ up and including time $t$, i.e. $[T^*]_t = \{ \overline{T}^*\mid T^*\equiv_t \overline{T}^*\}$.
\end{definition}

The next step is to reduce truth of a formula in a maximal consistent set to membership of that set, which is the content of the truth lemma. We first present a lemma that we need in the proof of the valuation lemma, which follows after that.

\begin{lemma}
\label{lemma:introbox}
Let $\Sigma=\{\varphi_1,\ldots,\varphi_n\}$ be some set of \logic{}-formulas and abbreviate $\{\Box_t\varphi_1,\ldots,\Box_t\varphi_n\}$ with $\Box_t\Sigma$. If $\Sigma\vdash \varphi$, then $\Box_t\Sigma \vdash\Box_t\varphi$
\end{lemma}

\begin{proof}
Suppose $\{\varphi_1,\ldots,\varphi_n\}\vdash\varphi$. By the deduction lemma, $\vdash (\varphi_1\wedge\ldots\wedge\varphi_{n})\rightarrow\varphi$. Applying necessitation gives $\vdash \Box_t((\varphi_1\wedge\ldots\wedge\varphi_{n})\rightarrow\varphi)$, and from the K-axiom it follows that $\vdash \Box_t(\varphi_1\wedge\ldots\wedge\varphi_{n})\rightarrow\Box_t\varphi$. Since $\Box_t(\varphi_1\wedge\ldots\wedge\varphi_n)\equiv \Box_t\varphi_1\wedge\ldots\Box_t\varphi_n$, we obtain (1) $\vdash (\Box_t\varphi_1\wedge\ldots\Box_t\varphi_n)\rightarrow\Box_t\varphi$. Finally, since (2) $\{\Box_t\varphi_1,\ldots,\Box_t\varphi_n\}\vdash\Box_t\varphi_1\wedge\ldots\wedge\Box_t\varphi_n$ holds as well, we combine (1) and (2) and conclude that $\{\Box_t\varphi_1,\ldots,\Box_t\varphi_n\}\vdash\Box_t\varphi$.
\end{proof}

\begin{lemma} 
\label{lemma:mcs:introbox}
$T_t^*\vdash\Box_t T_t^*$
\end{lemma}

\begin{proof}
We show that for all $\varphi\in T_t^*$ we have $T_t^*\vdash\Box_t\varphi$ by induction on the depth of the proof. Take an arbitrary $\varphi\in T_t^*$. We distinguish two base cases, one where $\varphi$ is a proposition, and another where $\varphi$ is an atomic ``do'' formula.\\
\begin{enumerate}
\item[(Base case 1)] Suppose $\varphi=\chi_{t'}$ with $\chi\in\prop$ and $t'\le t$. $\Box_{t'} \chi_{t'}$ follows by applying Axiom A1. Then, apply Axiom A3 repeatedly until $\Box_t \chi_{t'}$ follows. 

\item[(Base case 2)] Suppose $\varphi=do(a)_{t'}$ with $t'< t$ (note that $do(a)_t$ cannot occur in $T_t^*$ because it does not occur in $Past(t)$). Using Axiom A4 and then repeatedly Axiom A3 we obtain $\Box_t do(a)_{t'}$.

\item[(Conjunction)] Suppose $\varphi=\psi\wedge\chi$. The induction hypothesis is $T_t^*\vdash \Box_t\psi$ and $T_t^*\vdash \Box_t\chi$, so therefore from $T_t^*\vdash \psi\wedge \chi$ we obtain $T_t^*\vdash \Box_t\psi\wedge \Box_t\chi$. Since $\Box_t\psi\wedge\Box_t\chi$ is equivalent to $\Box_t(\psi\wedge\chi)$, it follows directly that $T_t^*\vdash\Box_t(\psi\wedge\chi)$.

\item[(Box)] Suppose $\varphi=\Box_{t'}\psi$. By transitivity (which is not an axiom of our logic, but it holds in KT5): $\Box_{t'}\psi\rightarrow\Box_{t'}\Box_{t'}\psi$. Next, apply Axiom A3 repeatedly to obtain $\Box_t\Box_{t'}\varphi$.

\item[(Negation)] We make another case distinction on the negated formula. That is, we assume $T_t^*\vdash\neg\varphi$ and we show $T_t^*\vdash\Box_t\neg\varphi$, again by induction on the depth of the formula.\\

\begin{enumerate}
\item[(Base case 1)] Suppose $\varphi= \neg \chi_{t'}$ with $\chi\in\prop{}$ and $t'\le t$, then $\Box_t\neg \chi_t$ follows from Axiom A2 and A3 as described before.
\item[(Base case 2)] Suppose $\varphi=\neg do(a)_{t'}$ with $t'<t$. We apply Axiom A5 and A3 repeatedly until we have $\Box_t\neg do(a)_{t'}$. 
\item[(Conjunction)] Suppose $\varphi=\neg(\psi\wedge\chi)$. The induction hypothesis is $T_t^*\vdash \Box_t\neg \psi$ and $T_t^*\vdash \Box_t\neg\chi$, which implies $T_t^*\vdash \Box_t\neg\psi\vee\Box_t\neg\chi$, which again implies $T_t^*\vdash \Box_t(\neg\psi\vee\neg\chi)$. Using De Morgan we obtain $T_t^*\vdash \Box_t\neg(\psi\wedge\chi)$.
\item[(Box)] Suppose $\varphi=\neg\Box_{t'}\psi$, which is equivalent to $\Diamond_{t'}\neg\psi$. From Axiom 5 we obtain $\Box_{t'}\Diamond_{t'}\neg\psi$, and by again applying Axiom A3 repeatedly we obtain $\Box_t\Diamond_{t'}\neg\psi$, which is equivalent to $\Box_t\neg\Box_{t'}\psi$, and this is what we had to show. 
\end{enumerate}
\end{enumerate}
\end{proof}

\begin{lemma}[Valuation lemma]
\label{lemma:valuation}
For any maximal consistent set $T^*$, the following are true
\begin{enumerate}[nolistsep]
\item $T^*$ is deductively closed: $T^*\vdash\varphi$ implies that $\varphi\in T^*$;
\item $\varphi\in T^*$ iff $\neg\varphi\not\in T^*$;
\item $\varphi\wedge\psi\in T^*$ iff $\varphi\in T^*$ and $\psi\in T^*$;
\item $\Box_t\varphi\in T^*$ iff for all $\overline{T}^*$ s.t. $T^*\equiv_t \overline{T}^*: \varphi\in \overline{T}^*$.
\end{enumerate}
\end{lemma}

\begin{proof}
\begin{enumerate}
\item Because $T^*$ is maximally consistent, either $\varphi\in T^*$ or $\neg\varphi\in T^*$. Suppose that $T^*\vdash \varphi$, and suppose for contradiction that $\neg\varphi\in T^*$. From this it follows that $T^*\vdash \neg \varphi$ and therefore $T^*\vdash \perp$, which would contradict consistency of $T^*$. Hence $\varphi\in T^*$.
\item Follows directly from the definition of a maximally consistent set.
\item Follows directly as well.
\item $\Rightarrow$: Suppose $\Box_t\varphi\in T^*$. Take arbitrary $\overline{T}^*$ with $T^*\equiv_t \overline{T}^*$. From Def.~\ref{def:maxconsistentset:equiv} (equivalent mcs) it follows that $\Box_t\varphi\in\overline{T}^*$. Therefore, by Axiom T we obtain $\varphi\in\overline{T}^*$.

$\Leftarrow$: We show this by contraposition. Therefore, suppose $\Box_t\varphi\not\in T^*$. We will show that there exists some $\overline{T}^*$ with $T^*\equiv_t \overline{T}^*$ and $\varphi\not\in\overline{T}^*$. 

Suppose for contradiction that $\neg\varphi$ is not consistent with $T^*_t$, i.e. $T_t^*\cup\{\neg\varphi\}\vdash\perp$, so by the Deduction Theorem (Lemma~\ref{lemma:deductiontheorem}), $T^*_t\vdash\varphi$ holds. By Lemma~\ref{lemma:introbox}, we have (1) $\Box_t T^*_t\vdash\Box_t\varphi$. From Lemma~\ref{lemma:mcs:introbox} it follows that (2) $T_t^*\vdash\Box_t T^*_t$, so combining (1) and (2) gives $T_t^*\vdash \Box_t\varphi$. But this contradicts with our initial assumption that $\Box_t\varphi\not\in T^*$. Thus, the assumption is invalid so $T_t^*\cup\{\neg\varphi\}$ is consistent.

By Lindenbaum's lemma, $T_t^*$ can be extended to a mcs $\overline{T}^*$, and since $T_t^*\subseteq \overline{T}^*$, it follows directly that $\overline{T}^*\equiv_t T^*$. Therefore, there exists a mcs $\overline{T}^*$ with $\overline{T}^*\equiv_t T^*$ and $\varphi\not\in \overline{T}^*$, and this is what we had to show.
\end{enumerate}
\end{proof}

We construct the canonical model by naming the states in our model as equivalence classes of mcs's, which are parameterized by a time point. For instance, the state $s=[T^*]_t$ is named as the set of mcs's equivalent to the mcs $T^*$ up to and including time $t$. We then define accessibility relation between states named after equivalence classes up to and including subsequent time points of the same mcs. Finally, the valuation function assigns the set of propositions that are true in an equivalence class to the corresponding state.

\begin{definition}[Canonical Tree]
\label{def:canonicalmodel}
Given a mcs $T^*$, we obtain a \logic{}-canonical tree $Tree_{T^*}=(S,R,v,act)$, where
\begin{enumerate}[nolistsep]
\item $S = \bigcup_{t\in\mathbb{N}}S_t$ where $S_t=\{[\overline{T}^*]_t\mid \overline{T}^*\equiv_0 T^*\}$ 
\item $sRs'$ iff $\ex{\overline{T}^*, t\in\mathbb{N}}{s=[\overline{T}^*]_t\wedge s'=[\overline{T}^*]_{t+1}}$
\item $p\in v(s)$ iff $\ex{\overline{T}^*,t\in\mathbb{N}}{s= [\overline{T}^*]_t\wedge p_t\in \overline{T}^*}$.
\item $a= \vala((s,s'))$ iff $\ex{\overline{T}^*}{s=[\overline{T}^*]_t\wedge s'=[\overline{T}^*]_{t+1}\wedge do(a)_t\in \overline{T}^*}$
\end{enumerate}
\end{definition}

Note that the existential quantifier in (3) of Def.~\ref{def:canonicalmodel} could equivalently be replaced by a universal quantifier, because of the definition of equivalence classes (Def.~\ref{def:equivclass}): All mcs's in $[\overline{T}^*]_t$ are equivalent up to time $t$, so if some timed proposition $p_t$ is an element of some mcs in this set, then it is necessarily an element of any other mcs in this set as well.

\begin{lemma}
\label{lemma:itsatree}
Given a mcs $T^*$, $Tree_{T^*}$ is a tree.
\end{lemma}

\begin{proof}
Suppose some $T^*$ and let $Tree_{T^*}=(S,R,v,act)$. We have to show that $R$ is serial, linearly ordered in the past, and connected.
\begin{itemize}
\item \emph{serial}: Suppose some $s\in S$ s.t. $s=[T^*]t$. We have to show that there exists some $s'\in S$ such that $sRs'$, i.e. there exists some $\overline{T'}^*$ s.t. $s=[\overline{T'}]_t$ and $s'=[\overline{T'}]_{t+1}$. This directly follows for $\overline{T'}=\overline{T}$ and by the fact that $T^*$ is a mcs.
\item \emph{linearly ordered in the past:}  Suppose some $s\in S$ s.t. $s=[T^*]t$ and $t>0$ ($t=0$ is the root of the tree). We show that there exists exactly one $s'$ s.t. $s'Rs$. Suppose for contradiction that there exist $s',s''\in S$ with $s'=[\overline{T'}]_{t-1}$ and $s''=[\overline{T''}]_{t-1}$ such that $s'\not=s''$, i.e. $\overline{T'}\not\equiv_{t-1}\overline{T''}$. However, then $\overline{T'}\not\equiv_t\overline{T''}$ holds as well, but this contradicts with $s'Rs$ and $s''Rs$. Thus, $s'\not=s''$ is not possible.
\item \emph{connected:} Suppose $s,s'\in S$ with $s=[\overline{T}^*]_t$ and $s'=[\overline{T'}^*]_{t'}$. We show that there exists some $s''$ such that $s''R^*s$ and $s''R^*s'$, where $R^*$ is the transitive closure of $R$. This directly holds for $s''=[T^*]_0$, since then $s''\in\{[\overline{T}^*]_0\mid \overline{T}^*\equiv_0 T^*\}$.
\end{itemize}
\end{proof}

Given a mcs $T^*$, we construct a path $\pi_{T^*}=(s_0,s_1,\ldots)$ from it by letting $s_t=[T^*]_t$. So $p\in v_p(\pi_{T^*t})$ iff $p_t\in T^*$ and $a=\vala(\pi, T^{*t})$ iff $do(a)_t\in T^*$.

Given a path $\pi$ in a canonical tree $Tree_{T^*}$ and a $t\in\mathbb{N}$, we denote
\begin{align*}
T_{\pi| t} := \overline{T}^*\cap Past(t),
\end{align*}
where $\overline{T}^*\in \pi_t$. Note that the definition is correct: $T_{\pi| t}$ does not depend on the choice of the element from $T_\pi$ by Definition 2 and 3. We construct the set $T_\pi$ from a path $\pi$ as follows:
\begin{align*}
T_\pi = \bigcup_{t\in\mathbb{N}} T_{\pi|t}.
\end{align*}

The next two lemmas show that for each $\pi$ in the canonical tree $Tree_{T^*}$, $T_\pi$ is a mcs.
\begin{lemma}
\label{lemma:constr:subset}
Given a mcs $T^*$, For any path $\pi$ in the canonical tree $Tree_{T^*}$: $T_{\pi|t}\subseteq T_{\pi|t+1}$.
\end{lemma}

\begin{proof}
Suppose some mcs $T^*$ and somoe arbitrary path $\pi$ in the canonical tree $Tree_{T^*}$. From the construction of the canonical tree we have that $\pi_tR\pi_{t+1}$ iff there is some $\overline{T}^*$ with $\pi_t=[\overline{T}^*]_t$ and $\pi_{t+1}=[\overline{T}^*]_{t+1}$. Clearly, we have that $\overline{T}^*_t\subseteq \overline{T}^*_{t+1}$, and since $T_{\pi|t}\in [\overline{T}^*]_t$ and $T_{\pi|t}\in[\overline{T}^*]_{t+1}$, we also have that $T_{\pi|t}\subseteq T_{\pi|t+1}$. 
\end{proof}

\begin{lemma}
\label{lemma:pathtomcs:mcs}
Given a mcs $T^*$, for any path $\pi$ in the canonical tree $Tree_{T^*}$: $T_\pi$ is a mcs.
\end{lemma}

\begin{proof}
(Consistent) From Lemma~\ref{lemma:constr:subset} we have that $T_{\pi|0}\subseteq\ldots\subseteq T_{\pi|t}$. Moreover, $T_{\pi|t}\subseteq \overline{T}^*$ where $\overline{T}^*$ is a mcs, which is consistent by definition, so $T_\pi$ is consistent as well.

(Maximal) Suppose an arbitrary \logic{}-formula $\varphi$. Then  there is a maximal $t$ that appears in $\varphi$, therefore, $\varphi\in Past(t)$. By definition, $\varphi\in T_{\pi|t}$ or $\neg\varphi\in T_{\pi|t}$. Since $T_{\pi|t}\subseteq T_\pi$, we have that $\varphi\in T_\pi$ or $\neg\varphi\in T_\pi$. Hence $T_\pi$ is maximal.
\end{proof}





The following three lemmas are  direct consequences of the construction of $T_\pi$ and $\pi_T$.

\begin{lemma}
\label{lemma:path:equiv:mcs}
Given a mcs $T^*$, two paths $\pi$ and $\pip$ in the canonical tree $Tree_{T^*}$ and some time point $t$: $\pi\sim_t\pip$ iff $T_\pi\equiv_t T_{\pip}$.
\end{lemma}

\begin{lemma}
\label{lemma:mcs:equiv:paths}
Given two mcs $T^*$ and $\overline{T}^*$ and some time point $t$, $T^*\equiv_t\overline{T}^*$ iff $\pi_{T^*}\sim_t\pi_{\overline{T}^*}$.
\end{lemma}

\begin{lemma}
\label{lemma:mcs:functionsinverted}
Given a mcs $T^*$, in the canonical tree $Tree_{T^*}$, 
\begin{enumerate}
\item For each $\pi$, $\pi_{(T_\pi)}=\pi$.
\item For each $\overline{T}^*$, $T_{(\pi_{\overline{T}^*})}=\overline{T}^*$.
\end{enumerate}
\end{lemma}

Note that by the previous lemma, for every path $\pi$ in the canonical tree $Tree_{T^*}$, there exists a unique mcs $\overline{T^*}$ such that $\pi=\pi_{\overline{T^*}}$.

\begin{lemma}
Given a mcs $T^*$: $(Tree_{T^*},\pi_{\overline{T}^*})$ is a model.
\end{lemma}

\begin{proof}
Suppose some $T^*$. From Lemma~\ref{lemma:itsatree} we have that $Tree_{T^*}$ is a tree. In order to show that $(Tree_{T^*},\pi_{\overline{T}^*})$ is a model we prove the three conditions on a model. Recall that $$\pi_{T^*}=(s_0,s_1,\ldots)\text{ where }s_t=[T^*]_t.$$
\begin{enumerate}
\item Suppose $\vala(s, t)=a$. By the truth definition we have that $do(a)\in s_t$, so by the construction of $s_t$: $do(a)_t\in T^*$. By Axiom A9, $post(a)_{t+1}\in \overline{T}^*$, so $post(a)\in s_{t+1}$
\item Suppose $pre(\actseq{},b)_t\in s_t$, so similarly we have $pre(\actseq{},b)\in s_t$, and hence $pre(\actseq{},b)_t\in T^*$. By Axiom A11, $pre(\actseq{})_t\in T^*$ and so $pre(\actseq{})\in s_t$.
\item We can prove this condition in the same way.
\end{enumerate}
\end{proof}

\begin{lemma}[Truth lemma]
\label{lemma:truth}
Given a mcs $T^*$: for every maximally \logic{}-consistent set of formula $\overline{T^*}$ and for every formula $\varphi:$
\begin{align*}
(Tree_{T^*},\pi_{\overline{T}^*}) \models \varphi \text{ iff }\varphi\in \overline{T}^*
\end{align*}
\end{lemma}

\begin{proof}
By induction on the depth of the proof.

(Base case) Suppose $\varphi=\chi_t$, for some atomic proposition $\chi\in\lang{}$. From the truth definition we have $Tree_{T^*},\pi_{\overline{T}^*}\models \chi_t$ iff $\chi\in v(\pi_{\overline{T}^*t})$. From the construction of $\pi_{\overline{T}^*}$ it follows then directly that $\chi_t\in \overline{T}^*$. Suppose $\varphi=do(a)_t$. From the truth definition we have $Tree_{T^*},\pi_{\overline{T}^*}\models do(a)_t$ iff $\vala(\pi, \overline{T}^{*t})=a$. Again, from the construction of $\pi_{\overline{T}^*}$ we obtain $do(a)_t\in \overline{T}^*$.

(Negation) Suppose $\varphi=\neg \psi$. From the valuation lemma we know that $\neg\psi\in \overline{T}^*$ iff $\psi\not\in \overline{T}^*$. By the induction hypothesis, $\psi\not\in \overline{T}^*$ is equivalent to $Tree_{T^*},\pi_{\overline{T}^*}\not\models\psi$. According to the truth definition that is equivalent to $Tree_{T^*},\pi_{\overline{T}^*}\models\neg\psi$. Hence, $\neg\psi\in \overline{T}^*$ is equivalent to $Tree_{T^*},\pi_{\overline{T}^*}\models \neg\psi$.

(Conjunction) Suppose $\varphi=\psi\wedge\chi$. From the valuation lemma we know that $\psi\wedge\chi\in \overline{T}^*$ iff $\psi\in \overline{T}^*$ and $\chi\in \overline{T}^*$. By the induction hypothesis, that is equivalent to $Tree_{T^*},\pi_{\overline{T}^*}\models\psi$ and $Tree_{T^*},\pi_{\overline{T}^*}\models\chi$, respectively. Lastly, applying the truth definition, this is equivalent to $Tree_{T^*},\pi_{\overline{T}^*}\models\psi\wedge\chi$. Therefore, $\psi\wedge\chi\in \overline{T}^*$ iff $Tree_{T^*},\pi_{\overline{T}^*}\models \psi\wedge\chi$.

(Necessity) Suppose $\varphi=\Box_t\psi$. We show both directions of the bi-implication separately. 

$\Rightarrow$: Suppose that $Tree_{T^*},\pi_{\overline{T}^*}\models\Box_t\psi$, i.e. for all $\pi'$ with $\pi_{\overline{T}^*}\sim_t\pip: Tree_{T^*},\pip\models \psi$. Pick such $\pip$ arbitrarily. From Lemma~\ref{lemma:mcs:functionsinverted} we have that there is a unique mcs $\overline{\overline{T}}^*$ such that $\pip=\pi_{\overline{\overline{T}}^*}$. Thus, $Tree_{T^*},\pi_{\overline{\overline{T}}^*}\models\psi$ holds, and by the induction hypothesis, $\psi\in\pi(\overline{\overline{T}}^*)$ holds as well. Since $\pi(\overline{T}^*)\sim_t\pi(\overline{\overline{T}}^*)$, from Lemma~\ref{lemma:mcs:equiv:paths} we obtain $\overline{T}^*\equiv_t \overline{\overline{T}}^*$. Thus, by the valuation lemma we obtain $\Box_t\psi\in \overline{T}^*$.

$\Leftarrow$: Suppose that $\Box_t\psi\in \overline{T}^*$. By the valuation lemma, for all $\overline{\overline{T}}^*$ with $\overline{T}^*\equiv_t\overline{\overline{T}}^*: \psi\in\overline{\overline{T}}^*$. Take such $\overline{\overline{T}}^*$ arbitrarily. From the induction hypothesis we have that $Tree_{T^*},\pi_{\overline{\overline{T}}^*}\models\psi$. Since $\overline{T}^*\equiv_t \overline{\overline{T}}^*$, it follows from Lemma~\ref{lemma:mcs:equiv:paths} that $\pi_{\overline{T}^*}\sim^t\pi_{\overline{\overline{T}}^*}$. Since $\overline{\overline{T}}^*$ was chosen arbitrarily, we have that for all $\pi'$ with $\pi_{\overline{T}^*}\sim_t\pi'$ it holds that $Tree_{\overline{T}},\pi'\models\psi$. Therefore, $Tree_{T^*},\pi_{\overline{T}^*}\models\Box_t\psi$.
\end{proof}

We can now prove that the logic \logic{} is strongly complete:
\begin{proof}[Theorem 1, Completeness]
We prove this by contraposition, showing that $T\not\vdash\varphi$ implies $T\not\models\varphi$. If $T\not\vdash\varphi$, then $T\cup\{\neg\varphi\}$ is consistent, so there is a mcs $T^*\supset T$ containing $\neg\varphi$, as the Lindenbaum lemma shows. By the Truth lemma we have that $Tree_{T^*},\pi_{\overline{T}^*}\models\neg \varphi$ iff $\neg\varphi\in \overline{T}^*$. And thus $Tree_{T^*},\pi_{\overline{T}^*}\models \neg \varphi$, since $\neg\varphi\in \overline{T}^*$. Hence there is a model, namely $Tree$, and a path, namely $\pi_{\overline{T}^*}$, where all the formulas from $T$ hold, and $\varphi$ doesn't  hold. Therefore, $T\not\models\varphi$, and that is what we had to show.
\end{proof}

\begin{proposition}
\label{decidability}
The logic \logic{} is decidable.
\end{proposition}

\begin{proof}
Consider a formula $\varphi\in \lang$. Since  $\varphi$
 contains a finite number of time indexes,  there is a maximal time index $t$ appearing in $\varphi$. By Definition \ref{def:truthdef}, for checking if $\varphi$ is satisfied in a model $m$ it is enough to check the states and actions in  $m$ up to time $t+1$. We define the set $\prop (\varphi)$ as the smallest set such that:
if $p_\ell$ appears (for some $\ell$) in $\varphi$, then $p\in\prop (\varphi)$;
if $post(a)_\ell$ appears in $\varphi$, then $pre(a),post(a)\in\prop (\varphi)$;
  if $pre(a_1,\ldots,a_n)\ell$ appears in $\varphi$ and $(a_j,\ldots,a_k)$ is a subsequence of $(a_1,\ldots,a_n)$, then $pre(a_j,\ldots,a_k), post(a_j)\in\prop (\varphi)$. By Definition \ref{def:truthdef}, satisfiability of the formula $\varphi $ in any considered  model will not depend on the value of any proposition outside of $\prop (\varphi)$, in any state.

Thus, we can consider the models restricted to time $t+1$ such that  the evaluation functions $v$ are also restricted to $\prop (\varphi)$. Clearly, $\varphi$ is satisfiable, if it is satisfiable in such a restricted model. 

Since there are finitely many deterministic actions, and finitely many  restricted evaluation functions $v$ (to $\prop (\varphi)$) there are only finitely many 
possible  trees with paths whose length is $t+1$ and in each state $s$, $v(s)\subseteq \prop (\varphi)$.

For each of them we can check if they can be extended to a model by checking if it satisfies the four conditions of Definition \ref{def:model} and, if it does, if $\varphi$ is satisfied in it.
Thus, the satisfiability problem for the logic PAL is decidable.
\end{proof}

\section{Representation Theorem Proofs}

Recall that $Mod(\varphi)$ is the set of models of $\varphi$. Similarly, given some $t$-restricted \logic{} formula $\varphi$, we define $\restmod{}(\varphi)$ as the set of $t$-restricted models of $\varphi$. Recall from the paper that $Ext(\reststr)$ is the set of all possible extensions of a set of bounded model of strong beliefs $\reststr$ to models, i.e. $$Ext(\reststr) = \{m\in\mods{}\mid \restm\in\reststr{}\}.$$ 
Note that $Ext$ is defined on the sets of bounded models of \textbf{strong beliefs} only. In order to simplify notation in the proof of the representation theorem, we define the following abbreviation:\\

\noindent
\textbf{Given a set of restricted models $\{\restmsub{1},\ldots,\restmsub{n}\}$, with $\restmsub{i}=(\resttsub{i},\restpsub{i})$, we introduce  $Ext(\restmsub{1},\ldots,\restmsub{n})$ as:}  
\begin{eqnarray}
\label{abr ext} Ext(\restmsub{1},\ldots,\restmsub{n}):=Ext(\{(\restt,\restp) \ | \ \bigvee_{k=1}^{n}\restt=\resttsub{i}\}).
\end{eqnarray}

\begin{lemma}
\label{lemma form}
Given a set of $t$-bounded models of strong beliefs $\reststr$, there exists a strong belief formula $form(\reststr)$ such that $Mod(form(\reststr))=Ext(\reststr)$.
\end{lemma}

\begin{proof}
For a given $\restt{}$ we define the strong belief formula $$form(\restt) = \bigwedge_{\restp{}'\in\restt{}}\Diamond_0\alpha_{\restp{}'}\wedge\bigwedge_{\restp{}'\not\in\restt{}}\neg \Diamond_0 \alpha_{\restp{}'},$$ where $$\alpha_{\restp{}} = \bigwedge_{n=0}^t\left(\bigwedge_{\chi\in \valp{}(\restp{}_n)}\chi_n\wedge\bigwedge_{\chi\not\in \valp{}(\restp{}_n)}\neg \chi_n\wedge \bigwedge_{\vala{}(\restp{}, n)=a}do(a)_n\right).$$
Intuitively, $form(\restt{})$ is a strong belief formula describing all of the paths of $T$ up to $t$. Each $\alpha_{\restp}$ is a formula describing the path $\pi$ up to $t$: It contains all propositions that are true and false at each time moment, and all actions that are executed. Note that Axiom A7 of PAL-P ensures that only one action can be executed per time moment. 

Let $T'$ be a tree. From the construction of the formula $form(\restt)$ it follows that if $\restt'=\restt$
, then for every $\pi'\in T'$ we have $(T',\pi')\models form(\restt)$. On the other hand, if  $\restt'\neq\restt$, then there is $\pi$ such that either $\restp\in\restt\setminus\restt'$ or  $\restp\in\restt'\setminus\restt$. Suppose that  $\restp\in\restt\setminus\restt'$. Then  for any $\pi\in T'$ we have $(T',\pi')\not\models \Diamond_0 \alpha_{\restp{}}$, so  $(T',\pi')\not\models form(\restt)$. Similarly, if  $\restp\in\restt'\setminus\restt$, then $(T',\pi)\not\models \neg\Diamond_0 \alpha_{\restp{}}$, so again we have $(T',\pi)\not\models form(\restt)$.
Thus, we proved $$Mod(form(\restt))=Ext(\{(\restt,\restp)\in \restms\mid \restp\in\restt\}).$$

Now we define $$form(\restM_{SB})=\bigvee \{form(\restt)\mid (\restt,\restp)\in\restM_{SB}\}.$$ 
Note that the set of propositional letters from $\lang{}^{|t}$ is finite, and that we have finitely many deterministic actions, so $\restM_{SB}$ is a finite set. Consequently, the above disjunction is finite. 

Finally, we have
\begin{align*}
&\ Mod(form(\restM_{SB}))\\
=&\ \bigcup \{Mod(form(\restt))\mid (\restt,\restp)\in\restM_{SB}\}\\
=&\ \bigcup Ext(\{(\restt,\restp)\in \restM_{SB}\mid \restp\in\restt\})\\
=&\ Ext(\restM_{SB}).
\end{align*}
\end{proof}

Note that $form$ is defined on the set of bounded models of \textbf{strong beliefs} only. In order to simplify notation in the proof of the representation theorem, we define the following abbreviation:\\

\noindent
\textbf{Given a set of models $\{m_1,\ldots,m_n\}$, with $m_i=(T_i,\pi_i)$, we introduce $form(m_1,\ldots,m_n) $ as:}
\begin{eqnarray}
\label{abr form}form(m_1,\ldots,m_n) := form(\{(\restt,\restp)\mid\bigvee_{i}^n \restt=\resttsub{i}\}).
\end{eqnarray}

\setcounter{lemma}{17}
\begin{lemma}
\label{lemma strong}
If $\varphi$ is a formula of strong beliefs bounded up to $t$,  and $\resttsub{1}=\resttsub{2}$, then $(T_1,\pi_1)\models\varphi$ iff $(T_2,\pi_2)\models\varphi$.
\end{lemma}

\begin{proof}
By induction on the complexity of $\varphi$.
\end{proof}

\begin{corollary}
\label{cor:strongbeliefformula}
Given a $t$-bounded strong belief set $B$, there exists a $t$-bounded formula of strong beliefs $\psi$ such that $B=\{\varphi\mid\psi\vdash\varphi\}$.
\end{corollary}

\begin{proof}
\label{coherence restricted}
For a given belief set $B$, from Lemma \ref{lemma strong} follows that $\restmod{}(B)$ is a set of $t$-bounded models of a strong beliefs such that $Ext(\restmod(B))=Mod(B)$.  If $\psi=form(\restmod(B))$, then form Lemma \ref{lemma form} we obtain $Mod(\psi)=Mod(B)$, and by the completeness theorem, $B=Cl(\psi)$.
\qed
\end{proof}

Note that for a formula $\varphi\in B_t$, the satisfiability of the formula in a model $m$ depends only on the paths in its restricted counterpart $\restm$, for a set of intentions bounded up to $t$ so we can write that 
\begin{eqnarray}
(\restM,I) \rm{ \ is  \ coherent \  iff \ } (M,I) \rm{\ is \ coherent.}
\end{eqnarray}

We next repeat the definition of faithful assignment from the paper and we restate the representation theorem. After that, we prove the representation theorem.

\setcounter{definition}{24}

\begin{definition}[Faithful assignment]
A \emph{faithful assignment} is a function that assigns to each strong belief formula $\psi\in \sbel\rest{}$ a total pre-order $\lept$ over $\mods{}$ and to each intention database $I\in \mathbb{D}\rest{}$ a selection function $\sel$ and satisfies the following conditions:
\begin{enumerate}
\item If $m_1,m_2\in Mod(\psi)$, then $m_1\lept m_2$ and $m_2\lept m_1$.
\item If $m_1\in Mod(\psi)$ and $m_2\not\in Mod(\psi)$, then $m_1< m_2$.
\item If $\psi\equiv\phi$, then $\lept = \le_\phi^t$.
\item If $\restt=T_2\rest{}$, then $(T,\pi)\lept (T_2,\pi_2)$ and $(T_2,\pi_2)\lept (T,\pi)$.
\end{enumerate}
\end{definition}

\begin{theorem}[Representation Theorem]
The function  $\agr:\bis{}\times (\sbel{}\times \ib{})\rightarrow \bis$ is a belief-intention revision operator  
iff there exists a faithful assignment that maps each $\psi$ to a total pre-order $\lept$ and each $I$ to a selection function $\sel$ such that if $(\psi,I)*_t(\varphi,i)=(\psi',I')$, then:
\begin{enumerate}
\item $Mod(\psi')=\min(Mod(\varphi),\lept)$
\item $I' = \sel(Mod(\psi'),i)$
\end{enumerate}
\end{theorem}

\begin{proof}
The belief part of the representation proof is almost identical to the existing proofs of Katsuno and Mendelzon, apart from dealing with the new postulate for strong beliefs.

$``\Rightarrow'':$ Suppose that $\agr:\bis{}\times (\sbel{}\times \ib{})\rightarrow \bis$ is a belief-intention revision operator. Then there exist   a strong belief revision function $\brt$ (which satisfies (R1)--(R6)) and  an intention revision function $\agri$ (which satisfies the postulates (P1)--(P5) from Definition \ref{def:faithful-assignment}) such that $\agr$ can be represented as follows: 
$(\psi,I)\agr(\varphi,i) =(\psi\brt \varphi,I)\agri i.$ 

Given models $m_1$ and $m_2$, let $\psi\brt form(m_1,m_2)=\psi'$ (note that we use the abbreviation (\ref{abr form}) for $form$). We define $\lept$ by $m_1\lept m_2$ iff $m_1\models\psi$ or $m_1\models\psi'$. We also define $\sel$ by $\sel(\reststr,i)=I'$, where $(form(\reststr),I)\agri i=(form(\reststr),I')$. 

In the first part of the proof, modifying  the proof technique of  Katsuno and Mendelzon, we  show that 1) $\le_\psi^t$ is a total pre-order, 2) the assignment $\psi$ to $\lept$ is faithful, 3)  $Mod(\psi')=\min(Mod(\varphi),\lept)$. Then we show that 4) $\sel$ is a selection function and 5) $I' = \sel(\restmod(\psi'),i)$.
\begin{enumerate}
\item To show: $\lept$ is a total pre-order.
\begin{itemize}
\item To show: Totality and reflexivity. From (R1) and (R3): $Mod(\psi\circ_t form(m_1,m_2))$ is a nonempty subset of $Ext(\restmsub{1},\restmsub{2})$ (note that we use the abbreviation (\ref{abr ext}) for $Ext$). Therefore, for each $m\in Ext(\restmsub{1})$ and $m'\in Ext(\restmsub{2})$, we have that either $m\lept m'$ or $m'\lept m$. We now show, without loss of generality, that for each $m,m'\in Ext(\restmsub{1})$, both $m\lept m'$ and $m'\lept m$ hold. Therefore, let $m,m'\in Ext(\restmsub{1})$, so $\restm{}=\restm{}'$. By Lemma 3 of the paper, $Mod(form(\restm{})) = Ext(\restm{}) = Ext(\restm{}') = Mod(form(\restm{}'))$. Hence, $form(m)\equiv form(m')$, so $form(m,m')\equiv form(m)$. By (R4): $Mod(\psi\circ_t form(\restm{}))=Mod(\psi\circ_t form(m,m')))$. By (R1), $m\in Mod(\psi\circ_t form(m)$, so $m\in Mod(\psi\circ_t form(m,m'))$. Hence, by the definition of $\le_\psi^t$: $m\le_\psi^t m'$. We can prove $m'\le_\psi^t m$ similarly. This proves that $\lept$ is total, which implies reflexivity.
\item To show: Transitivity. Assume $m_1\lept m_2$ and $m_2\lept m_3$. We show $m_1\lept m_3$. There are three cases to consider:
\begin{enumerate}
\item $m_1\in Mod(\psi)$. $m_1\lept m_3$ follows from the definition of $\lept$.
\item $m_1\not\in Mod(\psi)$ and $m_2\in Mod(\psi)$. Since $Mod(\psi\wedge form(m_1,m_2))=Ext(\restmsub{2})$ holds, $Mod(\psi\circ_t form(m_1,m_2)) = Ext(\restmsub{2})$ follows from (R2). Thus $m_1\not\lept m_2$ follows from $m_1\not\in Mod(\psi)$. This contradicts $m_1\lept m_2$, so this case is not possible.
\item $m_1\not\in Mod(\psi)$ and $m_2\not\in Mod(\psi)$. By (R1) and (R3), $Mod(\psi\circ_t form(m_1,m_2,m_3))$ is a nonempty subset of $Ext(\restmsub{1},\restmsub{2},\restmsub{3})$. We now consider two subcases.
\begin{enumerate}
\item $Mod(\psi\circ_t form(m_1,m_2,m_3))\cap Ext(\restmsub{1},\restmsub{2})=\emptyset$. In this case, $Mod(\psi\circ_t form(m_1,m_2,m_3))=Ext(\restmsub{3})$ holds. If we regard $\varphi$ and $\varphi'$ as $form(m_1,m_2,m_3)$ and $form(m_2,m_3)$ respectively in Conditions (R5) and (R6), we obtain
\begin{align*}
Mod(\psi\circ_t form(m_1,m_2,m_3)) \cap Ext(\restmsub{2},\restmsub{2}) = Mod(\psi\circ_t form(m_2,m_3)).
\end{align*}
Hence, $Mod(\psi\circ_t form(m_2,m_3)) = Ext(\restmsub{3})$. This contradicts $m_2\lept m_3$ and $m_2\not\in Mod(\psi)$. Thus, this subcase is not possible.
\item $Mod(\psi\circ_t form(m_1,m_2,m_3))\cap Ext(\restmsub{1},\restmsub{2})\not=\emptyset$. Since $m_1\lept m_2$ and $m_1\not\in Mod(\psi)$, $m_1\in Mod(\psi\circ_t form(m_1,m_2))$ holds. Hence, by regarding $\varphi$ and $\varphi'$ as $form(m_1,m_2,m_3)$ and $form(m_1,m_2)$ respectively in Conditions (R5) and (R6), we obtain
\begin{align*}
Mod(\psi\circ_t form(m_1,m_2,m_3))\cap Ext(\restmsub{1},\restmsub{2}) = Mod(\psi\circ_t form(m_1,m_2)).
\end{align*}
Thus, 
\begin{align*}
m_1\in Mod(\psi\circ_t form(m_1,m_2,m_3))\cap Ext(\restmsub{1},\restmsub{2})
\end{align*}
holds. By using conditions (R5) and (R6) again in a similar way, we can obtain $m_1\in Mod(\psi\circ_t form(m_1,m_3))$. Therefore, $m_1\lept m_3$ holds.
\end{enumerate}
\end{enumerate}
\end{itemize}
\item To show: The assignment mapping $\psi$ to $\lept$ is faithful. We prove the four conditions separately
\begin{enumerate}
\item The first condition follows from the definition of $\lept$. 
\item For the second condition, assume that $m\in Mod(\psi)$ and $m'\not\in Mod(\psi)$. Then $Mod(\psi\circ_t form(m,m'))=Ext(\restm{})$ follows from (R2). Therefore, $m<_\psi^t m'$ holds. 
\item The third condition follows from (R4).
\item For the fourth condition, for $m_1=(T_1,\pi_1)$ and $m_2=(T_2,\pi_2)$ such that $\resttsub{1}=\resttsub{2}$, let $\psi'$ be as above. Since $\psi,\psi'\in B_t$, by Lemma \ref{lemma strong} we obtain  $m_1\models\psi$ iff $m_2\models\psi$ and $m_1\models\psi'$ iff $m_2\models\psi'$, so $m_1\lept m_2$ and $m_2\lept m_1$. 
\end{enumerate} 
\item To show:  $Mod(\psi')=\min(Mod(\varphi),\lept)$. Note that this can be equivalently rewritten as $Mod(\psi\circ_t\varphi)=\min(Mod(\varphi),\lept)$. If $\varphi$ is unsatisfiable then both sets are empty. So we assume $\varphi$ is satisfiable. We show both containments separately.
\begin{itemize}
\item To show: $Mod(\psi\circ_t\varphi)\subseteq \min(Mod(\varphi),\lept)$. Assume for contradiction that $m\in Mod(\psi\circ_t\varphi)$ and $m\not\in \min(Mod(\varphi),\lept)$. By condition (R1), $m$ is a model of $\varphi$. Hence, there is a model $m'$ of $\varphi$ such that $m'<_\psi^t m$. We consider two cases:
\begin{enumerate}
\item $m'\in Mod(\psi)$. Since $m'\in Mod(\varphi)$, $\psi\wedge\varphi$ is satisfiable. Hence, by condition (R2), $\psi\circ_t\varphi\equiv\psi\wedge\varphi$ holds. Thus, $m\in Mod(\psi)$ follows from $m\in Mod(\psi\circ_t \varphi)$. Therefore, $m\lept m'$ holds. This contradicts $m'<_\psi^t m$. 
\item $Mod(\psi\circ_t form(m,m')) = Ext(\restm{})$. Since both $m$ and $m'$ are models of $\varphi$, $\varphi\wedge form(m,m') \equiv form(m,m')$ holds. Thus,
\begin{align*}
Mod(\psi\circ_t\varphi)\cap Ext(\restm{},\restm{}')\subseteq Mod(\psi\circ_t form(m,m'))
\end{align*}
follows from condition (R5). Since we assume $Mod(\psi\circ_t form(m,m'))=Ext(\restm{}')$, we obtain $m\not\in Mod(\psi\circ_t\varphi)$. This is a contradiction.
\end{enumerate}
\item To prove: $\min(Mod(\varphi),\lept)\subseteq Mod(\psi\circ_t\varphi)$. Assume for contradiction that $m\in \min(Mod(\varphi),\lept)$ and $m\not\in Mod(\psi\circ_t\varphi)$. Since we also assume that $\varphi$ is satisfiable, it follows from condition (R3) that there is an interpretation $m'$ such that $m'\in Mod(\psi\circ_t\varphi)$. Since both $m$ and $m'$ are models of $\varphi$, $form(m,m')\wedge\varphi\equiv form(m,m')$ holds. By using conditions (R5) and (R6), we obtain
\begin{align*}
Mod(\psi\circ_t\varphi)\cap Ext(\restm{},\restm{}')=Mod(\psi\circ_t form(m,m')).
\end{align*}
Since $m\not\in Mod(\psi\circ_t\varphi)$, $Mod(\psi\circ_t form(m,m')) = Ext(\restm{}')$ holds. Hence, $m'\lept m$ holds. On the other hand, since $m$ is minimal in $Mod(\varphi)$ with respect to $\lept$, $m\lept m'$ holds. Since $Mod(\psi\circ_t form(m,m'))=Ext(\restm{}')$, $m\in Mod(\psi)$ holds. Therefore, $m\in Mod(\psi\circ_t\varphi)$ follows from condition (R2). This is a contradiction.
\end{itemize}
\item To show: $\sel$ is a selection function. This is direct consequence of the completeness theorem and  the postulates (P1)-(P5), taking into account (\ref{coherence restricted}). For example, if (P1) holds, then $\psi' $ is consistent with $Cohere(I')$, so by completeness there is a model of both  $\psi'$ and $Cohere(I')$. since $Mod(\psi')=\min(Mod(\varphi),\lept)$, we obtain that $(Mod(\psi'),I')$ is coherent. 
\item To show: $I' = \sel(\restmod(\psi'),i)$. This is true since from our definition of $\sel$ we have that 
$(\psi',I)\agri i=(\psi',\sel(\restmod(\psi'),i))$. 

\end{enumerate}

\item[``$\Leftarrow$'':] Assume that there is a faithful assignment that maps $\psi$ to a total pre-order $\lept$ and $I$ to a selection function $\sel$. 
We define the $t$-bounded revision operator $*_t$ as follows:
$$(\psi,C)*_t(\overline{\varphi},c)=(form(\min(\restmod(\varphi),\lept),\sel( \min(\restmod(\varphi),\lept),i)).$$
First we show that the operator is correctly defined, i.e. that $\min(\restmod(\varphi),\lept)$ is a set of $t$-bounded models of strong beliefs.  Let $T^{|t}=T'^{|t}$.  
Since $\varphi\in \mathbb{SB}\rest$ , by Lemma \ref{lemma strong} we obtain that  $(T,\pi)\models\varphi$ iff $(T',\pi')\models\varphi$. Now suppose that   $(T,\pi)\in \min(Mod(\varphi),\lept)$. If $(T' ,\pi' )\notin \min(Mod(\varphi),\lept)$, then   $(T,\pi)<(T',\pi')$, which is impossible by the definition of faithful assignment. Thus, $\min(\restmod(\varphi),\lept)$ is a set of $t$-bounded models of strong beliefs. 

Let us now prove that $*_t$ can be represented as a composition of  a strong belief revision function $\brt$ which satisfies the postulates  (R1)--(R6), and  an intention revision function $\agri$ which satisfies the postulates (P1)--(P5) (from Definition \ref{def:faithful-assignment}), in the following way:
$$(\psi,I)\agr(\varphi,i) =(\psi\brt \varphi,I)\agri i.$$
Namely, 
we define the operator $\circ_t$ by $\psi\circ_t\varphi =form(\min(\restmod(\varphi),\lept)),$ and the operator $\agri$ by $(\psi',I)\agri i = (\psi',\sel( \restmod(\psi'),i)).$
First we show that $\circ_t$ satisfies conditions (R1)--(R6). It is obvious that condition (R1) follows from the definition of the revision operator $\circ_t$. It is also obvious that conditions (R3) and (R4) follow from the definition of the faithful assignment. What remains to show is condition (R2), (R5), and (R6).
\begin{itemize}
\item To prove: (R2). It suffices to show if $Mod(\psi\wedge\varphi)$ is not empty then $Mod(\psi\wedge\varphi)=\min(Mod(\varphi),\lept)$. $Mod(\psi\wedge\varphi)\subseteq \min(Mod(\varphi),\lept)$ follows from the conditions of the faithful assignment. To prove the other containment, we assume that $m\in \min(Mod(\varphi),\lept)$ and $m\not\in Mod(\varphi\wedge\varphi)$. Since $Mod(\psi\wedge\varphi)$ is not empty, there is a model $m'\in Mod(\psi\wedge\varphi)$. Then $m\not\lept m'$ follows from the conditions of the faithful assignment. Moreover, $m'\lept m$ follows from the conditions of the faithful assignment. Hence, $m$ is not minimal in $Mod(\varphi)$ with respect to $\lept$. This is a contradiction. 
\item To prove: (R5) and (R6). It is obvious that if $(\psi\circ_t\varphi)\wedge\varphi'$ is unsatisfiable then (R6) holds. Hence, it suffices to show that if $\min(Mod(\mu),\lept)\cap Mod(\varphi')$ is not empty then
\begin{align*}
\min(Mod(\varphi),\lept)\cap Mod(\varphi') = \min(Mod(\varphi\wedge\varphi'),\lept)
\end{align*} 
holds. Assume that $m\in \min(Mod(\varphi),\lept)\cap Mod(\varphi')$ and $m\not\in \min(Mod(\varphi\wedge\varphi'),\lept)$. Then, since $m\in Mod(\varphi\wedge\varphi')$, there is an interpretation $m'$ such that $m'\in Mod(\varphi\wedge\varphi')$ and $m'<_\psi m$. This contradicts $m\in \min(Mod(\varphi),\lept)$. Therefore, we obtain
\begin{align*}
\min(Mod(\varphi),\lept)\cap Mod(\varphi') \subseteq \min(Mod(\varphi\wedge\varphi'),\lept).
\end{align*}
To prove the other containment, we assume that $m\not\in \min(Mod(\varphi),\lept)\cap Mod(\varphi')$ and $m\in \min(Mod(\varphi\wedge\varphi'),\lept)$. Since $m\in Mod(\varphi'), m\not\in \min(Mod(\varphi),\lept)$ holds. Since we assume that $\min(Mod(\varphi),\lept)\cap Mod(\varphi')$ is not empty, suppose that $m'$ is an element of $\min(Mod(\varphi),\lept)\cap Mod(\varphi')$. Then $m'\in Mod(\varphi\wedge\varphi')$ holds. Since we assume that $m\in \min(Mod(\varphi\wedge\varphi'),\lept)$ and $\lept$ is total, $m\lept m'$ holds. Thus, $m\in \min(Mod(\varphi),\lept)$ follows from $m'\in \min(Mod(\varphi),\lept)$. This is a contradiction.
\end{itemize}
Note that the  conditions (R1)-(R6)  imply the conditions (P1)-(P6). For example, suppose (R3) and let $(\psi,I)\agr(\varphi,i) = (\psi',I')$. Then $\psi' =form(\min(Mod(\varphi),\lept)\cap Mod(\varphi'))=\psi\circ_t\varphi$ so if $\varphi$ is satisfiable, $\psi' $ is satisfiable as well. Thus, (P3) holds.

Finally. the postulates (P1)-(P5) follow directly (using the completeness theorem and taking into account (\ref{coherence restricted})) from the conditions the conditions 1-5 of the definition of selection function.

\end{proof}
\section{Iterated Revision Proofs}

\begin{theorem}[Representation Theorem for iterated revision]

A function $\agr:\ebis{}\times (\sbel{}\times \ib{})\rightarrow \ebis{}$  is an epistemic belief-intention revision operator  iff there exists a faithful assignment for iterated revision that maps each $\Psi$ to a total pre-order $\le_\Psi^t $ 
%
 and each $I$ to a selection function $\sel$
  such that if $(\Psi,I)*_t(\varphi,i)=(\Psi',I')$, then:
\begin{enumerate}
\item $Mod(\Psi')=\min(Mod(\varphi),\le_{\Psi}^t)$
\item $I' = \sel(Mod(Bel(\Psi')),i)$
\end{enumerate}
\end{theorem}

\begin{proof}

The proof of this theorem consists from two parts.
In the first part, we show that the function  $\agr:\ebis{}\times (\sbel{}\times \ib{})\rightarrow \ebis$, which maps an epistemic belief-intention database, a strong belief formula, and an intention--- all bounded up to $t$--- to an epistemic belief-intention database bounded up to $t$, is a composition ($(\Psi,I)\agr(\varphi,i) =(\Psi\brt \varphi,I)\agri  $) of two functions, $\brt$  and $\agri$, that satisfy R1--R3, R*4, R5 and R6, and P1--P5, respectively,   \emph{if and only if}
   there exist (i) a 
mapping of each epistemic state $\Psi$ to a total pre-order $\le_\Psi^t $, such that 
\begin{enumerate}
\item[(1)] If $m_1,m_2\in Mod( \Psi)$, then $m_1\le_\Psi^t m_2$ and $m_2\le_\Psi^t m_1$
\item[(2)]  If $m_1\in Mod( \Psi)$ and $m_2\not\in Mod( \Psi)$, then $m_1<_\Psi^t m_2$
\item[(3)]   $\Psi=\Phi$ only if $\le_\Psi=\le_\Phi$
\item[(4)]  If $\restt=T_2\rest{}$, then $(T,\pi)\le_\Psi^t (T_2,\pi_2)$ and $(T_2,\pi_2)\le_\Psi^t (T,\pi)$,
\end{enumerate}
(in this proof, we call this mapping  \emph{restricted faithful assignment}), and (ii) a mapping of each $I$ to a selection function $\sel$
  such that if $(\Psi,I)*_t(\varphi,i)=(\Psi',I')$, then:
\begin{enumerate}
\item $Mod(\Psi')=\min(Mod(\varphi),\le_{\Psi}^t)$
\item $I' = \sel(Mod(Bel(\Psi')),i)$
\end{enumerate}
Note that this statement is a simplification of the representation theorem which does not consider the postulates  (C1)-(C4) from Definition \ref{def:iterated-revision-postulates} nor the conditions 5--8 of Definition \ref{def:tbounded:faithfulassigment:epst}.
The proof of this statement is symmetric to the proof of Theorem \ref{thm:singlestep}, so we will not repeat it here.

In the second part, we complete the proof, modifying the proof technique of Darwiche and Pearl, showing  that $\brt$ in addition satisfies (C1)-(C4) iff the operator and its corresponding restricted faithful assignment (introduced above) 
 satisfy the conditions 5--8 of Definition \ref{def:tbounded:faithfulassigment:epst}.
Let us   assume
 
 $\Psi\brt\varphi=\Psi'$,
 
$\Psi\brt\varphi' = \Psi''$, and 

$\Psi''\brt\varphi=\Psi'''$.
  
 Now we prove the following statements:

\begin{itemize}

\item (C1) is equivalent to the condition 5 of Definition \ref{def:tbounded:faithfulassigment:epst}:

(1) Assume that if $m_1\in Mod(\varphi')$ and $m_2 \in Mod(\varphi')$, then $m_1\le_\Psi^t m_2$ iff $m_1\le_{\Psi''}^tm_2$ (the condition 5 applied to $\varphi'$). Assume also $\varphi\models\varphi'$. In order to infer (C1), we need to prove  $\Psi'\equiv \Psi'''$. From condition 5 we obtain that the restrictions of $\le_\Psi^t$ and $\le_{\Psi''}^t$ to the set $(Mod(\varphi))^2$ coincide.  Therefore, from  $Mod(\Psi')=\min(Mod(\varphi),\le_{\Psi}^t)$ and $Mod(\Psi''')=\min(Mod(\varphi),\le_{\Psi''}^t)$ we obtain $Mod(\Psi')=Mod(\Psi''')$, i.e.,  $\Psi'\equiv \Psi'''$. 

(2) For the other direction, assume that $\varphi\models\varphi'$ always imply  $\Psi'\equiv \Psi'''$ (C1). Assume also that $m_1\in Mod(\varphi')$ and $m_2 \in Mod(\varphi')$. We need to show that $m_1\le_\Psi^t m_2$ iff $m_1\le_{\Psi''}^tm_2$.
Let us chose $\varphi= form(m_1,m_2)$  (note that we use the abbreviation (\ref{abr form}) for $form$). Then 
$$Mod(\varphi)=Ext(\restmsub{1})\cup Ext(\restmsub{2})\subseteq Mod(\varphi'),$$ 
i.e., $\varphi\models\varphi'$. By (C1) we have $\Psi'\equiv \Psi'''$. 
Note that the condition (4) of the restricted faithful assignment ensures that  for each $m_1'\in Ext(\restmsub{1})$ we have $m_1\le_\Psi^t m_1'$ and $m_1'\le_\Psi^t m_1$ and, similarly, for each $m_2'\in Ext(\restmsub{2})$ we have $m_2\le_\Psi^t m_2'$ and $m_2'\le_\Psi^t m_2$.
Consequently, 
$\min(Mod(\varphi),\le_{\Psi}^t)=\min(Mod(\varphi),\le_{\Psi''}^t)$; and we showed $m_1\le_\Psi^t m_2$ iff $m_1\le_{\Psi''}^tm_2$.

\item The proof that (C2) is equivalent to the condition 6 of Definition \ref{def:tbounded:faithfulassigment:epst} is symmetric to the proof that (C1) is equivalent to the condition 5.

\item (C3) is equivalent to the condition 7 of Definition \ref{def:tbounded:faithfulassigment:epst}:

(1) Assume that if $m_1\in Mod(\varphi')$, $m_2\not \in Mod(\varphi')$ and $m_1<_\Psi^t m_2$, then $m_1<_{\Psi''}^tm_2$ (the condition 7 applied to $\varphi'$). Assume also  $\Psi'  \models\varphi'$. In order to infer (C3), we need to prove  $\Psi''' \models  \varphi'$. We will use the following fact:

\begin{itemize}

\item[(i)] $\Psi' \models \theta$ iff there is a model $m\in Mod(\varphi\wedge \theta)$ s.t. for every $m'\in Mod(\varphi\wedge \neg\theta)$ we have $m \le_{\Psi}^t m'$.

\end{itemize}
Then there exists $m\in Mod(\varphi\wedge \varphi')$ such that for every $m'\in Mod(\varphi\wedge \neg\varphi')$ we have $m \le_{\Psi}^t m'$. Then, by (C3), there exists $m\in Mod(\varphi\wedge \varphi')$ such that for every $m'\in Mod(\varphi\wedge \neg\varphi')$ we have $m \le_{\Psi''}^t m'$. Using $(i)$ we obtain $\Psi''' \models  \varphi'$.

(2) For the other direction, assume that $\Psi'  \models\varphi'$ always imply   $\Psi''' \models  \varphi'$ (C*3). Assume also that $m_1\in Mod(\varphi')$, $m_2\not \in Mod(\varphi')$ and $m_1<_\Psi^t m_2$. We need to show that $m_1<_{\Psi''}^tm_2$.
Let us chose $\varphi= form(m_1,m_2)$  (again, we use the abbreviation (\ref{abr form}) for $form$), i.e., $Mod(\varphi)=Ext(\restmsub{1})\cup Ext(\restmsub{2})$.
Note that  $\Psi'  \models\varphi'$ by  $(i)$, because $m_1\in Mod(\varphi \wedge\varphi')$, $Mod(\varphi \wedge\neg\varphi')= Ext(\restmsub{2})$  and  $m_1<_\Psi^t m_2$.  Consequently, by (C3), $\Psi''' \models  \varphi'$. Then 
$Mod(\varphi \wedge\varphi')Ext(\restmsub{1})$, $Mod(\varphi \wedge\neg\varphi')= Ext(\restmsub{2})$ and $(i)$ imply $m_1<_{\Psi''}^tm_2$.

\item The proof that (C4) is equivalent to the condition 8 of Definition \ref{def:tbounded:faithfulassigment:epst} is similar  to the proof that (C3) is equivalent to the condition 7.

\end{itemize}
Thus, we proved that an operator $\brt$, which 
 satisfies the postulates  R1--R3, R*4, R5 and R6, also satisfies (C1)-(C4) iff the operator and its corresponding restricted faithful assignment 
 satisfy the conditions 5--8 of Definition \ref{def:tbounded:faithfulassigment:epst}, which completes the proof.
\end{proof}

\begin{proposition}
\label{prop:extspohn}
If $\kappa_t$ is a Spohn ranking function, then $\kappa_t'$ from Definition \ref{def:spohn-iterated-revision-operator} is also a  Spohn ranking function.
\end{proposition}

\begin{proof}
Let $\kappa_t$ be a Spohn ranking function and let 
 $m_1=(T_1,\pi_1)$ and $m_2=(T_2,\pi_2)$ such that $T_1\rest{} = T_2\rest{}$. Then 
\begin{itemize}

\item $\kappa_t(m_1)=\kappa_2(m_2)$, by Definition \ref{def:spohnA},

\item   for a strong belief formula $\varphi$ bounded up to $t$, we have $m_1\in Mod(\varphi)$ iff $m_2\in Mod(\varphi)$.
\end{itemize}
Therefore, by definition of $\kappa_t'$ we have $\kappa_t'(m_1)=\kappa_t'(m_2)$.
\end{proof}

\begin{theorem} The function  $\spohn$ is an epistemic belief-intention revision operator.
\end{theorem}

\begin{proof}

First we define the function  $\sel:\msbsets{}\times \ib \rightarrow \idb$, which maps an msb set (Definition~\ref{def:msbset}) and an intention to an updated intention database---all bounded up to $t$--- as follows: $\sel(\msbsetrest,i)=I'$,  where  $I'$ is defined iteratively (below, $n$ is the maximal time instance that occurs in $I$):
\begin{enumerate}
\item $I_{-1} =\{ i\}$, if $(\msbsetrest,\{ i \})$ is coherent, otherwise $I_{-1} = \emptyset$.
\item for every $\ell\in\{0,1,\dots,n\}$

\begin{enumerate}
\item if there is no $a$ such that $(a,\ell)\in I$, $I_{\ell} =I_{\ell-1}$.

\item otherwise, if $a$ is  the unique action such that $(a,\ell)\in I$, then  $I_{\ell} =I_{\ell-1}\cup\{ (a,\ell)\}$, if 
$(\msbsetrest,I_{\ell-1}\cup\{ (a,\ell)\})$ is coherent, otherwise $I_{\ell} =I_{\ell-1}$.

\end{enumerate}
\item $I'=I_n$.
\end{enumerate}

Now we show that   $\sel:\msbsets{}\times \ib \rightarrow \idb$ is a selection function.
We need to show that the conditions 1--5 of Definition \ref{def:selection-function}  hold:
\begin{enumerate}
\item $(\msbsetrest,I_{-1})$ is obviously coherent, and in the step 2.b a new intention is added only if coherence is preserved, so    $(\msbsetrest,I')$ is coherent.
\item If $(\msbsetrest,\{i\})$ is coherent, then  $I_{-1} =\{ i\}$, so   $i\in I'$.
\item If $(\msbsetrest,I\cup\{i\})$ is coherent, then  $I_{-1} =\{ i\}$ and every intention from $I$ will be added in 2.b, so   $I\cup\{i\}= I'$.
\item $I'\subseteq I\cup\{i\}$ by the construction of $I'$.
\item Let us consider $I''$ such that $I'\subset I''\subseteq I\cup\{i\}$. Then there is $(a,k)\in I''\setminus I'$. Since $(a,k)\notin  I'$, by step 2.b we have that $(\msbsetrest,I_{k-1}\cup\{ (a,k)\})$ is not coherent, and since  $I_{k-1}\cup\{ (a,k)\}\subseteq I'\cup\{ (a,k)\}\subseteq I''$, 
we obtain that
$(\msbsetrest,I'')$ is not coherent.
\end{enumerate}
For any Spohn ranking function $\kappa_t$, we define, similarly as Darwiche and Pearl, the  pre-order $\leq_{\kappa_t}$ on the set of  models in the following way:
$m_1 \leq_{\kappa_t} m_2$ iff $\kappa_t(m_1)\leq \kappa_t(m_2)$.

Let $(\kappa_t,I)\spohn(\varphi,i) = (\kappa_t',I')$. 
Then,  by Definition \ref{def:spohn-iterated-revision-operator},
\begin{align*}
\kappa_t'(m) = \begin{cases}
\kappa_t(m)-\kappa_t(\varphi), &\text{ if } m\models \varphi;\\
\kappa_t(m)+1,&\text{ if } m\models \neg \varphi.
\end{cases}
\end{align*}

Assume that $m_1\in Mod(Bel(\kappa_t'))$. Then $\kappa_t'(m_1)=0$. Also, by the above definition we have that $m_1\in Mod(\varphi)$. Then  $\kappa_t(m_1)=\kappa_t(\varphi)$  since $\kappa_t'(m_1)=\kappa_t(m_1)-\kappa_t(\varphi)$ and we know  $\kappa_t'(m_1)=0$. From $\kappa_t(m_1)=\kappa_t(\varphi)=\min_{m\models\varphi}\kappa_t(m)$, we obtain that $m_1 \leq_{\kappa_t} m$ for every $m\in Mod(\varphi)$, so $m_1 \in \min(Mod(\varphi), \leq_{\kappa_t})$. Thus, we proved that $ Mod(Bel(\kappa_t'))\subseteq \min(Mod(\varphi), \leq_{\kappa_t})$.

Now assume that $m_1 \in \min(Mod(\varphi), \leq_{\kappa_t})$. Then we have that: $m_1\in Mod(\varphi)$. Also, $m_1 \leq_{\kappa_t} m$ for every $m\in Mod(\varphi)$ and thus $\kappa_t(m_1) \leq \kappa_t(m)$ for every $m\in Mod(\varphi)$.
Consequently, $\kappa_t(m_1)=\kappa_t(\varphi)$, so $\kappa_t'(m_1)=\kappa_t(m_1)-\kappa_t(\varphi)=0$. Therefore, $m_1\in Mod(Bel(\kappa_t'))$, so $\min(Mod(\varphi), \leq_{\kappa_t})\subseteq  Mod(Bel(\kappa_t'))$.

Thus, we proved that $$ Mod(Bel(\kappa_t'))=\min(Mod(\varphi), \leq_{\kappa_t}).$$

It remains to prove that $\leq_{\kappa_t}$  satisfies the conditions 1--8 of Definition \ref{def:tbounded:faithfulassigment:epst}. Then, by Theorem \ref{thm:singlestepA},  $\spohn$ satisfies all the postulates. 

\begin{enumerate}
\item Assume that  $m_1,m_2\in Mod(Bel(\kappa_t))$. Then  $\kappa_t(m_1)=0$ and $\kappa_t(m_2)=0$, so 
 $m_1\leq_{\kappa_t} m_2$ and $m_2\leq_{\kappa_t} m_1$.
 
\item Assume that  $m_1\in Mod(Bel(\kappa_t))$ and  $m_2\notin Mod(Bel(\kappa_t))$.  Then  $\kappa_t(m_1)=0$ and $\kappa_t(m_2)\neq 0$, so  $m_1<_{\kappa_t} m_2$.

\item   Third condition follows directly from the definition of $\leq_{\kappa_t}$.

\item If $\restt=T_2\rest{}$, then $\kappa_t((T,\pi))=\kappa_t((T_2,\pi_2))$ by Definition \ref{def:spohnA}, so  $(T,\pi)\leq_{\kappa_t}  (T_2,\pi_2)$ and $(T_2,\pi_2)\leq_{\kappa_t}  (T,\pi)$.

\item -- $ \ $ 8.  $ \ $ follow directly from the definition of $\kappa_t'$.
\end{enumerate}
Now the result follows directly form Theorem \ref{thm:singlestepA}.
\end{proof}



\end{document}